\documentclass[a4paper,11pt]{article}
\usepackage{epsfig}
\usepackage{amsmath, amsfonts,amssymb,euscript,bbm}
\usepackage{cite}
\usepackage{ifthen}
\usepackage{graphicx}
\usepackage{sidecap}
\usepackage{color}
\usepackage{float}
\usepackage[section]{placeins}
\usepackage{wrapfig}
\usepackage{subcaption}
\usepackage{pbox}
\usepackage{bbm}
\usepackage[font=small,labelfont=bf]{caption}
\usepackage{framed}
\usepackage{multirow}
\usepackage{tensor}
\usepackage{array}
\usepackage{rotating}
\usepackage[]{siunitx}
\usepackage{hyperref}

\makeatletter
\def\spher#1{%
  \vbox{\hbox{%
    \offinterlineskip
    \valign{&\hb@xt@2\p@{\hss$##$\hss}\vskip.2ex\cr#1\crcr}%
  }\vskip-.36ex}%
}

\let\gsh\spher@harm
\makeatother
\DeclareMathAlphabet{\mathpzc}{OT1}{pzc}{m}{it}
\DeclareFontFamily{OT1}{pzc}{}
\DeclareFontShape{OT1}{pzc}{m}{it}{<-> s * [0.900] pzcmi7t}{}
\DeclareMathAlphabet{\mathpzc}{OT1}{pzc}{m}{it}
\DeclareFontFamily{U}{dutchcal}{\skewchar\font=45 }
\DeclareFontShape{U}{dutchcal}{m}{n}{<-> s*[1.0] dutchcal-r}{}
\DeclareFontShape{U}{dutchcal}{b}{n}{<-> s*[1.0] dutchcal-b}{}
\DeclareMathAlphabet{\mathlcal}{U}{dutchcal}{m}{n}
\SetMathAlphabet{\mathlcal}{bold}{U}{dutchcal}{b}{n}

\usepackage{colortbl}
\definecolor{summersky}{cmyk}{0.71,0.33,0,0.5}
\definecolor{flamingo}{cmyk}{0,0.51,0.71,0.5}
\definecolor{rp}{cmyk}{0.2, 1, 0.6, 0}
\definecolor{pacificblue}{cmyk}{0.95,0.3,0, 0.5}
\definecolor{gray60}{cmyk}{0.4,0.4,0,0.8}

\hypersetup{
    pdftoolbar=true,        
    pdfmenubar=true,        
    pdffitwindow=true,     
    pdfstartview={FitH},    
    pdfnewwindow=true,      
    colorlinks=true,       
    linkcolor=pacificblue,          
    citecolor=flamingo,        
    filecolor=magenta,      
    urlcolor=pacificblue           
}

\newcommand{\be}{\begin{eqnarray}}
\newcommand{\ee}{\end{eqnarray}}
\newcommand{\beq}{\begin{equation}}
\newcommand{\eeq}{\end{equation}}

\newcommand{\intk}{\int_{\bf k}}

\renewcommand{\v}[1]{\ensuremath{\mathbf{#1}}} 
\renewcommand\L{\mathcal{L}}
\renewcommand\H{\mathcal{H}}

\DeclareMathOperator{\Tr}{Tr}
\newcommand{\bin}[1]{\ensuremath{\mathlcal}{#1}} 

\newcommand{\Bth}{B^\text{th}}
\newcommand{\Ber}{B^\text{er}}

\def\d{{\rm d}}

\def\Tr{{\rm Tr}}

\def\d{{\rm d}}

\def\H{{\cal H}}

\def\0{{\boldsymbol 0}}
\def\k{{\boldsymbol{k}}}
\def\q{{\boldsymbol{q}}}

\def\x{{\boldsymbol{x}}}

\def\d{{\rm d}}

\def\Tr{{\rm Tr}}

\newcommand{\vev}[1]{\langle #1 \rangle}

\begin{document}

\begin{titlepage}

\setcounter{page}{1} \baselineskip=15.5pt \thispagestyle{empty}

\bigskip\

\vspace{1cm}
\begin{center}

{\fontsize{20}{28}\selectfont  \sffamily \bfseries  Lifting Primordial Non-Gaussianity\\[10pt] Above the Noise }

\end{center}

\vspace{0.2cm}

\begin{center}
{\fontsize{13}{30}\selectfont Yvette Welling$^{\clubsuit}$, Drian van der Woude$^{\spadesuit}$ and Enrico Pajer$^{\spadesuit}$}
\end{center}

\begin{center}

\textsl{$^\clubsuit$ Leiden Observatory, Universiteit Leiden, \\ Niels Bohrweg 2, 2333 CA Leiden, The Netherlands}

\textsl{$^\spadesuit$ Institute for Theoretical Physics, Utrecht University, \\
Princetonplein 5, 3584 CC Utrecht, The Netherlands}
\vskip 7pt

\end{center}

\vspace{1.2cm}
\hrule \vspace{0.3cm}
\noindent {\sffamily \bfseries Abstract} \\[0.1cm]
Primordial non-Gaussianity (PNG) in Large Scale Structures is obfuscated by the many additional sources of non-linearity. Within the Effective Field Theory approach to Standard Perturbation Theory, we show that matter non-linearities in the bispectrum can be modeled sufficiently well to strengthen current bounds with near future surveys, such as Euclid. We find that the EFT corrections are crucial to this improvement in sensitivity. Yet, our understanding of non-linearities is still insufficient to reach important theoretical benchmarks for equilateral PNG, while, for local PNG, our forecast is more optimistic. We consistently account for the theoretical error intrinsic to the perturbative approach and discuss the details of its implementation in Fisher forecasts.

\vskip 10pt
\hrule

\vspace{0.6cm}
 \end{titlepage}
\newpage


\tableofcontents

\section{Introduction and summary}
Primordial deviations from Gaussianity are key to understand inflation and will soon be tested via a number of ambitious Large Scale Structure (LSS) surveys. It is therefore imperative to understand how late-time LSS observations can be related to the parameters that characterize \textit{primordial non-Gaussianity} (PNG). This relation is complicated and non-linear. The degree to which we will be able to collect further primordial information from LSS survey will eventually be determined by our ability to model this non-linear relation. In this work, we focus on specific source of non-linearities, namely perturbative \textit{matter non-linearities}. These are generated by the sub-horizon gravitational evolution of small initial matter inhomogeneities into larger ones, until the perturbations compete locally with the homogeneous background expansion. For concreteness, we study local, equilateral and quasi-single field non-Gaussianity, since these are well-motivated theoretically and represent signals that are complementary from the point of view of observations. 
Additional sources of non-linearity are also important, such as for example bias and redshift space distortion. In case of equilateral and quasi single field PNG, these are expected to further worsen our ability to constraint primordial parameters. In this sense, our results can be interpreted as \textit{lower bounds} on the precision of future constraints.
For local PNG, it is possible that non-linearities encapsulated in the biasing of tracers, if very well understood, might eventually help us improve on the bounds we find here (see \cite{Dalal:2007cu} and e.g. \cite{Dore:2014cca,Tellarini:2016sgp} for a recent estimate). We will discuss this possibility in subsection \ref{subsubsec:currentandupcomingsurveys}.

In our analysis, we will use the Effective Field Theory of Large Scale Structures (EFT of LSS) \cite{Baumann:2010tm}, which builds on Standard Perturbation Theory (SPT) \cite{Bernardeau:2001qr}, and provides a consistent perturbative approach to describe the evolution of matter distribution. We focus exclusively on the matter bispectrum, since it is a primary probe of PNG that is affected by matter non-linearities. Recent work on PNG and the bispectrum includes \cite{Scoccimarro:2003wn,Sefusatti:2007ih,Baldauf:2010vn,Dore:2014cca,Jeong:2009vd,Tasinato:2013vna,Roth:2012yy}. Within the EFT approach, the bispectrum generated by the late-time gravitational evolution from otherwise Gaussian initial conditions has been studied in \cite{Baldauf:2014qfa,Angulo:2014tfa}). This contribution plays the role of background noise in PNG searches. The signal, namely the primordial bispectrum, is also affected by gravitational non-linearities. This has been recently studied in \cite{Assassi:2015jqa}. Here, we use these two results and present a Fisher forecast for constraints on PNG. A key element of our forecast is the inclusion of theoretical error, employing and further developing the recent proposal of \cite{Baldauf:2016sjb}. 

For the convenience of the reader, we collect here our major findings with references to where they are discussed in the rest paper.
\begin{itemize}
\item When using the EFT of LSS, the perturbative approach to model matter non-linearities will not prevent upcoming LSS surveys to \textit{improve upon the current bounds} from CMB anisotropies \cite{Ade:2015ava} (see Table \ref{table:fullresult}).
\item Our limited perturbative understanding of matter non-linearities limits the achievable bounds on equilateral non-Gaussianity from planned galaxy surveys to $ \sigma(f_{NL}^{eq})\gtrsim 10 $ (see Table \ref{table:fullresult}), \textit{far from the theoretically interesting benchmark} $ \sigma(f_{NL}^{eq})\sim 1 $ (see e.g. \cite{Baumann:2014cja,Alvarez:2014vva} and references therein). We estimate that this remains true even if one included the full two-loop corrections (see Table \ref{table:2loopvs3loop}).
Local non-Gaussianity is more promising, and we find e.g. for Euclid $ \sigma(f_{NL}^{loc})\gtrsim 1 $.
\item The consistent treatment of short-scale effects within the EFT approach allows one to \textit{improve PNG constraints by a factor of about 3} (see Table \ref{table:sptversuseft}). This relies on two facts. First, the EFT parameters provide a better description of the late-time gravitational non-linearities (the ``background'' discussed in \cite{Baldauf:2014qfa}). Second, for the specification of most upcoming experiments, the EFT parameters are only weakly correlated with PNG, and so marginalizing over them hardly degrades the constraints (see subsection \ref{subsec:correlations}).
\item Both the SPT loops and the EFT corrections to primordial non-Gaussianity (the ``signal'' discussed in \cite{Assassi:2015jqa}) are small and their inclusion does \textit{not} improve the PNG constraints appreciably (see first and second line of Table \ref{table:sptversuseft}). 
\item We discuss several aspects of the method proposed in \cite{Baldauf:2016sjb} to \textit{model the theoretical error} inherent to the perturbative approach. We show that a wrong shape for the theoretical error can lead to a biased estimate for $ f_{NL} $ . This happens when it partly underestimates the error. Conversely, a conservatively large theoretical error leads to correct unbiased results. 
We thoroughly analyze the dependence of the Fisher forecast on the correlation length used in \cite{Baldauf:2016sjb}, and explain our results with a toy model.
\end{itemize}

This paper is organized as follows. In section \ref{sec:theoreticaldescriptionbispectrum}, we review the results for the matter bispectrum in the EFT of LSS accounting for PNG. In section \ref{sec:fisheranalysis}, we discuss the details of the Fisher forecast with particular emphasis on a consistent treatment of theoretical uncertainties. Section \ref{sec:results} is devoted to a discussion of the results of the Fisher forecast on PNG constraints from LSS surveys. We conclude in section \ref{sec:conclusionanddiscussion}. Several appendices contain more technical results. In Appendix \ref{app:eftresults}, we summarize all relevant formulae to compute the bispectrum in the EFT of LSS. In Appendix \ref{app:theoreticalnoise}, we present a detailed discussion of how to consistently account for theoretical errors. Appendix \ref{app:choicebinning} discusses the issue of binning the data for the Fisher forecast and finally, for the convenience of the reader, we collected all symbols used in this paper and their meaning in Appendix \ref{app:tableparameters}.

\paragraph{Conventions} Redshift $z$ and conformal spatial coordinates $\v{x}$ are used as measures of time and position. We use the following convention for the Fourier transform
\begin{equation}
F(\v{x})=\intk \tilde{F}(\v{k})e^{i\v{k}\cdot \v{x}}, \quad \text{where we use the shorthand} \quad \intk\equiv  \int \frac{d^3\v{k}}{(2\pi)^3}.
\end{equation}
In particular, this implies that we have the following relation between any $N$-point equal-time correlation function and its spectrum
\begin{equation}
\langle \delta(\v{k}_1)\ldots\delta(\v{k}_N)\rangle = (2\pi)^3\delta_D(\v{k}_1+\ldots+\v{k}_N)S(\v{k}_1, \ldots, \v{k}_N),
\end{equation}
where we suppressed the time dependence. 

For the numerical analysis, we compute the linear power spectrum with CAMB \cite{Lewis:2002ah}, where we assume a standard cosmological model with $\Omega^0_\Lambda = 0.728$, $\Omega^0_m = 0.272$, $h=0.704$, $n_s = 0.967$ and $A_\zeta = 2.46\times 10^{-9}$. 


\section{Analytical predictions for the bispectrum}
\label{sec:theoreticaldescriptionbispectrum}
In this section, we review the analytical predictions for the late-time matter bispectrum within the Effective Field Theory of Large Scale Structures (EFT of LSS), accounting for non-Gaussian initial conditions. In subsection \ref{subsec:bispectruminEFToLSS} we collect the contributions to the bispectrum up to first order in primordial non-Gaussianity and up to `one-loop' order in perturbation theory. In subsection \ref{subsec:shapesPNG}, we specify the types of primordial non-Gaussianity (PNG) we study in this paper. In subsection \ref{subsec:theoreticalerror}, we discuss the theoretical errors, which are intrinsic to the perturbative approach. 


\subsection{The bispectrum in the EFT of LSS}
\label{subsec:bispectruminEFToLSS}

Despite being almost collisionless, cold dark matter on large scales behaves approximately as a fluid. This relies on the fact that the primordial universe is locally in (thermodynamical) equilibrium \textit{and} that, during the age of the universe, dark matter particles move only over a distance that is small compared with the scales of interest. This displacement plays the same role as the mean free path in the more familiar interacting fluids. As long as we consider scales much larger than this displacement, an effective fluid description can be applied \cite{Baumann:2010tm}. Here we follow the Effective Field Theory approach to Large Scale Structures (EFT of LSS). The dark matter bispectrum induced by gravity was discussed in \cite{Baldauf:2014qfa, Angulo:2014tfa}. Non-Gaussian initial condition were subsequently accounted for in \cite{Assassi:2015jqa}. The EFT of LSS allows to perturbatively compute non-linear correlators of the matter density contrast $\delta(\v{x}, z)$ \cite{Carrasco:2012cv} and velocity $ v(\v{x},z) $ \cite{Mercolli:2013bsa}, taking into account the effect of short-scale non-perturbative physics on the large-scale dynamics. In practice, one can use the results of Standard Perturbation Theory (SPT) \cite{Bernardeau:2001qr}, and correct them with additional effective terms, which will be denoted with the subscript `EFT'. We differentiate between contributions to the bispectrum coming from primordial non-Gaussianities (superscript ``NG'') and those coming from the late-time gravitational evolution (superscript ``G''). Schematically, the perturbative theoretical prediction for the bispectrum is
\begin{equation}
B^{\text{th}} = B^{\text{G}}_{\text{SPT}} + B^{\text{G}}_{\text{EFT}} + f_{\text{NL}} \left(B^{\text{NG}}_{\text{SPT}} + B^{\text{NG}}_{\text{EFT}}\right).
\label{eq:Btot}
\end{equation}
As we will see in section \ref{sec:fisheranalysis}, for a Fisher forecast we do not need to specify\footnote{On the other hand, we do need to specify the SPT contributions to the power spectrum to compute the cosmic variance. Assuming it is dominated by the linearly evolved matter power spectrum, we do not have to specify additional `EFT' parameters.} $B^{\text{G}}_{\text{SPT}}$. The leading order counterterms for Gaussian initial conditions have been computed in \cite{Baldauf:2014qfa,Angulo:2014tfa} and read
\begin{equation}
B^{\text{G}}_{\text{EFT}}= \xi B_\xi^{\text{G}} +  \sum_{i = 1}^3\epsilon_i B^{\text{G}}_{\epsilon_i}. \label{eq:BGEFT}
\end{equation}
For non-Gaussian initial conditions, short modes and long modes are already correlated at the initial time. This leads to additional contributions to the matter bispectrum. To leading order, these are given by \cite{Assassi:2015jqa}
\begin{equation}
B^{\text{NG}}_{\text{EFT}}  = \xi B_\xi^{\text{NG}} + \gamma B^{\text{NG}}_\gamma + \sum_{i = 1}^2\gamma_i B^{\text{NG}}_{\gamma_i}.\label{eq:BNGEFT}
\end{equation}
For convenience, we have adopted the notation of \cite{Assassi:2015jqa} and collected in appendix \ref{app:eftresults} all the explicit expressions for the terms appearing in this subsection.


\subsection{Primordial non-Gaussianity}
\label{subsec:shapesPNG}
To evaluate the non-Gaussian contributions to \eqref{eq:Btot}, we need to specify the primordial bispectrum. In this paper, we study the constraints on three types of primordial non-Gaussianity: local \cite{Bartolo:2001cw}, equilateral \cite{Creminelli:2003iq} and quasi-single field \cite{Chen:2009zp}. In terms of the primordial potential $\phi$, the primordial bispectra are given by the following shapes
\begin{subequations}
\begin{align}
  B_\phi^{\text{loc}}(k_1, k_2, k_3) &= 2f_{\text{NL}}^{\text{loc}}\left(P_\phi(k_1)P_\phi(k_2) + \text{perm}\right),
\label{eq:localshape}\\
  B_\phi^{\text{eq}}(k_1, k_2, k_3) &= 162f_{\text{NL}}^{\text{eq}}A^2_\phi\frac{1}{k_1k_2k_3 K^3},
\label{eq:equilateralshape}\\
 B_\phi^{\text{qsf}}(k_1, k_2, k_3) &= 18\sqrt{3}f_{\text{NL}}^{\text{qsf}}A^2_\phi\frac{1}{k_1k_2k_3 K^3}
\label{eq:qsfshape}\frac{N_\nu(8\kappa)}{\sqrt{\kappa}N_\nu(8/27)}.
\end{align} 
\end{subequations}
Here we define  $K = k_1+k_2 +k_3$, and $\kappa = k_1k_2k_3/K^3$. Moreover, $N_\nu$ is the Neumann function of order $\nu$ and we choose $\nu = \tfrac{1}{2}$. The normalization of the primordial power spectrum\footnote{Note that we define $A_\phi = 2\pi^2\tfrac{9}{25}A_\zeta$.} is given by $A_\phi =1.72\cdot 10^{-8}$. To linearly evolve these to the late time matter bispectrum\footnote{See appendix \ref{app:eftresults} for relevant notation.} $B_{111}$, we use the transfer function $M(k,z)$, defined by
\begin{equation}
\delta_1(k,z) = M(k,z)\phi(k), \quad \text{with} \quad M^2(k,z) \equiv\frac{k^3 P_{11}(k,z)}{A_\phi \left(\frac{k}{k_\star}\right)^{n_s-1}}.
 \label{eq:transferfunction}
\end{equation}
Here $k_\star = 0.0028\ h \text{Mpc}^{-1}$ and $n_s = 0.967$. This means we have
\begin{equation}
B_{111}(k_1, k_2, k_3, z) = M(k_1, z)M(k_2, z)M(k_3, z)B_\phi(k_1, k_2, k_3).
\end{equation}
We collect all relevant higher order non-Gaussian contributions to the bispectrum in appendix \ref{app:eftresults}.


\subsection{Theoretical error}
\label{subsec:theoreticalerror}
\begin{figure}[h]
\begin{subfigure}{0.4\textwidth}
\includegraphics[width=\linewidth]{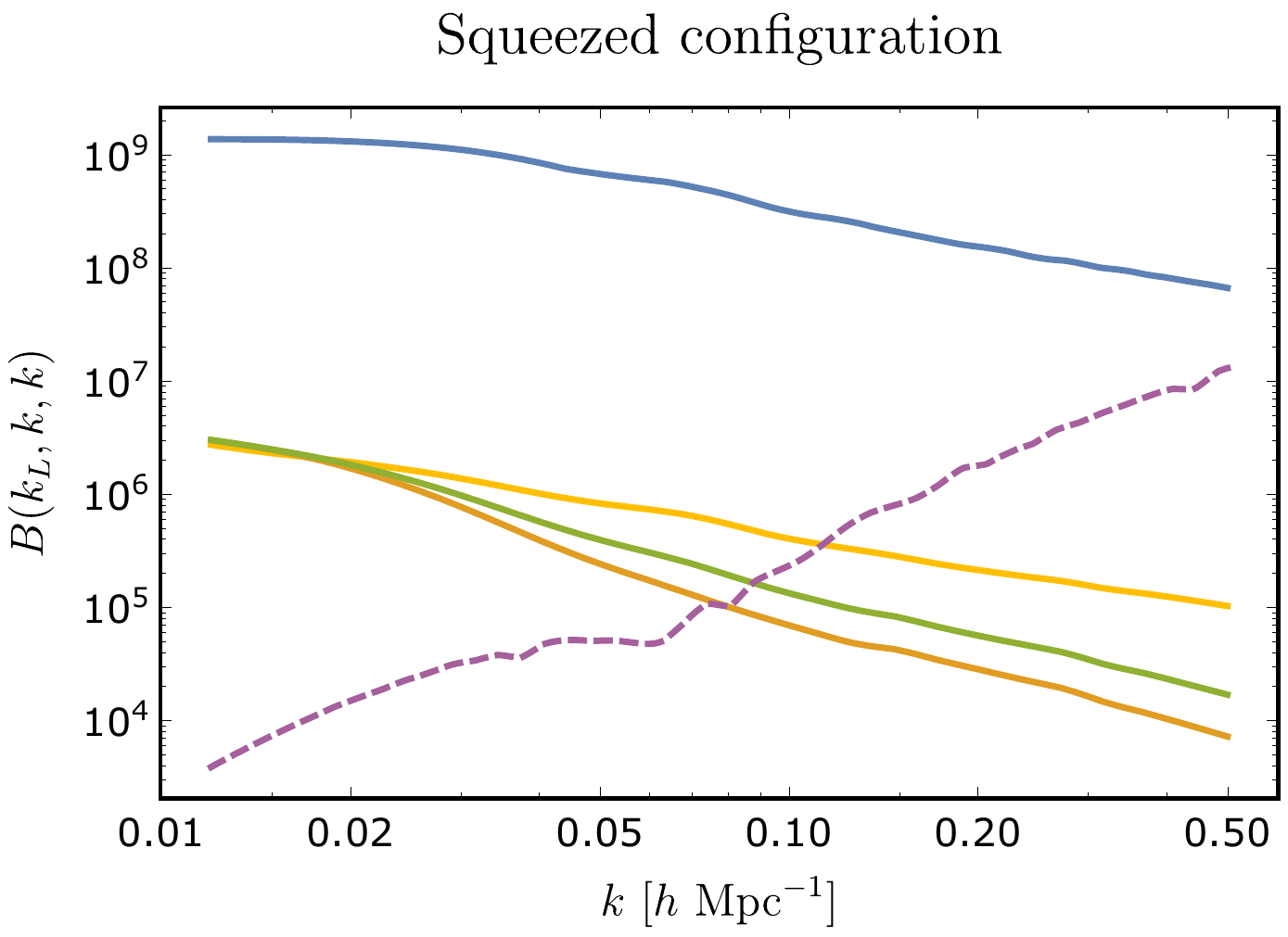} 
\end{subfigure}
\begin{subfigure}{0.4\textwidth}
\includegraphics[width=\linewidth]{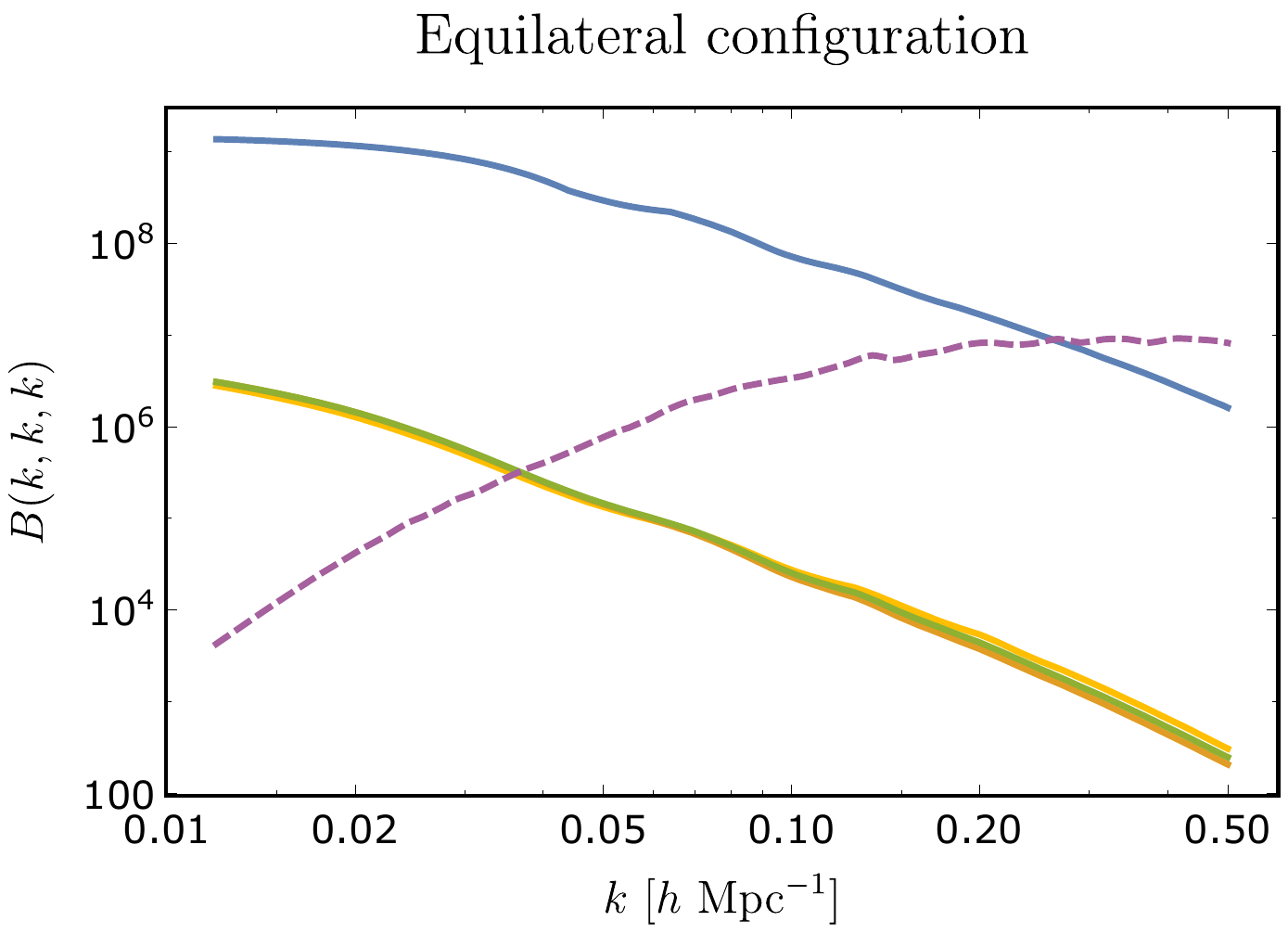} 
\end{subfigure}
\begin{subfigure}{0.18\textwidth}
\includegraphics[width=\linewidth]{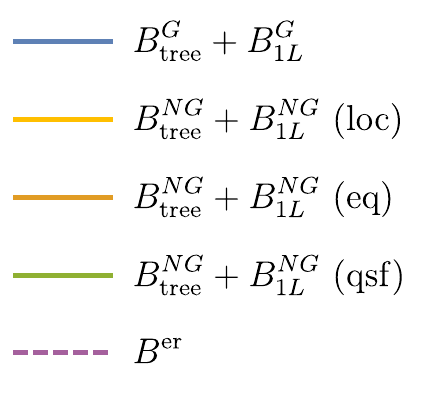}
\end{subfigure}
\caption{The SPT contributions to the bispectrum in the squeezed (left) and equilateral (right) configurations. The blue solid line denotes the Gaussian tree-level and one-loop contributions. The yellow, orange and green lines denote the one loop non-Gaussian contribution for $f_{\text{NL}}=1$ for local, equilateral and quasi-single-field PNG, respectively. The dashed purple line corresponds to our order-of-magnitude estimate for the Gaussian two-loop correction $B_{332}$. In the squeezed configurations (left), we chose $k_L = 0.012\ h \text{Mpc}^{-1}$.}
\label{fig:bispectrumplot}
\end{figure}

By definition, any results from perturbation theory are approximate - there is always an intrinsic theoretical error, typically estimated within perturbation theory itself. The true bispectrum is therefore given by 
	\begin{align}
	B^{true}=B^{th}+B^{er},
	\end{align}	
where $B^{th}$ is the perturbative theoretical prediction given in \eqref{eq:Btot}, and $B^{er}$ represents the theoretical error. The strength of a well defined perturbation theory is of course that the error can be estimated within the perturbation theory itself. \\
In our case, there are in principle two perturbative schemes employed. First, we assume perturbative primordial non-Gaussianity. This means we assume the primordial potential can be schematically expanded as 
	\begin{align}
	\varphi_{p}=\varphi_{p}^{\text{G}}+f_{\text{NL}}\varphi_{p}^{\text{G}}\star \varphi_{p}^{\text{G}}+\ldots\,.
	\end{align} 
Here $\varphi_{p}^{G}$ is a Gaussian field and $\star$ denotes a convolution in Fourier space. This means we are effectively expanding in $f_{\text{NL}}\varphi_{p}\sim f_{\text{NL}} \sqrt{A_\phi}$, which is indeed very small according to current bounds. Hence we will not worry about corrections to this approximation for the rest of the paper. \\
Second, the EFT of LSS relies on the smallness of density perturbations on large scales, consistently taking into account our ignorance of short scale physics. Effectively, this comes down to an expansion in $k/k_{\text{NL}}$ \cite{Baumann:2010tm}. As argued in \cite{Baldauf:2014qfa}, the most relevant correction to $B^{th}$ is the two-loop bispectrum. Since we have not computed the full two-loop bispectrum, we are forced to make an educated guess about its size and shape. One way to do this was proposed in \cite{Baldauf:2016sjb}, and relies on the scaling universe results of \cite{Pajer:2013jj}. Here we use instead a different estimate. Unless indicated otherwise, we estimate the two-loop bispectrum by adding up the absolute values of the two two-loop diagrams we can compute, namely the so-called reducible diagrams, which we indicate by $B_{332}$. An explicit expression for $B_{332}$ is again given in appendix \ref{app:eftresults}. We compare our estimate to the scaling estimate of \cite{Baldauf:2016sjb} in appendix \ref{appsub:comparisonansatze}.\\
The importance of keeping track of the theoretical error for forecasts has recently been stressed in \cite{Baldauf:2016sjb}, and we build on their approach. Qualitatively, one expects not to be able to learn much about $f_{\text{NL}}$ from bispectrum configurations for which $B^{er}$ is larger than the non-Gaussian signal. To get an idea of the configurations for which this is the case, we plot the one-loop expressions for the three types of non-Gaussianities we consider (with $f_{\text{NL}}=1$) and $B_{332}$ as a function of scale for both squeezed and equilateral configurations in Figure \ref{fig:bispectrumplot}. For reference we also plotted the one-loop Gaussian contribution to the bispectrum. As expected, for local PNG we can push to smaller scales in the squeezed configuration than for equilateral PNG. Note also that the naive $k_{max}$, beyond which we do not expect to gain any more signal, is \textit{configuration dependent}. A detailed discussion on how to incorporate this theoretical error in a Fisher analysis is given in section \ref{subsec:theoreticalerror}, which proceeds along the lines of \cite{Baldauf:2016sjb}. In appendix \ref{app:theoreticalnoise}, we present further investigation of the validity of this method of treating the theoretical error.

\section{Fisher analysis}
\label{sec:fisheranalysis}
In this section, we outline our method to forecast constraints on primordial non-Gaussianity. We have in mind a Gedankenexperiment that provides us with the matter distribution in space and time up to some maximal redshift. In this highly idealized scenario, we determine to what extent our inability to analytically describe the non-linear gravitational collapse of matter limits the information we can extract on primordial perturbations. We proceed along the lines of \cite{Scoccimarro:2003wn,Sefusatti:2006pa,Sefusatti:2007ih}. The outcome of the analysis for various surveys is presented in the next section.


\subsection{Assumptions and approximations}
\label{subsec:assumptions}

For the convenience of the reader, we summarize the assumptions and approximations we make in the Fisher analysis. 
\begin{itemize}
\item We assume we are given an idealized \textit{survey of the late time dark matter density field}, instead of that of some biased tracer. This allows us to answer the question of whether further progress is needed in the modeling of the dark matter distribution to strengthen current bounds on PNG using upcoming LSS surveys.
\item The idealized dark matter survey is characterized by a redshift range and the fraction of the sky covered. We divide the survey in redshift bins, to which we assign a fixed time that is equal to the mean redshift of the bin. Hence, we only need to know $z_{\text{bin}}$ to predict the power spectrum and bispectrum. Observational redshift errors are neglected. 
\item  We assume that each redshift bin can be approximated by a cube. Then we just need the volume of the bin $V(z_{\text{bin}})$ to account for cosmic variance.
\item We compute correlation functions only within each bin. This does not seem to be a big drawback in the case of equilateral PNG. Instead, for local PNG, this might cause an unnecessary loss of information. We will discuss this issue elsewhere.
\item  We include shotnoise in the analysis to correctly remove weight from the higher redshift bins. For this, we use the specifications of specific upcoming surveys. We discuss this in section \ref{subsubsec:currentandupcomingsurveys}.
\item We assume that the bispectra for different configurations are uncorrelated with each other. This means that we approximate the \textit{bispectrum covariance matrix as diagonal}. In \cite{Sefusatti:2006pa} it has been checked that this approximation works fine for the scales $k \leq 0.3 h \text{Mpc}^{-1}$ at redshift zero. We assume it holds up to $k \leq 0.4 h \text{Mpc}^{-1}$, since for local PNG we still gain signal up to this scale, as we see in Figure \ref{fig:freezeoutsigma}. Moreover, we assume that only the linear power spectrum determines the covariance matrix (see subsection \ref{subsec:covariancematrix} for more details). Finally, we neglect covariance due to observational effects, such as survey geometry and mask. 
\end{itemize}
Importantly, we parameterize the theoretical error by treating the higher loop corrections to the bispectrum as a source of noise, which we integrate out. This contributes to the effective covariance matrix. Our estimate of the typical size of the two-loop corrections is given by $B_{332}$, defined in Appendix \ref{app:eftresults}. \\
The time-dependence of the counterterms has been chosen to match the one loop diagrams they are supposed to renormalize \cite{Baldauf:2015aha} (see also \cite{Foreman:2015lca} for a related discussion). All the time dependence is absorbed in the contributions to the bispectrum, so that all the theoretical parameters become time-independent (see appendix \ref{app:eftresults}).  This means that we are measuring the same theoretical parameters in each redshift bin.\\
We need to discretize the bispectrum in order to compute the Fisher matrix. We will use logarithmic bins instead of linear bins, since we do not fully trust linear binning. We refer the reader to the appendix \ref{app:choicebinning} for more details. Finally, we do not marginalize over the standard cosmological parameters, but fix their values.


\subsection{Fisher matrix}
\label{subsec:fishermatrix}
In a Fisher forecast, one computes the expected curvature around the maximum of the likelihood. The likelihood is given by
\begin{eqnarray}
\L(\text{data} | \Theta, \text{ priors}) &=& \frac{1}{\sqrt{(2\pi)^N\det(C_B)}} P_\text{prior}(\Theta)\\
&&\quad \times \exp\left(-\frac{1}{2}\sum_{\bin{k}\bin{p}}\Delta B(\bin{k},\Theta)C^{-1}_B(\bin{k}, \bin{q},\Theta)\Delta B(\bin{p},\Theta)\right)\,,\nonumber
\end{eqnarray}
where $\Theta$ denote the set of theoretical parameters and $N$ is the number of datapoints. We suppressed the time dependence. Here we use the Dutch calligraphic lower case symbols $ \bin{k} $ as a shortcut for a triplet of wavenumbers on which the bispectrum depends, i.e. $\bin{k}=(k_1, k_2, k_3)$. Furthermore, $\Delta B$ corresponds to the difference between estimator and theoretical prediction $\Delta B(\bin{k},\Theta) = \hat{B}(\bin{k}) - B(\bin{k},\Theta)$ and $C_B$ is the covariance matrix of the bispectrum $C_B =\langle\Delta B\Delta B\rangle$. 
 
Neglecting the theoretical error for the moment, the theoretical prediction for the bispectrum is given in equation \eqref{eq:Btot}. The 8 parameters we include in the Fisher analysis are therefore $\{f_{\text{NL}},\ \xi,\ \epsilon_1,\ \epsilon_2,\ \epsilon_3,\ \tilde{\gamma},\ \tilde{\gamma}_1,\ \tilde{\gamma}_2\}$\footnote{Here we denote $\tilde{\gamma} =  f_{\text{NL}} \cdot \gamma$ and similarly for $\gamma_i$, so that the bispectrum is linear in all parameters. This is convenient for the Fisher analysis, as this makes the result independent of the fiducial values of the parameters. On the other hand, we effectively assume that the one-loop non-Gaussian counterterms have amplitudes independent of $f_{NL}$. Later in this paper we will find that these counterterms are negligibly small, therefore, this will not affect our results. }. The parameter $\xi$ also appears in the power spectrum and has been measured to be\footnote{Previous measurement gave $(1.62 \pm 0.03)\ h^{-2}\text{Mpc}^2 $\cite{Carrasco:2013mua} and $(1.5 \pm 0.03)\ h^{-2}\text{Mpc}^2 $ \cite{Baldauf:2014qfa}, but neglected two-loop corrections.}  $0.98 \ h^{-2}\text{Mpc}^2 $ \cite{Baldauf:2015aha}. 
Therefore, we can put a sharp prior on this parameter. The other parameters are unknown, but we expect them to be of the same order of magnitude (see \cite{Assassi:2015jqa,Baldauf:2014qfa} for naive numerical estimates). Therefore, we can use a fiducial value of zero and a Gaussian prior with variance of 10. 

For simplicity, we assume that all priors are Gaussian with covariance matrix $C_\Theta$. Then the Fisher matrix is given by (see e.g. \cite{Heavens:2009nx})
\begin{equation}
F_{ij} \equiv -\left\langle \frac{\partial^2 \log(\L)}{\partial \Theta_i \partial\Theta_j}\right\rangle = \frac{1}{2}\Tr\left[C_B^{-1}C_{B,i}C_B^{-1}C_{B,j}\right] + B_i^TC_B^{-1}B_j + \left(C^{-1}_{\Theta}\right)_{ij}.
\label{eq:fishermatrix}
\end{equation}
As we will see in a moment, our approximation for the covariance matrix does not depend on the theoretical parameters. Writing out the time dependence explicitly, the Fisher matrix simplifies to
\begin{equation}
F_{ij}(z) =  \sum_{\bin{k}, \bin{k'}} B_i^T(\bin{k}, z) C_B^{-1}(\bin{k},\bin{k'}, z)B_j(\bin{k'}, z)+ \left(C^{-1}_{\Theta}\right)_{ij}(z)
\label{eq:fishermatrixlinear}
\end{equation}
for each redshift bin. Since the bispectrum is linear in all parameters - taking into account that $\xi$ has been measured - the Fisher matrix is independent of the fiducial value of the theoretical parameters\footnote{To be more precise, the Fisher matrix is independent of the fiducial values of the parameters \textit{to good approximation}. We choose a fiducial value for $\xi$ of zero and in Section \ref{sec:results} we either specify a prior with $\sigma =1 $ for $\xi$ or no prior at all. Therefore, the Fisher matrix has some dependence on the choice of $\xi$, but it will come exclusively from the non-Gaussian counterterm proportional to $\xi$. Again, since the non-Gaussian counterterms turn out to be extremely small, we expect this not to affect the results. Moreover, we have checked this explicitly by changing its fiducial value to 1.}, which is very convenient for the analysis. To combine the results from the different redshift bins, we use that the parameters are the same in each bin, since we have fixed their time dependence. This time dependence is chosen to match the time dependence of the divergences they are supposed to cancel, motivated by \cite{Baldauf:2015aha}. The explicit expression can be found in Appendix \ref{app:eftresults}. 
Therefore, we can compute the constraints on $f_{\text{NL}}$ by summing the Fisher matrices and then marginalizing over the EFT parameters i.e.,
\begin{equation}
 \sigma(f_{\text{NL}}) = \sqrt{\left(\sum_z F_{ij}(z)\right)^{-1}_{11}},
\end{equation}
where we assumed that the entry of the Fisher matrix belonging to $f_{\text{NL}}$ is the first. Note that we have not included cross correlations between bins, which means we might be throwing away valuable information. This effect will be studied elsewhere \cite{Pajer:tobepublished}.

\subsection{Covariance matrix}
\label{subsec:covariancematrix}
To evaluate the Fisher matrix, we need to know the covariance matrix. 
Let us shortly review the derivation of the bispectrum covariance matrix. The estimator of the bispectrum is given by \cite{Scoccimarro:1997st}
\begin{equation}
\hat{B}(k_1,k_2,k_3,z) = \frac{1}{ V(z) V_{123}}\int_{\v{q}_1}\int_{\v{q}_2}\int_{\v{q}_3}\ \delta(\v{q}_1,z)\delta(\v{q}_2,z)\delta(\v{q}_3,z)\delta_D^{3}(\v{q}_1+\v{q}_2+\v{q}_3),
 \label{eq:estimatorbispectrum}
\end{equation}
with $V(z)$ the volume of the bin. The integration is over logarithmic bins centered around the given wavenumbers, i.e. $\ln(|\v{q}_i|) \in [\ln(k_i)-\tfrac{1}{2}\Delta\ln(k),\ln(k_i)+\tfrac{1}{2}\Delta\ln(k)]$. Moreover, $V_{123}$ corresponds to the following $k$-space volume squared
\begin{equation}
V_{123}= \int_{\v{q}_1}\int_{\v{q}_2}\int_{\v{q}_3}\ \delta_D(\v{k}_1+\v{k}_2+\v{k}_3) \approx \frac{8 \pi^2}{(2\pi)^9}  k_1^2 k_2^2 k_3^2 \sinh^3(\Delta \ln k).
\label{eq:V123}
\end{equation}
This approximation becomes exact when we consider `internal' bins, but it fails on the `edge' bins. In the numerical analysis we compute the exact value of $V_{123}$ for each bin, see appendix \ref{app:choicebinning} for more details.
This allows us to compute the bispectrum covariance matrix, namely 
\begin{eqnarray}
C_B(\bin{k}, \bin{k'}, z) &=& \left\langle\left(\hat{B}(\bin{k}, z)-B(\bin{k}, z)\right)\left(\hat{B}(\bin{k'}, z)-B(\bin{k'}, z)\right)\right\rangle \\
&\approx& \frac{s_{123}}{(2\pi)^3V(z) V_{123}}P(k_1, z)P(k_2, z)P(k_3, z)\delta_{\bin{k},\bin{k'}}.
\label{eqn:covariancematrix}
\end{eqnarray}
There is a factor $s_{123}$ which counts the number of non-vanishing contractions when computing $\langle\hat{B}(\bin{k})\hat{B}(\bin{k})\rangle$, which depends on the type of triangle that the triplet $\bin{k}$ forms. As each contraction comes with a delta function, this counting factor equals 6, 2 or 1 for equilateral, isosceles and scalene triangles respectively.
If we include shotnoise in the covariance matrix, we replace $P(k_i, z)$ with $P(k_i, z) + \frac{1}{\bar{n}}$  in equation \eqref{eqn:covariancematrix}, where $\bar{n}$ is the effective number density for the density contrast. This will be explained more when we include shotnoise in Section \ref{subsec:constraintsasfunctionofzmax}.

In this expression for the covariance matrix, we completely neglected higher order corrections beyond the power spectrum, making it approximately diagonal. In \cite{Sefusatti:2006pa} it has been checked that this approximation works fine for the scales we are considering. The off-diagonal terms become important exactly when the higher order corrections to the power spectrum become important, since they are of the same order. Therefore, in order to be consistent, we only take into account the linear contribution $P_{11}$  to the power spectrum $P(k_i, z)$. In particular, this means that the covariance matrix is independent of the theoretical parameters.  


\subsection{Theoretical error as nuisance parameters}
\label{subsec:theoreticalnoise} 
To account for the theoretical error inherent to the perturbative expansion, we parameterize the bispectrum as
\begin{eqnarray}
B(\bin{k}) &=& \Bth(\bin{k})+n(\bin{k}) \Ber(\bin{k})\,, \\
\Bth(\bin{k}) &=& B^{\text{G}}_{\text{SPT}}(\bin{k}) + B^{\text{G}}_{\text{EFT}}(\bin{k}) + f_{\text{NL}} \left[B^{\text{NG}}_{\text{SPT}}(\bin{k}) + B^{\text{NG}}_{\text{EFT}}(\bin{k})\right]\,,
\end{eqnarray}
where $ \Bth $ represents the theoretical prediction up to some order in perturbation theory as before, and $ \Ber $ is the estimate of the theoretical error. Following \cite{Baldauf:2016sjb}, we introduce a series of nuisance parameters $ n(\bin{k}) $, one per bin in $ k $-space. The reason we implement the theoretical error this way, instead of proposing some $k_{max}$ is that, as discussed in subsection \ref{subsec:theoreticalerror}, $k_{max}$ depends on where the theoretical error and the signal become comparable. This complicates the analysis in two ways. First, $k_{max}$ is configuration dependent, and second, it depends on the fiducial value of $f_{\text{NL}}$, which makes finding the error on $f_{\text{NL}}$ a recursive problem. In the approach we take, the set of theoretical parameters thus becomes $\Theta = \{n(\bin{k})\}_{\bin{k}}\cup \{f_{\text{NL}},\ \xi,\ \epsilon_1,\ \epsilon_2,\ \epsilon_3,\ \tilde{\gamma},\ \tilde{\gamma}_1,\ \tilde{\gamma}_2\}$. Since the bispectrum remains linear in all parameters, expression \eqref{eq:fishermatrixlinear} for the respective block of the Fisher matrix still applies. 

We assume that the true corrections to the bispectrum are of similar size as $\Ber(\bin{k})$. Therefore, we put a Gaussian prior on the parameters $n(\bin{k})$, with mean zero and variance one. Moreover, we expect the correction to have a smooth shape, which varies not too rapidly within the contours defined by $\Ber(\bin{k})$. Therefore, the coefficients should have non-negligible cross correlations. Since we would like to have an increasing correlation for nearby points, we include cross-correlations with a typical correlation length as follows
\begin{equation}
N_{\alpha\beta} = \exp\left(-\frac{\sum_i\ln(| k^i_\alpha/k^i_\beta |)}{l}\right).
 \label{eq:covariancetheoreticalnoise}
\end{equation}
We replaced the label $\bin{k}_\alpha$ of the nuisance parameter of a given bin with the index $\alpha$, so that we can reserve latin indices for the other theoretical parameters. Moreover, in order not to confuse this covariance matrix with the covariance matrix of the bispectrum $C_B$, we denote it as $N_{\alpha\beta}$. Note that we choose $\sigma_\alpha =1$ for all $\alpha$'s. Here, $l$ denotes the logarithmic correlation length. We could have also chosen a quadratic correlation length, similar to modeling it as a random field \cite{randomfield_notes}. The reason we opted for this form is that here the inverse matrix is very sparse, which is convenient for numerical purposes. Since our final results are quite insensitive to the correlation length (see \ref{app:correlations}), we do not believe this choice affects the results very much. 

Since we introduced a set of new nuisance parameters, we should write down the full Fisher matrix $F_{\mu\nu}$ and invert it
\begin{equation}
 F_{\mu\nu}^{-1} =
\begin{pmatrix}
 F_{\alpha\beta} & F_{\alpha j} \\
  F_{i \beta} & F_{ij}
 \end{pmatrix}^{-1} 
 = 
 \begin{pmatrix}
 \bullet &   \bullet \\
 \bullet & (F_{ij}  -  F_{i \gamma} F^{-1}_{\gamma\delta} F_{\delta j})^{-1}
 \end{pmatrix},
 \label{eq:fullfishermatrix}
\end{equation}
where we use latin indices for the parameters $\{f_{\text{NL}},\ \xi,\ \epsilon_1,\ \epsilon_2,\ \epsilon_3,\ \tilde{\gamma},\ \tilde{\gamma}_1,\ \tilde{\gamma}_2\}$ and greek indices from the early alphabeth for the theoretical error parameters $\{n(\bin{k}_\alpha)\}_\alpha$. We did not write out explicitly the other entries, since we are only interested in the effective Fisher matrix, after marginalizing over the nuisance parameters coming from the theoretical uncertainty. To compute the effective Fisher matrix, we need to know $F_{\alpha\beta}$ and $F_{\alpha i}$. 
Since the derivative of the bispectrum with respect to the nuisance parameters $\Theta_\alpha$ is only non-zero for the corresponding bin, these contributions to the Fisher matrix are particularly simple. We have
\begin{equation}
F_{\alpha \beta } =  \Ber(\bin{k}_\alpha) C^{-1}_B(\bin{k}_\alpha,\bin{k}_\beta)  \Ber(\bin{k}_\beta)  + N^{-1}_{\alpha\beta} \equiv D_{\alpha\beta}+ N^{-1}_{\alpha\beta}\,,
\label{eq:fisheralphabeta}
\end{equation}
and similarly
\begin{equation}
F_{i \beta } = \sum_{\bin{k}} B_{i}(\bin{k}) C_B^{-1}(\bin{k},\bin{k}_\beta)  \Ber(\bin{k}_\beta)  \quad \text{and} \quad F_{\alpha j } = \sum_{\bin{k}}  \Ber(k_\alpha)(\bin{k}_\alpha) C_B^{-1}
(\bin{k}_\alpha,\bin{k})  B_{j}(\bin{k}).
\label{eq:fisheralphai}
\end{equation}
This allows us to compute $F_{ij}^{\text{eff}}$ by using \eqref{eq:fullfishermatrix}. After some algebraic manipulations, we can rewrite it in the simple form
\begin{equation}
F_{ij}^{\text{eff}} = \sum_{\bin{k}, \bin{p}} B_i(\bin{k}) \left( N^{\text{eff}}(\bin{k}, \bin{p}) +C_B(\bin{k}, \bin{p})  \right)^{-1} B_j(\bin{p})+\left(C^{-1}_{\Theta}\right)_{ij},
\label{eq:effectivefishermatrix}
\end{equation}
with $N_{\alpha\beta}^{\text{eff}} =  \Ber(k_\alpha) N_{\alpha\beta} \Ber(k_\beta)$. 
Again, the time dependence has been suppressed. In Appendix \ref{app:theoreticalnoise} we present two alternative derivations of\eqref{eq:effectivefishermatrix} and provide further detail.

In the next subsection we show the effectiveness of the current treatment of the theoretical error. However, we believe the interpretation of this method, and its relation to the actual situation, is a subtle matter. In particular, in Appendix \ref{appsub:toymodel} we argue by means of a simple toy model that this way of dealing with the theoretical error is certainly not the right way in the extremes of zero and maximal correlation among the parameters. Namely, on the one hand the theoretical error acts as shot noise per bin for zero correlation length, whereas for maximal correlation it acts as a a single coefficient multiplying a fixed shape, effectively reducing the uncertainty about its shape to one number. Neither of these cases correspond to the way we believe the theoretical error should act. At the same time, Appendix \ref{app:correlations} shows that the effect of the correlation length on the results is very mild. This suggests that the main reason our method works so well is that our ansatz for the error is a much steeper function of $k$ than the signal, so that the size of the error is much more important than its shape. Thus, even though our treatment of the theoretical error seems to work for the current case, we recommend a conservative use of the method. In this spirit, we use the correlation lenght that gives the most pessimistic results for the analysis, which we found to be $l\sim 0.5$.


\subsection{Testing the effect of the theoretical error}
\label{subsec:integratingouttheoreticalerror}
To test the method of integrating out the theoretical error, we study its effect on the constraints on $f_{NL}$ in a $\chi^2$-analysis. To that end, we compare two types of analyses, one which includes the theoretical error as outlined above, and one which does not. We generate a fake dataset with no primordial non-Gaussianity to test the theory. Our datapoints are given by
\begin{equation}
D(\bin{k}) = B_{112}(\bin{k})+ E_b(\bin{k})+\text{cosmic noise},
\label{eqn:datavector}
\end{equation}
where we add some random noise, with variance equal to the cosmic variance, to each point. We consider a survey at redshift $z=0$ with volume $V=10\ (h^{-1}\text{Gpc})^3$, and restrict $(k_1,\ k_2,\ k_3)$ to be the central values of the binned range $[0.001, 1]\ h\text{Mpc}^{-1}$, where we take 27 logarithmic sized bins. The additional contribution to the bispectrum is given by 
\begin{equation}
 E_b(k_1, k_2,k_3) = 3 B_{112}(k_1, k_2,k_3)\left( \frac{k_1+k_2+k_3}{3 k_{NL}}\right)^{(3 + n)l}\,,
\label{eqn:theoreticalnoise2loopansatz}
\end{equation}
with $n = -1.4$, $k_{NL} = 0.45$ and $l=2$. This is exactly the ansatz for the two-loop contribution to the bispectrum used in \cite{Baldauf:2016sjb} and it is based on scaling universes \cite{Pajer:2013jj}. In appendix \ref{appsub:comparisonansatze}, we compare the ansatz for the higher loop corrections $E_b$ with our ansatz $B_{332}$.  As theoretical model for the bispectrum, we use 
\begin{equation}
B^{\text{th}}(\bin{k}) = f_{NL}\cdot B_{111}(\bin{k})+B_{112}(\bin{k}).
\label{eqn:theoryvector}
\end{equation}

We now consider two analyses. In the first analysis, we neglect theoretical errors and take only cosmic variance into account. In the second analysis, we use our ansatz for the higher order corrections, namely $B_{332}$, and we account for both theoretical error and cosmic variance. In order to find the best fit value for $f_{NL}$, we minimize $\chi^2_B$, which is given by
\begin{equation}
\chi^2_B = \left(D(\bin{k})-B^{\text{th}}(\bin{k})\right)\left(C_B(\bin{k},\bin{p})+N^{\text{eff}}(\bin{k},\bin{p})\right)^{-1}\left(D(\bin{p})-B^{\text{th}}(\bin{p})\right) + \text{const.},
\end{equation}
see \eqref{eqn:effectivelikelihood} in Appendix \ref{app:theoreticalnoise}. In the first case, we set $N^{\text{eff}}$ to zero. Minimizing $\chi^2_B$ yields
\begin{equation}
\text{Est}(f_{NL}) = \frac{B_{111}(\bin{k})\left(C_B(\bin{k},\bin{p})+N^{\text{eff}}(\bin{k},\bin{p})\right)^{-1}\left(D(\bin{p})-B_{112}(\bin{p})\right)}{B_{111}(\bin{k})\left(C_B(\bin{k},\bin{p})+N^{\text{eff}}(\bin{k},\bin{p})\right)^{-1}B_{111}(\bin{p})},
\label{eqn:fnlfromchisquared}
\end{equation}
and taking another derivative with respect to $f_{NL}$ allows us to compute the standard deviation
\begin{equation}
\sigma(f_{NL}) = \left(B_{111}(\bin{k})\left(C_B(\bin{k},\bin{p})+N^{\text{eff}}(\bin{k},\bin{p})\right)^{-1}B_{111}(\bin{p})\right)^{-1/2}.
\label{eqn:sigmafnlfromchisquared}
\end{equation}
With the best-fit value of $f_{NL}$, we can evaluate $\left(\chi^2_B\right)_\text{{red}}$ and the $p$-value, which are given by
\begin{equation}
\left(\chi^2_B\right)_\text{{red}}=\frac{\chi^2_B}{N} \quad \text{and} \quad p\text{-value} = 1- \text{CDF}_{\chi^2}(N,\chi^2_B),
\label{eqn:pvalue}
\end{equation}
with $\text{CDF}_{\chi^2}$ the cumulative distribution function of the $\chi^2$-distribution, and $N=N_{\text{bins}}-N_{\text{dofs}}-1$ the number of datapoints minus one minus the number of fitting parameters. The $p$-value takes values between 0 and 1. It gives the probability of finding a higher value for $\chi^2_B$ if it was drawn from a $\chi^2$-distribution. Therefore, it should take values around to $0.5$. If the $p$-value is very close to zero, then the proposed theory vector is not a good description of the data. If the $p$-value is close to one, then either one is overfitting the data, or  the estimate for the noise is too pessimistic. 

\begin{figure}[t]
  \centering
    \includegraphics[width=\textwidth]{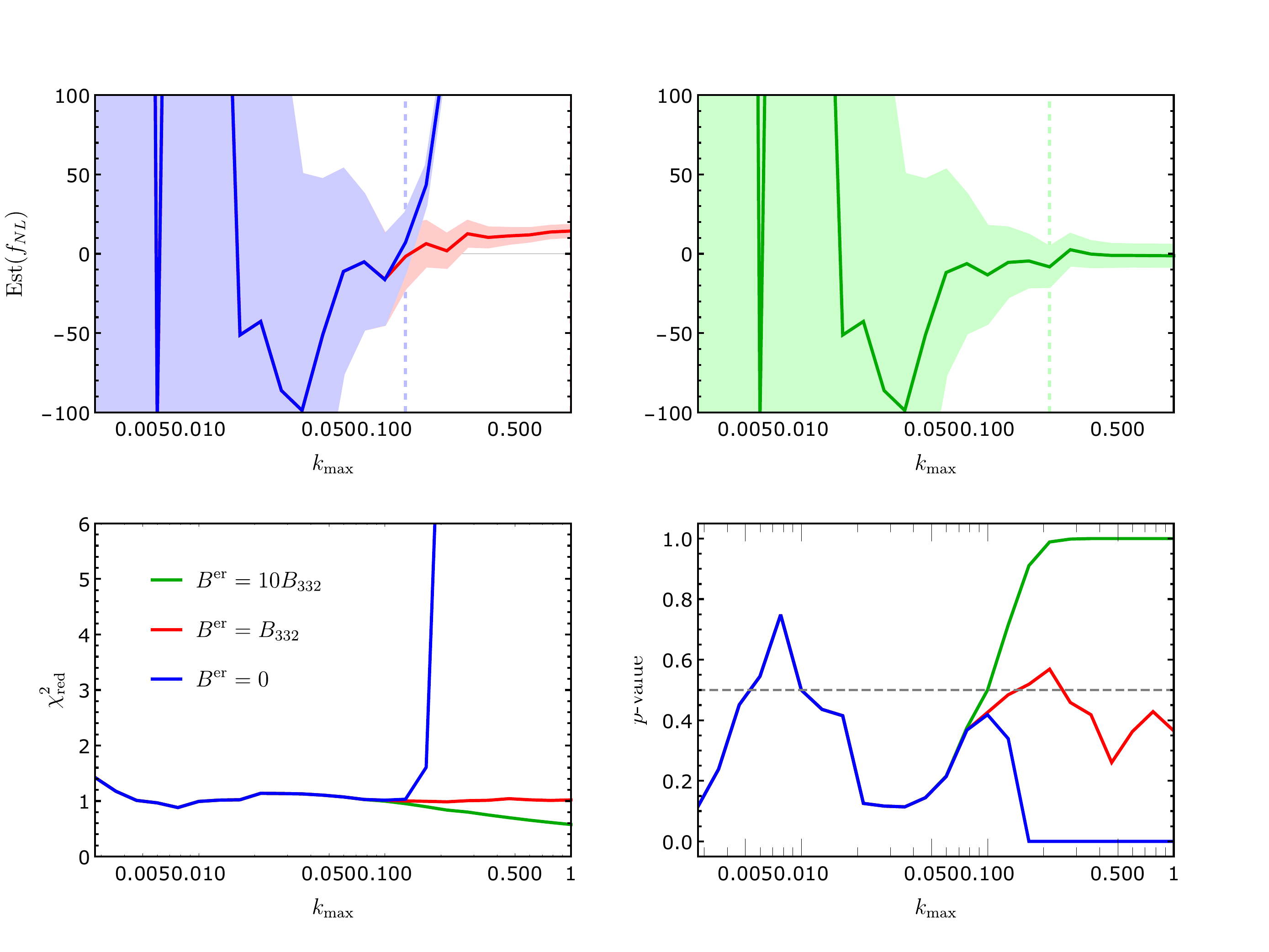}
  \caption{The figure shows the results from the $\chi$-squared analysis for the data and theory given in equations \eqref{eqn:datavector} and \eqref{eqn:theoryvector}. In the left panel, we show the results for both the analysis in which the higher order corrections are neglected (blue) and the one in which we use $B_{332}$ as ansatz (red). In the right panel, we use $10 B_{332}$ as ansatz (green) instead. In the upper panels, we show the best fit value for $f_{NL}$ (solid line) as function of $k_{max}$ and the lighter coloured regions correspond to the 2$\sigma$ errorbars. The dashed vertical line corresponds to the largest value for $k_{max}$ where the $p$-value is still between $0.01$ and $0.99$. The second and third row show $\left(\chi^2_B\right)_\text{{red}}$ and the $p$-value as function of $k_{max}$. }
\label{fig:pvalue}
\end{figure}

In Figure \ref{fig:pvalue}, we plot the estimate for $f_{NL}$ with errorbars, $\left(\chi^2_B\right)_\text{{red}}$ and the $p$-value, as given in equations \eqref{eqn:fnlfromchisquared}, \eqref{eqn:sigmafnlfromchisquared} and \eqref{eqn:pvalue} respectively, for the two analyses. In the left panel, we show both the results for the analysis in which the higher order corrections are neglected, and the analysis in which we use $B_{332}$ as an ansatz. In the right panel, we use $10\times B_{332}$ as error estimate to make sure that our ansatz is always bigger than the true value of the higher order corrections.  
One can check that $E_b$ has a different shape than $B_{332}$. For instance, in the equilateral configuration, $E_b$ is smaller than $B_{332}$ on small scales (more optimistic). On the other hand, on large scales in the equilateral configuration, and in the squeezed limit, it tends to be larger than $B_{332}$ (more pessimistic). Upon multiplying the latter by a factor 10, we find a robust, conservative estimate  (see \ref{appsub:comparisonansatze}).

In the left panel of Figure \ref{fig:pvalue}, we see that if we neglect the theoretical error (blue lines and contours), we get the wrong value for the best fit value for $f_{NL}$, because higher order corrections are mistakingly interpreted as signal. Fortunately, the $p$-value singles out where the theoretical description fails. Taking this into account, we get a reliable estimate for $f_{NL}$, albeit with larger errorbars, since we have to stop already at a relatively small $k_{max}$. From the analysis that accounts for the theoretical error (red lines and contours), it seems we can continue the analysis to a higher $k_{max}$. However, the result we get for $f_{NL}$ is biased, i.e. it is more than $5\sigma$ away from the actual value. The problem is that, in certain configurations of the bispectrum, our ansatz takes smaller values than the actual value in the data. This tends to decrease the $p$-value. At the same time, the $p$-value increases in the configurations where the theoretical error is overestimated. The interplay of these two effects can lead to a $p$-value, which is neither too small or too large, and this gives rise to a biased estimate. Hence, if one wants to use the $ p $-value as diagnostic for $ k_{max} $ and avoid biased results, it is important to have a fairly good understanding of the form of the theoretical noise.  
Alternatively, one can work with an ansatz which is consistently \textit{underestimating} the theoretical error. In this case, the $p$-value should go to zero rapidly, as soon as the theoretical error kicks in. As a double check, we did the analysis using $0.1 E_b$ as ansatz instead, which indeed gives unbiased results similarly to the case where we neglect the theoretical error altogether. In general, when performing a datafit, if one has insufficient information about the higher order corrections, it is therefore safer not to integrate out the theoretical error at all. Summarizing, \textit{assuming the wrong shape for the theoretical error might lead to a false detection of primordial non-Gaussianity}.

In the right panel (green lines and contours), we show the same results, where now the ansatz is always more pessimistic than the actual theoretical error ($  10\times B_{332}>E_{b}$). In this case, the estimate for $f_{NL}$ is equal to the real value within $2\sigma$. The $p$-value now is very \textit{large} and it would naively tells us to stop at some smaller $k_{max}$. However, since we are obviously not overfitting the data, this reflects the fact that our ansatz for the theoretical error is too pessimistic. Therefore, we can safely evaluate the estimate for $f_{NL}$ and the corresponding errorbar at the highest value of $k_{max}$, where the errorbar is frozen to a finite value. As expected, we find better errorbars than in the case we neglect the theoretical error. This shows that integrating out the theoretical error helps constraining $f_{NL}$, as long as one is careful to take a conservative enough ansatz. In the next section, we take $ B_{332}$ as ansatz for the theoretical error\footnote{We checked that our results for $\sigma(f_{NL})$ change with less than a factor of 2, when we use $10 B_{332}$ instead.}. 


\section{Results}
\label{sec:results}

In this section, we present the main results of our analysis. First, we give $\sigma(f_{\text{NL}})$ for various surveys, comparing our results to \cite{Baldauf:2016sjb} and \cite{Tellarini:2016sgp}. Next, we study the correlations among the EFT parameters for relevant surveys. Furthermore, we address the question of how much better the constraints would be, if we were able to compute the two-loop bispectrum. Finally, we show that the EFT of LSS clearly outperforms SPT in the constraints on $f_{\text{NL}}$, where we assume the EFT contributions to the bispectrum are part of the theoretical error in SPT.

\subsection{Constraints as function of $z_{\text{max}}$}
\label{subsec:constraintsasfunctionofzmax}
In this subsection, we compute the constraints on $f_{NL}$ as function of maximum redshift for surveys similar to the ones studied in \cite{Baldauf:2016sjb} and \cite{Tellarini:2016sgp}. This allows us to study the effects of shotnoise and to compare our results with theirs. Furthermore, we show the effect of marginalizing over the EFT parameters for these surveys. 

\subsubsection{A large redshift survey (comparison with Baldauf et al. 2016)}
\label{subsubsec:alargeredshiftsurvey}

First, to compare with \cite{Baldauf:2016sjb}, we focus on local and equilateral PNG. In the following, we list the specifications of the survey and the particular assumptions we make (in addition to general assumptions for the Fisher analysis discussed in subsection \ref{subsec:assumptions}).
\begin{itemize}
	\item We assume a survey with maximum redshift of $z = 5$. We divide the survey in redshift bins of equal volume, where the first bin runs from $z=0$ to $z=1$. We assume the survey covers $20000\ \text{deg}^2$, which means that, with our choice of cosmological parameters, the volume of each bin is given by $V = 26.5 (h^{-1}\text{Gpc})^3$. We approximate each bin to be a cube, so that $k_{\text{min}} \approx 0.002\ h \text{Mpc}^{-1}$. The maximum redshifts of all the redshift bins are given by\footnote{Here, and in the next section, we make a particular choice of redshift bins. Larger redshift bins imply that we can include more configurations of the bispectrum in the analysis, in particular more configurations in the squeezed limit. At the same time, we fix the redshift of the bin to be the mean redshift, therefore, larger redshift bins imply a smaller maximum redshift. Therefore, the choice of binning might affect the final result. We will study a bin-independent approach in \cite{Pajer:tobepublished}.}
	$$\{1.00, 1.39, 1.71, 2.02, 2.31, 2.60, 2.89, 3.19, 3.49, 3.80, 4.11, 4.44, 
	4.78, 5.13\}.$$ 
	\item We restrict $(k_1,\ k_2,\ k_3)$ to the values in the binned range $[0.002, 1]\ h\text{Mpc}^{-1}$, where we use 15 logarithmic bins per decade\footnote{This is a little less than 45 bins over the full range, corresponding to $O(3000)$ 
	triangles.}.
\item At high redshift, the late-time non-Gaussian background is particularly small and so the PNG signal becomes comparatively more pronounced. One the other hand, at high redshift, there are also much fewer tracers and this degrades our ability to measure the distribution of matter. To be able to capture this fact, we introduce a shotnoise that mimics what happens for example in galaxy surveys. For the purpose of comparison, we adopt the same convention for shotnoise as in \cite{Baldauf:2016sjb}, namely
\begin{equation}
\bar{n} = b_1^2 n_0 (1 + z)^\alpha,
\label{eq:shotnoise}
\end{equation}
where we correct for the galaxy bias $b_1 = 2$, since the shotnoise in \cite{Baldauf:2016sjb} applies to galaxies, while here it has been translated to the dark matter density field. We choose $n_0 = 10^{-3} h^3 \text{Mpc}^{-3}$ and $\alpha = -1$.
	\item When we marginalize, we assume for each EFT coefficient a Gaussian prior with $\sigma = 10$, except for the EFT parameter $\xi$ for which we take $\sigma = 1$.
	\item For $\Ber$ we consider both $B_{332}$ and the ansatz $E_b$ given in \cite{Baldauf:2016sjb} (see \eqref{eqn:theoreticalnoise2loopansatz}).
\end{itemize}

\begin{figure}[!t]
  \centering
    \includegraphics[width=1\textwidth]{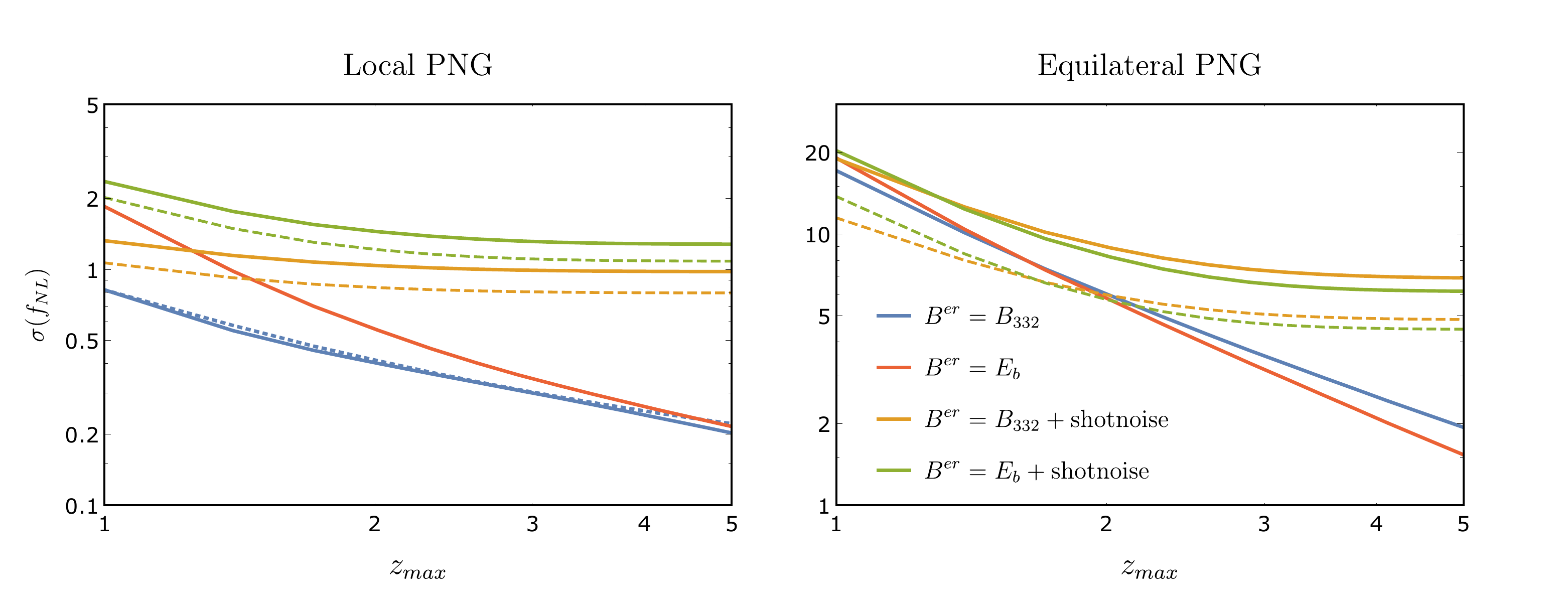}
  \caption{In these plots we show $\sigma(f_{NL})$ as function of maximum redshift $z_{\text{max}}$ for local PNG (left) and equilateral PNG (right). We include shotnoise with the same specifications as \cite{Baldauf:2016sjb} (orange and green lines). For the green lines we used $E_b$ as ansatz for $B^G_{2L}$ whereas for the other lines we used $B_{332}$. The blue solid line shows the result without shotnoise. The solid lines show the marginalized results assuming Gaussian priors for the EFT parameters. The dashed lines correspond to the unmarginalized result with the same color. Finally, the dotted blue line in the left panel corresponds to the curve $\sigma(z_1)/\sqrt{N}$ with $N$ the number of redshift bins.}
\label{fig:comparison1}
\end{figure}

The results for $\sigma(f_{NL})$ are shown in Figure \ref{fig:comparison1}. We show the effects of the ansatz for $\Ber$, shotnoise, and marginalization over the EFT parameters.

We can compare our results with those found in \cite{Baldauf:2016sjb} by looking at the unmarginalized results, using their ansatz $ E_{b} $ for $\Ber$. Thus, we should compare our dashed green lines with their dotted red lines in the last columns of their Figure 6 and 7. We indeed find a reasonably good agreement, given the fact that our analyses are not identical (different sky coverage, redshift bins and cosmological parameters, and moreover the translation of their shotnoise to ours is not perfect as we only took into account $b_1$). This check confirms that our code runs as expected.

Let us now study the effect of the EFT parameters. The solid lines in Figure \ref{fig:comparison1} correspond to marginalization over the EFT parameters, where we assume a Gaussian prior with $\sigma = 10$ for each EFT coefficient, except for the EFT parameter $\xi$, for which we take $\sigma = 1$. We see that \textit{the results for local PNG are almost unaffected by marginalizing over the EFT parameters}. The constraints on equilateral PNG weaken by a factor of about two, however.

Using $B_{332}$ as an ansatz for $\Ber$, we find slightly more optimistic results for local PNG, and slightly worse results for equilateral PNG, as compared with \cite{Baldauf:2016sjb}. This can be understood from the comparison between $B_{332}$ and $E_b$, shown in appendix \ref{appsub:comparisonansatze}. Local PNG peaks in the squeezed limit, and $B_{332}$ is more optimistic than $E_b$ in this configuration. On the other hand, equilateral PNG peaks in the equilateral configuration, and in this configuration $E_b$ is more optimistic. 

If we neglect shotnoise, we find that the differences at low redshifts are even bigger for the two ans\"atze. The difference is largest for local PNG, since $B_{332}$  is an order of magnitude bigger than $E_b$ in the squeezed configuration, whereas the difference in the equilateral configuration is only of the order of a few. However, at higher $z_\text{max}$ the differences disappear. This might seem a bit strange at first. However, we should stress that what we find here is not the true result in case of zero shotnoise. It turns out that with no shotnoise, we can always gain information in the ultra squeezed limit, and at higher redshift we can actually go to higher $k$ than our choice $k_\text{max} = 1 h \text{Mpc}^{-1}$. Since $\sigma(f_{NL})$ does not freeze before we reach $k_\text{max}$, this explains why the red and blue curves can get close for high $z_{\text{max}}$. In appendix \ref{appsub:comparisonansatze}, we show these statements explicitly. 

For comparison, note that if we were to always gain signal up to the same $k_\text{max}$ in each redshift bin, we would find roughly the same $\sigma(f_{NL})$ for each redshift bin\footnote{To good approximation: the entry of the Fisher matrix, corresponding to $f_{NL}$, will scale as $F\sim D(a)^6/D(a)^6 \sim 1$, if we neglect the loop corrections to $B_{NG}$. Then, forgetting about the marginalization over the EFT parameters, we find the same $\sigma(f_{NL})$ in each redshift bin, since we took the bins so that they have the same volume.}. Therefore, combining all the redshift bins, we should find a $1/\sqrt{N_\text{bins}}$ behavior if we neglect shotnoise. The dotdashed blue line in the figure corresponds to $\sigma(z_1)/\sqrt{N_\text{bins}}$, which indeed resembles the blue solid line quite well. This provides another indication that we can go up to higher $k_\text{max}$.  Interestingly, we find that we can also go to much smaller scales for equilateral PNG than suggested by scaling estimates for $k_\text{max}$ (see for instance \cite{Baldauf:2016sjb}). The squeezed limit allows us to extract more information, also for equilateral PNG.

If we include shotnoise, it correctly cuts off the signal before we reach $k_\text{max}$, so these results are reliable. However, one should keep in mind that, for more optimistic galaxy number densities, we might still extract more information from the ultra-squeezed limit. For the number densities we consider at the higher redshifts, shotnoise is the dominant source of noise. This is why we do include shotnoise in our analysis.  


\subsubsection{Current and upcoming surveys (comparison with Tellarini et al. 2016)}
\label{subsubsec:currentandupcomingsurveys}

\begin{figure}
  \centering
    \includegraphics[width=0.8\textwidth]{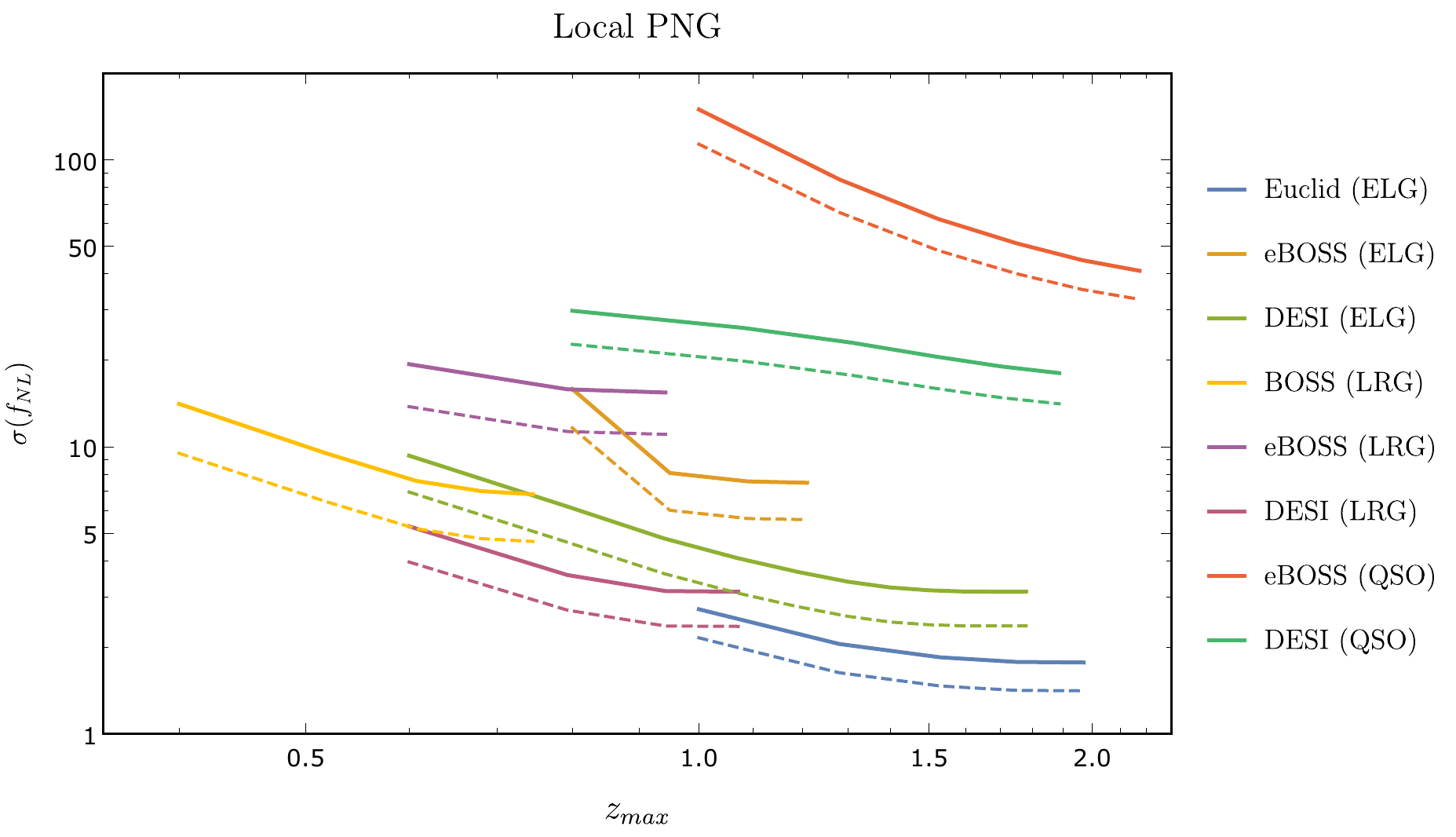}
  \caption{We show $\sigma(f_{NL})$ as function fo $z_\text{max}$ for local PNG. We use the specification from Euclid (blue), BOSS (yellow), eBOSS (orange, purple and red) and DESI (green/yellow, pink and green). We show both the marginalized (solid lines) and unmarginalized results (dashed lines). }
\label{fig:comparison2}
\end{figure}

Next, we compare with \cite{Tellarini:2016sgp}, using the specifications of the surveys Euclid \cite{Laureijs:2011gra}, BOSS \cite{Dawson:2012va}, eBOSS \cite{Dawson:2015wdb} and DESI \cite{Levi:2013gra}. We have to consider the emission line galaxies (ELG), the luminous red galaxies (LRG) and quasars (QSO) separately, as they are measured at different redshifts, and have different number densities and bias coefficients. The specifications and assumptions are as follows. 
\begin{itemize}
	\item For the precise number densities and  bias coefficients as function of redshift we refer to appendix D of \cite{Tellarini:2016sgp}. Moreover, one can find here the fraction of sky covered by each survey. 
	\item As before, we divide each survey in equal sized redshift bins. The boundaries of all the redshift bins are given by
	\begin{align*}
	\text{eBOSS (ELG)}:\   & \{0.6, 0.8, 0.95, 1.09, 1.21\}&  & \ V_{\text{bin}} = 2.8 (h^{-1}\text{Gpc})^3, \\
	\text{DESI (ELG)}:\   & \{0.1, 0.6, 0.79, 0.94, 1.07, \\
	&\quad 1.19, 1.3, 1.4, 1.5, 1.59, 1.69, 1.78\}&  &  \ V_{\text{bin}} = 5.4 (h^{-1}\text{Gpc})^3, \\
	\text{Euclid (ELG)}:\   & \{0.6, 1., 1.28, 1.53, 1.75, 1.97\}&  & \ V_{\text{bin}} = 14.0 (h^{-1}\text{Gpc})^3 ,\\
	\text{eBOSS (LRG)}:\   & \{0.6, 0.75, 0.87, 0.98\}&  & \ V_{\text{bin}} = 2.0 (h^{-1}\text{Gpc})^3, \\
	\text{DESI (LRG)}:\   & \{0.1, 0.6, 0.79, 0.94, 1.07\}&  &  \  V_{\text{bin}} = 5.4 (h^{-1}\text{Gpc})^3, \\
	\text{BOSS (LRG)}:\   & \{0., 0.4, 0.52, 0.61, 0.68, 0.75, 0.8\}&  & \ V_{\text{bin}} = 1.3 (h^{-1}\text{Gpc})^3 ,\\
	\text{eBOSS (QSO)}:\   & \{0.6, 1., 1.28, 1.53, 1.75, 1.97, 2.17\}&  & \ V_{\text{bin}} = 6.6 (h^{-1}\text{Gpc})^3, \\
	\text{DESI (QSO)}:\   & \{0.1, 0.8, 1.08, 1.31, 1.51, 1.7, 1.89\}&  &  \  V_{\text{bin}} = 11.0 (h^{-1}\text{Gpc})^3.\\
	\end{align*}
	\item We use $k_\text{min}$ determined by the volume of each bin. Moreover, we choose the same binning of the $k$-range as in Section \ref{subsubsec:alargeredshiftsurvey}. 
	\item The ansatz for shotnoise is now $\bar{n}(z) = b_1^2(z)n(z)$, where we correct for galaxy bias, similar as before.
	\item When we marginalize, we take the same prior as before. We assume for each EFT coefficient a Gaussian prior with $\sigma = 10$, except for the EFT parameter $\xi$ for which we take $\sigma = 1$.
	\item As ansatz for the higher order corrections we use $\Ber=B_{332}$. 
\end{itemize}
The results for local PNG are shown in Figure \ref{fig:comparison2}. We plot $\sigma(f_{NL})$ as function fo $z_\text{max}$ for the four surveys. We show both the marginalized and unmarginalized results. After combining the different galaxy catalogs of a survey, we get $\sigma(f_{NL})$  for each survey, as summarized in Table \ref{table:fullresultloc} for local PNG and in Tables \ref{table:fullresulteq} and \ref{table:fullresultqsf} for equilateral and quasi-single field PNG respectively. 

\begin{table}
	\begin{subtable}{.5\linewidth}
		\begin{tabular}{c|ccc}
			\text{ $\sigma(f^\text{loc}_{NL})$ } & \text{unmarg.} & \text{with prior} & \text{no prior}  \\
			\hline
			\text{BOSS} & 4.67 & 6.81 & 17.3 \\
			\text{eBOSS} & 4.91 & 6.6 & 14.15 \\
			\text{Euclid} & 1.41 & 1.77 & 3.66 \\
			\text{DESI} & 1.66 & 2.18 & 4.68 \\
		\end{tabular}
		\caption{Local PNG}
		\label{table:fullresultloc}
	\end{subtable} 
	\begin{subtable}{.5\linewidth}
		\begin{tabular}{c|ccc}
			\text{ $\sigma(f^\text{eq}_{NL})$ } & \text{unmarg.} & \text{with prior} & \text{no prior}   \\
			\hline
			\text{BOSS} & 16.89 & 29.86 & 37.99 \\
			\text{eBOSS} & 17.25 & 26.88 & 33.4 \\
			\text{Euclid} & 7.46 & 11.37 & 13.66 \\
			\text{DESI} & 7.18 & 11.4 & 13.48 \\
		\end{tabular}
		\caption{Equilateral PNG}
		\label{table:fullresulteq}
	\end{subtable} 
	\\ 
	\begin{subtable}{.5\linewidth}
		\centering
		\begin{tabular}{c|ccc}
			\text{ $\sigma(f^\text{qsf}_{NL})$ } & \text{unmarg.} & \text{with prior} & \text{no prior}  \\
			\hline
			\text{BOSS} & 12.57 & 23.65 & 27.26 \\
			\text{eBOSS} & 13.1 & 21.43 & 23.49 \\
			\text{Euclid} & 5.52 & 8.92 & 9.74 \\
			\text{DESI} & 5.37 & 8.98 & 9.66 \\
		\end{tabular}
		\caption{Quasi-single-field PNG}
		\label{table:fullresultqsf}
	\end{subtable}
	\caption{The final $\sigma(f_{NL})$ for local, equilateral and quasi-single field PNG for each survey, combining all expected galaxy catalogues. For the marginalized $\sigma(f_{NL})$, we put a Gaussian prior on each EFT coefficient with $\sigma = 10$, except for the EFT parameter $\xi$, for which we take $\sigma = 1$. In the last row, we also show the marginalized results, without prior on the EFT coefficients.}
	\label{table:fullresult}
\end{table}

We compare our results with \cite{Tellarini:2016sgp} by looking at our unmarginalized results in Table \ref{table:fullresultloc} and their results in the last column in Table 1 of their paper. We find much weaker constraints, varying from 4 to 8 times smaller. This can be explained by the fact that we account for the theoretical error, which freezes the errorbars. Therefore, including the theoretical error gives rise to more conservative constraints. Moreover, scale-dependent bias could actually help us constrain local PNG also in the bispectrum. In fact, redoing the analysis for Euclid up to $k_\text{max}(z) = 0.1 D(z)$, as used in \cite{Tellarini:2016sgp}, with the same specifications, except that we ignore the theoretical error, gives $\sigma(f_{NL})$ equal to $0.57$, $0.71$ and $1.3$ (unmarginalized, including and neglecting priors, respectively). This is roughly a factor three improvement from the results in Table \ref{table:fullresultloc}. The fact that this still does not challenge the results from \cite{Tellarini:2016sgp} seems to indicate that scale-dependent bias helps to improve the constraints on primordial non-Gaussianity.

From the combined catalogs, we ultimately find $\sigma(f^\text{loc}_{NL}) = 1.8$, $\sigma(f^\text{eq}_{NL}) = 11.4$ and $\sigma(f^\text{qsf}_{NL}) = 8.9$, with priors on the EFT parameters, assuming the surveys are not independent. These results do not change dramatically if we do not put priors on the EFT parameters. If the surveys are independent, the constraints improve approximately with a factor $1/\sqrt{2}$ upon combining Euclid and DESI.


\subsection{Correlation coefficients}
\label{subsec:correlations}
\begin{figure}[!ht]
  \centering
    \includegraphics[width=1\textwidth]{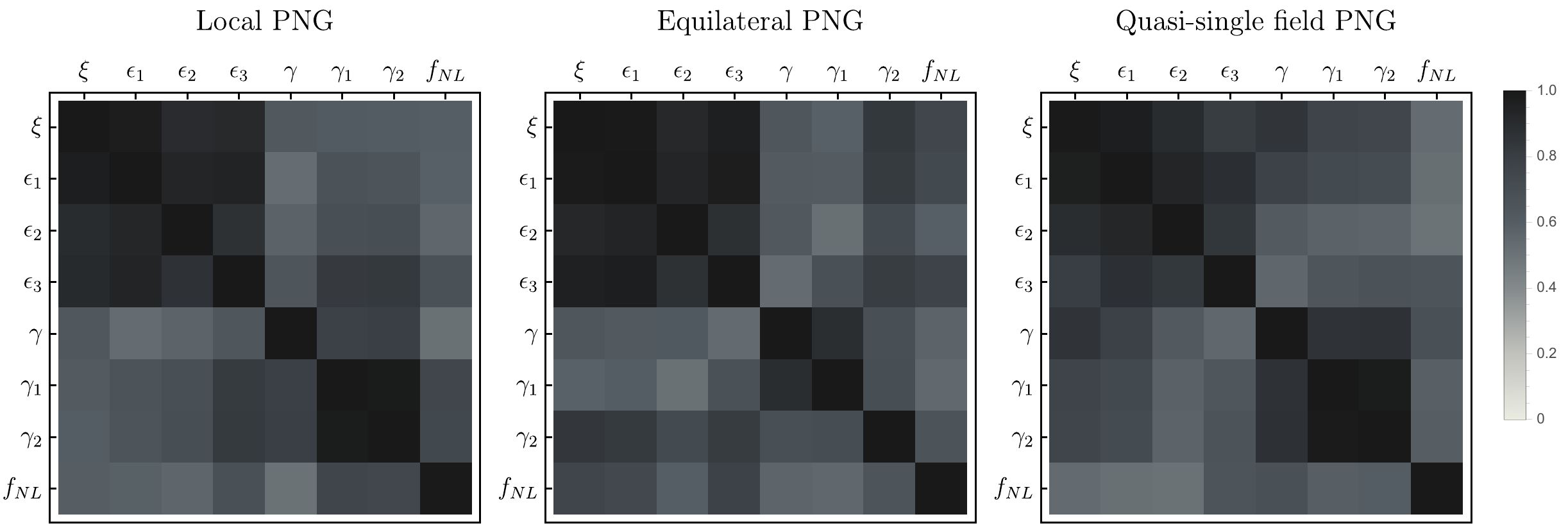}
  \caption{In these plots we show the correlation coefficients $r_{ij}$ for each pair of theoretical parameters. We include shotnoise with the same specifications as Euclid and included all information up to redshift $z_{\text{max}}=2$. A value of 1 (black) corresponds to perfectly correlated or anti-correlated. A value of 0 (white) corresponds to no correlation.   }
\label{fig:correlationcoefficients}
\end{figure}

To gain intuition on how much the EFT parameters affect the constraints on $f_{NL}$ for local, equilateral and quasi-single field PNG, we compute the correlation coefficients between parameters $\theta_i$ and $\theta_j$. These are defined as 
$$
r_{ij} = \frac{F_{ij}^{-1}}{\sqrt{F_{ii}^{-1}F_{jj}^{-1}}}.
$$
The correlation coefficient takes a value between 1 (perfectly correlated) and $-1$ (perfectly anti-correlated). In particular, the parameters are perfectly correlated with themselves. In Figure \ref{fig:correlationcoefficients}, we plot the absolute value of the correlation coefficients for each pair of parameters. We make the following assumptions
\begin{itemize}
	\item We use the same binning in $k$-space as in Section \ref{subsubsec:alargeredshiftsurvey}. 
	\item We use the redshift binning and shotnoise from Euclid, as given in Section \ref{subsubsec:currentandupcomingsurveys}.
	\item As ansatz for the theoretical error we use $\Ber=B_{332}$.
	\item We do not marginalize over the EFT parameters. The marginalized results are quoted in the text below.  
\end{itemize}

 We find that the groups of parameters $\{\xi,\ \epsilon_1,\ \epsilon_2,\ \epsilon_3\}$ and $\{\gamma,\ \gamma_1,\ \gamma_2\}$ have strong correlation among themselves. The correlation between $f_{NL}$ and the other parameters is, however, small. 

For local PNG, we find $f_{NL}$ is mainly correlated with $\gamma$, $\gamma_1$ and $\epsilon_3$, with correlation coefficients $0.44$, $-0.43$ and $0.29$ respectively. The other correlation coefficients are in absolute value smaller than $0.2$. If we include a Gaussian prior on the EFT parameters, with the same variances as before, the correlation coefficients become $0.12$, $0.08$ and $0.14$. 

In case of equilateral PNG, we find $f_{NL}$ has appreciable correlation with $\xi$, $\epsilon_1$, $\epsilon_3$ and $\gamma_2$ ,with correlationcoefficients $0.43$, $-0.39$, $-0.47$ and $-0.27$ respectively. Including priors on the EFT parameters, we find they become $0.14$, $-0.05$, $-0.30$ and approximately zero. This could motivate further study on the Gaussian EFT coefficients, see for instance \cite{Garny:2015oya}. It is surprising that $f^{\text{eq}}_{NL}$ is not extremely degenerate with $\xi$, since the latter comes with an additional $k^2$ scaling, similar to equilateral non-Gaussianity. It turns out, however, that the full shapes are sufficiently distinct. This will make it easier to constrain equilateral PNG from the bispectrum than naively thought. Then, for quasi-single-field PNG, we find that $ f_{NL}^{} $ is mostly correlated with $\epsilon_3$ and $\gamma$, with correlation coefficients $-0.26$ and $0.31$. Including the priors, they reduce to $-0.18$ and approximately zero. 

Summarizing, although the ignorance about EFT parameters does affect the final result, for reasonable priors this is only a small effect, especially for local PNG. This indeed agrees with what is seen in Figures \ref{fig:comparison1} and \ref{fig:comparison2}, and Table \ref{table:fullresultloc}. 

\subsection{Higher loop corrections}
\begin{SCfigure}
  \centering
    \includegraphics[width=0.6\textwidth]{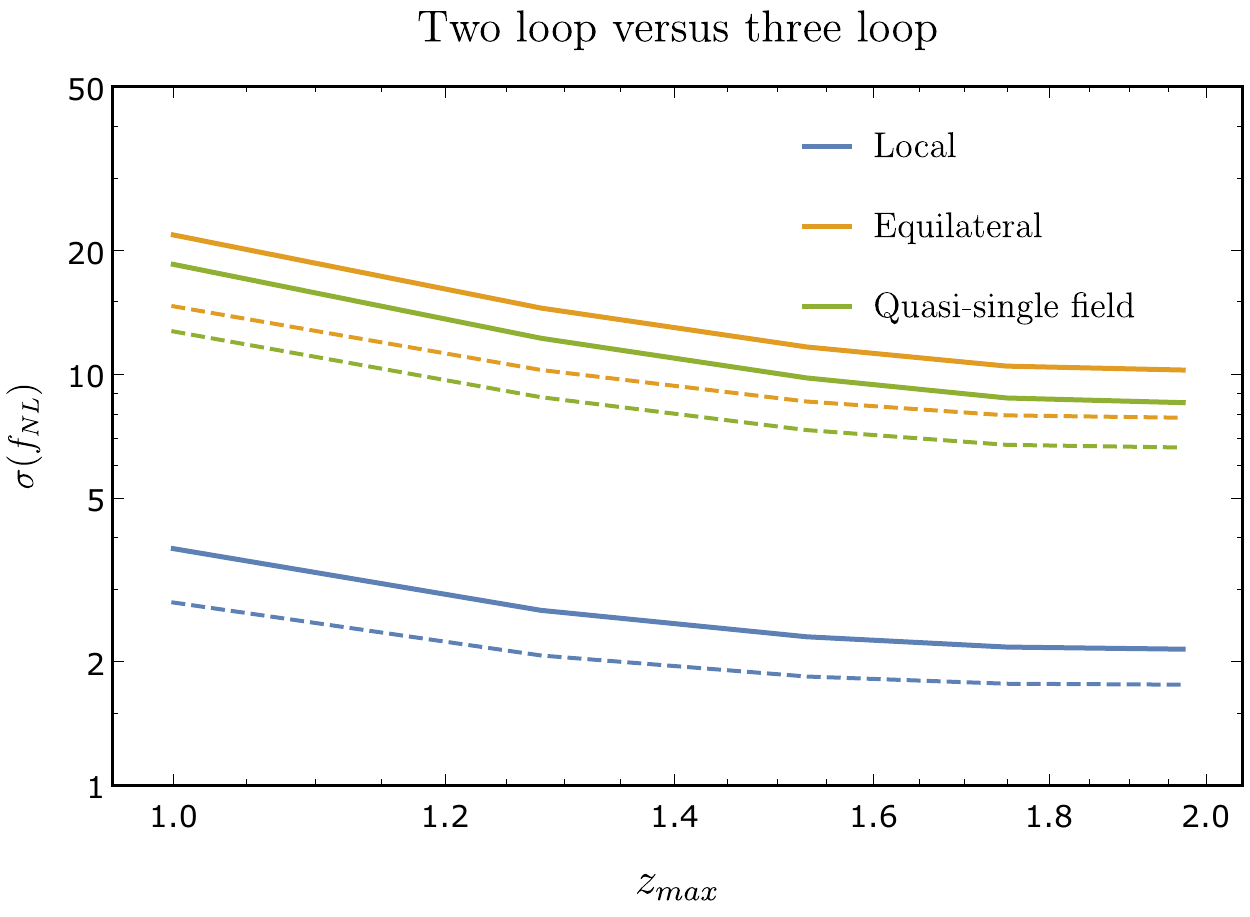}
  \caption{In this plot, we show $\sigma(f_{NL})$ as function of $z_{\text{max}}$ for local (blue), equilateral (orange) and quasi-single field (green) PNG. We use both $B_{2L}$ (solid lines) and  $B_{3L}$ (dashed lines) as order of magnitude estimation for the theoretical error. We use the specifications of Euclid in the analysis.}
\label{fig:2loopvs3loop}
\end{SCfigure}

We can ask ourselves whether it is useful to compute the bispectrum up to two loops in gravitational non-linearities. Note that our analysis does not depend on the actual value of the two-loop bispectrum, as there is no theoretical parameter in front. This means we can simply assume that we have computed all diagrams, neglect the counterterms, and assume the theoretical error is given by the SPT three loop bispectrum $B^G_{3L}$. Again, we do not know what it is, so we have to make an ansatz for it. Here we use the ansatz for the higher loop corrections from \cite{Baldauf:2016sjb}, given in \eqref{eqn:theoreticalnoise2loopansatz}, since it is easy to compute\footnote{An alternative - more in line with our two loop ansatz - would be to compute the reducible diagram of $B_{433}$ as order of magnitude estimate of $B_{3L}^{G}$. However, we point out that using only one diagram is dangerous. For instance, in \cite{Assassi:2015jqa} we considered only one of the two reducible two loop diagrams in our qualitative analysis, namely $B_{332}^I$. 
In the squeezed limit, the two loop contribution turns out to be much larger if we include $B_{332}^{II}$. }. For a rough estimate this should suffice. We estimate $B_{3L}^{G}$ using scaling universes \cite{Pajer:2013jj}. We choose\footnote{We choose a larger $k_{NL}$ than for the two loop estimate, since this scale determines when the three loop correction equals the lower order corrections. Since we are doing a perturbative expansion, we assume this happens at a smaller scale than when the two loop correction becomes equal to its lower order corrections. Moreover, each loop will scale as $k/k_{NL}$ to the power $3+n$, where $n$ will be of order of the scaling of the power spectrum at the non-linear scale $k_{NL}$. The power spectrum is steeper on smaller scales, therefore we take a more negative value for $n$. Each loop has this scaling, so we have to take $l=3$, in case of three loops.} $E_b$ with $n=-1.5$, $k_{NL} = 0.5\ h \text{Mpc}^{-1}$ and $l=3$  (see \eqref{eqn:theoreticalnoise2loopansatz}). Using the specifications from Euclid, we perform the Fisher analysis with both $B_{332}$ and  $B_{3L}$ as order of magnitude estimates for the noise. We collect the result in Figure \ref{fig:2loopvs3loop}. This shows that the constraints would improve if one computed the two-loop corrections to the bispectrum. The precise values are given in Table \ref{table:2loopvs3loop}, where we consider all surveys again. The tighest constraints on local, equilateral and quasi-single field PNG improve with a factor $1.2$, $1.3$ and $1.3$ respectively with this particular choice for $B_{3L}^{G}$. If it turns out we can get constraints on PNG close to the theoretical benchmarks, it would then be worth computing the two-loop corrections. It might be time consuming, but otherwise much cheaper than doubling the survey volume.

\begin{table}
\begin{subtable}{.33\linewidth}
\begin{tabular}{c|ccc}
 \text{ $\sigma(f^\text{loc}_{NL})$ } & \text{2 loop} & \text{3 loop} \\
\hline
 \text{BOSS} & 8.73 & 6.05 \\
 \text{eBOSS} & 7.12 & 6.07 \\
 \text{Euclid} & 2.14 & 1.75 \\
 \text{DESI} & 2.62 & 2.09 \\
\end{tabular}
\caption{Local PNG}
\label{table:2loopvs3looploc}
\end{subtable}
\begin{subtable}{.33\linewidth}
\begin{tabular}{c|ccc}
 \text{ $\sigma(f^\text{eq}_{NL})$ } & \text{2 loop} & \text{3 loop} \\
\hline
 \text{BOSS} & 27.8 & 19.14 \\
 \text{eBOSS} & 22.99 & 18.44 \\
 \text{Euclid} & 10.22 & 7.83 \\
 \text{DESI} & 10.2 & 7.81 \\
\end{tabular}
\caption{Equilateral PNG}
\label{table:2loopvs3loopeq}
\end{subtable}
\begin{subtable}{.33\linewidth}
\begin{tabular}{c|ccc}
 \text{ $\sigma(f^\text{qsf}_{NL})$ }& \text{2 loop} & \text{3 loop} \\
\hline
 \text{BOSS} & 23.66 & 16.65 \\
 \text{eBOSS} & 19.15 & 15.59 \\
 \text{Euclid} & 8.52 & 6.62 \\
 \text{DESI} & 8.46 & 6.6 \\
\end{tabular}
\caption{Quasi-single-field PNG}
\label{table:2loopvs3loopqsf}
\end{subtable}
\caption{The final $\sigma(f_{NL})$ for equilateral and quasi-single-field PNG (left and right), for each survey, combining all expected galaxy catalogues. For the marginalized $\sigma(f_{NL})$, we put a Gaussian prior on each EFT coefficient with $\sigma = 10$, except for the EFT parameter $\xi$, for which we take $\sigma = 1$. In the last row, we also show the marginalized results without prior on the EFT coefficients.}
\label{table:2loopvs3loop}
\end{table}

\subsection{EFT of LSS versus SPT} 
In the EFT of LSS we are forced to include free parameters over which we have to marginalize. Above, we saw that this marginalization weakens the constraints on $f_{\text{NL}}$, be it only mildly. One might therefore wonder how much the improvement actually is over a more conservative approach, in which one uses only SPT results for $\Bth$ and moves all other gravitational contributions to the theoretical error. In this section we confirm that the EFT approach always performs sizably better. We consider a couple of options, for different choices of $\Bth$ and $\Ber$, and compute the constraints. We use the specifications from Euclid as given in the previous section.   
	
Table \ref{table:sptversuseft} shows our results, which include the usual Gaussian priors for the EFT parameters whenever they are included in $\Bth$. The second and third columns give the theoretical description of the bispectrum $B^\text{th}$ and what we consider to be the unknown $\Ber$. For the latter, we sum the absolute values of all contributions indicated in the table. The EFT contributions, except for $\xi$, are multiplied by a factor 10, consistent with the priors we chose when we included them in $B^\text{th}$. The first row of Table \ref{table:sptversuseft} corresponds to the results we find in the `with prior' columns of the `Euclid' rows of Tables \ref{table:fullresultloc}, \ref{table:fullresulteq} and \ref{table:fullresultqsf}.

We find that including the non-Gaussian counterterms does not improve the bounds on $f_{NL}$. This could have been anticipated from the qualitative results in Figure 5 - 7 of \cite{Assassi:2015jqa}. We see that the counterterms are negligible in many configurations. Apparently, they are negligible in most configurations. However, using the EFT for the Gaussian part of the bispectrum performs significantly improves the results compared with just the SPT predictions. We find that the \textit{EFT of LSS improves the constraints on PNG approximately by a factor 3}. Finally, neglecting the one-loop non-Gaussian contribution to the bispectrum makes only about a $10\%$ difference. This is consistent with the observation that the non-Gaussian counterterms are not very important, as the non-Gaussian one loop correction itself is not very relevant.

\begin{table}
\begin{tabular}{lcc|ccc}
 \text{ Approach } &  $ \Bth=B^G_\text{0}+\dots$  &  $ \Ber=B_{332}+\dots$  & $\sigma(f^\text{loc}_{NL})$& $\sigma(f^\text{eq}_{NL})$ &$\sigma(f^\text{qsf}_{NL})$  \\
\hline
 \text{EFT (G+NG)} & $+B^{NG}_\text{0}+B^G_{EFT}+B^{NG}_{EFT}$ & $ $ &  1.77 & 11.37 & 8.92 \\
 \text{EFT G+SPT NG} & $+B^{NG}_\text{0}+B^G_{EFT}$ & $B^G_\text{2L} + B^{NG}_{EFT} $ & 1.78 & 11.37 & 8.92 \\
 \text{SPT (G+NG)} & $+B^{NG}_\text{0}$ & $ + B^{NG}_{EFT}+B^G_{EFT} $ & 6.11 & 27.61 & 21.76 \\
\text{SPT (G+NG tree)} & $+B^{NG}_\text{tree}$ & $ + B^{NG}_{EFT} +B^G_{EFT} + B^{NG}_{1L}$ & 7.17 & 30.58 & 24.23\\
\end{tabular}
\caption{We show the contraints on primordial non-Gaussianity of the local, equilateral and quasi-single-field type (last three columns). In the first row, we use the EFToLSS for both the Gaussian and non-Gaussian part of the bispectrum (`EFT (G+NG)'). In the second row, we only use the EFT for the Gaussian part of the bispectrum, and include the non-Gaussian counterterms in the higher order corrections (`EFT G + SPT NG'). Then, in the third row, we use only the SPT for describing the bispectrum (`SPT (G+NG)'). In the last row, we only include the tree level non-Gaussian contribution to the bispectrum (`SPT (G+NG tree)'). The second and third column denote all the contributions to the theoretical description of the bispectrum $B^\text{th}$, and the higher order corrections $\Ber$ respectively.}
\label{table:sptversuseft}
\end{table}

\section{Discussion and Outlook}
\label{sec:conclusionanddiscussion}
In this work, we have presented how the EFT of LSS helps us improve the constraints on primordial non-Gaussianities (PNG), using the matter bispectrum as observable. We have accounted for intrinsic theoretical uncertainties in the perturbative description, and studied in details their modeling in a Fisher forecast. 

Our main results are given in Table \ref{table:sptversuseft}. The forecasted values for $\sigma(f_{NL})$ for the local, equilateral and quasi-single field types of PNG are presented. Moreover, we show that the EFT approach improves the constraints on PNG by almost a factor 3 with respect to the results from SPT. 

\paragraph{Limitations} Let us first discuss the limitations of these results. We would like to compare the constraints we find with theoretically interesting benchmarks and constraints coming from the CMB. However, we should be careful in making a direct comparison, as there are other sources of non-linearities and noise that we have not accounted for in our analysis. First, we have modeled the matter bispectrum. To relate it to the observed galaxy bispectrum, we have to include galaxy bias and redshift space distortions. These introduce new uncertainties, leading to worse constraints. However, considering the results found in \cite{Dore:2014cca,Tellarini:2016sgp}, scale-dependent bias might actually \textit{improve} the constraints on (only) local PNG, since it enhances the non-Gaussian signal in the bispectrum. Second, except for shotnoise, we neglected all observational sources of noise. Survey geometry and survey mask may increase the errorbars as well. For instance the authors of \cite{Scoccimarro:2003wn} found that the errorbars increased by a factor of 4-5. Errors in determining the redshift of galaxies are another source of observational noise. Third, we made some simplifications in the Fisher analysis itself, such as neglecting the covariance between different points of the bispectrum. Combining this with the covariance induced by the survey geometry could further increase the errorbars by a factor of 8 \cite{Sefusatti:2004xz}. 

\paragraph{Improvements} On the other hand, there are also ways the constraints could be improved. First, we have used the specifications of Euclid to get a reasonable estimate for the limitations due to shotnoise. This determined our final forecasted result for $\sigma(f_{NL})$. It might well be possible, in a more futuristic survey, to optimize the number densities of galaxies and redshift range to be more suitable for constraining PNG (see for instance \cite{Dore:2014cca}). Moreover, we should perform a joint analysis of all large scale structure surveys. We have assumed for simplicity that we can do as well as the single best survey, which turned out to be Euclid for the four surveys we considered. In principle, we can do better if the surveys are not all precisely overlapping. Similarly, we should combine the results from different observables. For instance, we should perform a joint analysis of the power spectrum and bispectrum. This could improve the results by a factor of 2 for local PNG. Combining the results found in \cite{Tellarini:2016sgp}, and using the multitracer technique proposed in \cite{Seljak:2008xr} instead, which could improve upon the constraints from the power spectrum by a factor of about 7.  In addition, the trispectrum might turn out to be an important probe for non-Gaussianity, since linear theory works for a larger range of scales compared to the bispectrum \cite{Verde:2001pf}. The one loop corrections to the trispectrum in the EFT of LSS have recently been computed in \cite{Bertolini:2016bmt}. Last, we divided the full redshift range in smaller redshift bins, and only considered correlations within each redshift bin. If we also include correlations among galaxies separated by a larger distance along the line of sight, we might extract more information from a given survey. We will discuss the issue elsewhere \cite{Pajer:tobepublished}. Finally, our focus here was on near future galaxy surveys, but of course our results will be relevant in the future also for 21 cm survey (see e.g. \cite{Cooray:2006km}).

One of our main results is that the EFT approach helps constraining PNG. The improvement comes completely from the EFT corrections to the late time gravitational non-linear evolution of matter. Both the SPT loops and EFT corrections to the primordial non-Gaussian signal, discussed in \cite{Assassi:2015jqa}, do not help much improving the constraints. 

Comparing our results with the theoretically interesting benchmark $\sigma(f^\text{eq}_{NL}) \sim 1 $, we see that it does not look promising for equilateral PNG. Even with zero shotnoise, as we can see in Figure \ref{fig:comparison1}, we barely touch the theoretical targets. Our lack of understanding of matter non-linearities is already an important obstacle to reach $\sigma(f^\text{eq}_{NL}) \sim 1 $. The same applies to quasi-single field PNG. Additional sources of non-linearities such as bias and redshift space distortions will make things worse. On the other hand, for local primordial non-Gaussianity, things look more promising. Matter non-linearities can be modeled well enough to get close to $\sigma(f^\text{loc}_{NL}) \sim 1 $ from large scale structure experiments.

We can ask whether $N$-body simulations can help reaching better constraints on primordial non-Gaussianity. As pointed out in \cite{Baldauf:2016sjb}, using end to end simulations, without any perturbative input, will most likely be insufficient to reach $\sigma(f^\text{eq}_{NL}) \sim 1 $. The reason is that simulations do not solve the exact problem but make a series of approximations, such as for example the particle mesh and tree approaches to solve Poisson equation, finite size effects and approximate initial conditions. Currently, simulations reach approximately $ 1\% $ precision \cite{Heitmann:2008eq,Schneider:2015yka}. Heuristically, looking at our Figure \ref{fig:bispectrumplot}, we see that the PNG signal we are trying to extract is much smaller than that, so large improvements in the precision of simulations are needed. Alternatively, one can use $N$-body simulations to determine the EFT parameters\footnote{In fact, in our analysis, we assumed that a one EFT parameter, $ \xi $, was fixed by fitting the power spectrum to simulations.}. We can then look directly at the unmarginalized columns in Table \ref{table:fullresultqsf}. We see that, even in the very optimistic case that all relevant EFT parameters at one loop are fixed, $ \sigma(f^\text{eq}_{NL})$ still remains around 7.

\paragraph{Theoretical error} Another goal of our paper is to clarify some aspects of the modeling of theoretical uncertainties in forecasting observational constraints, and, eventually, in analyzing data (see  Section \ref{subsec:integratingouttheoreticalerror} and Appendix \ref{app:theoreticalnoise}). We introduced the concept of correlation length in Section \ref{subsec:theoreticalnoise}, along the lines of \cite{Baldauf:2016sjb}. In Appendix \ref{app:theoreticalnoise}, we argued that the choice of correlation length in integrating out the theoretical error is subtle and no ``right'' choice can be established a priori. However, in our particular analysis, we hardly find any dependence on the correlation length (see Figure \ref{fig:correlationlength}). In future studies, with different observables and different perturbative approaches, we believe that an analysis on the choice of correlation length should be always performed.

In Figure \ref{fig:pvalue} we have seen that assuming the wrong shape for the theoretical error can lead to biased results in a $\chi^2$-analysis. Therefore, if we want to fit to data, we need good estimates for the higher order corrections. For instance by using estimates from N-body simulations, or alternatively, by computing additional two-loop diagrams. 

\paragraph{Outlook} Our work can be extended and improved in different ways.
\begin{itemize}
\item Instead of dividing the survey volume in redshift bins and only consider correlations within these bins, it would be interesting to see how much we gain including all possible cross-correlations across redshift bins. This is work in progress \cite{Pajer:tobepublished}.
\item  It would be interesting to perform a similar Fisher analysis with an updated study of covariance effects due to geometry, masking and non-Gaussian gravitational evolution.
\item We should join all forces. It would be interesting to do a joint analysis of multiple observations, such as the CMB, LSS surveys and the 21 cm observations. Moreover, all the different LSS surveys should be combined to have maximum constraining power. Furthermore, the results from the power spectrum, bispectrum and trispectrum should be combined too. Finally, on the theory side, one should also try to use results from N-body simulations as soon as our perturbative description starts to break down, i.e. when the theoretical error becomes dominant. 
\end{itemize}

\section*{Acknowlegdements}

We thank Valentin Assassi and Daniel Baumann for initial collaboration on this project. We would like to thank Emiliano Sefusatti for useful discussions. We are especially thankful to Tobias Baldauf for the many useful discussions. We would like to thank Fabian Schmidt and Giovanni Cabass for feedback on the draft. Y.W. would like to thank Koen Kuijken and Ana Ach\'ucarro for providing a sounding board now and then. Y.W. is supported by a de Sitter Fellowship of the Netherlands Organization for Scienti c Research (NWO). E.P. is supported by the Delta-ITP consortium, a program of the Netherlands organization for scientic research (NWO) that is funded by the Dutch Ministry of Education, Culture and Science (OCW).

\appendix
\section{Explicit Results for the Bispectrum}
\label{app:eftresults}

The purpose of this appendix is to provide, and to some extent clarify, explicit expressions for the bispectrum in the presence of primordial non-Gaussianity. This is essentially a summary of \cite{Assassi:2015jqa}, so we refer the reader to that work for a thorough discussion and explanation of these results. We adopt the same notation, which we summarize in Appendix \ref{app:tableparameters}. 
\subsection{Perturbation Theory in the EFT of LSS}  

In the EFToLSS, the equations of motion for the density contrast $\delta$ and the velocity divergence $\theta=\partial_{i}v^{i}$ on large scales are
\begin{subequations}\label{EoM}
	\begin{align}
	\delta_\tau \delta + \theta &\ = \cal{S}_{\alpha} \ ,\\
	(\delta_\tau+\H)\theta + \frac{3}{2}\Omega_m \H^2\delta &\ = \cal{S}_{\beta} + \tau_\theta \ .
	\end{align} 
\end{subequations}
Here the source terms $\cal{S}_{\alpha,\beta}$ are the standard nonlinear terms in the Euler equations, which, in Fourier space, are given by the following convolutions,
\begin{subequations}
\begin{align}
{\cal S}_\alpha({\bf k},\tau) &\equiv - \int_{{\bf p}} \alpha({\bf p},{\bf k}-{\bf p}) \theta({\bf p},\tau) \delta({\bf k}-{\bf p},\tau)\ , \qquad \alpha({\bf k}_{1},{\bf k}_{2}) \equiv \frac{{\bf k}_{1}\cdot({\bf k}_{1}+{\bf k}_{2})}{{\bf k}_{1}^2} \ ,\\
{\cal S}_\beta({\bf k},\tau) &\equiv - \int_{{\bf p}} \beta({\bf p},{\bf k}-{\bf p}) \theta({\bf p},\tau) \theta({\bf k}-{\bf p},\tau) \ , \qquad \beta({\bf k}_{1},{\bf k}_{2}) \equiv \frac{({\bf k}_{1}+{\bf k}_{2})^2}{2}\frac{{\bf k}_{1}\cdot {\bf k}_{2}}{{\bf k}_{1}^2 {\bf k}_{2}^2} \ . 
\end{align}
\end{subequations}
Clearly, we neglected large scale vorticity and large scale velocity dispersion in \eqref{EoM}. 
However, the backreaction from unknown short scale physics is taken into account through the effective stress tensor $\tau_\theta$. A complete description and motivation of this term in the presence of primordial non-Gaussianity was the main purpose of \cite{Assassi:2015jqa}. Here we just quote the leading contributions for the types of non-Gaussianities we consider, to first order in $f_{\text{NL}}$,
\begin{align}\label{stresstensor}
\tau_\theta&= -d^2\hskip -1pt\bigtriangleup\hskip -2pt\delta-e_1\hskip -1pt\bigtriangleup\hskip -2pt(\delta^2) -e_2\hskip -1pt\bigtriangleup\hskip -2pt(s^2) -e_3\hskip 1pt\partial_i(s^{ij}\partial_j\delta)\nonumber\\[2pt]
&\hskip 12pt-f_{\text{NL}}\big[g\big(\hskip -1pt\bigtriangleup \hskip -2pt\Psi-\partial_i(\delta\partial^i\Psi)\big)+g_1 \hskip -1pt\bigtriangleup \hskip -2pt(\Psi\delta)+g_2\partial_{i}\partial_j(\Psi s^{ij})\big]\ ,
\end{align} 
where $\bigtriangleup$ denotes the Laplacian, and the coefficients in this expression are functions of time only. The equations are formally solved using a Green's function method
\begin{align}
\delta({\bf k},a)=D_{1}(a)\delta_{1}({\bf k})+\int_{a_{in}}^{a}G_{\delta}(a,a')\left[{\cal S}_\beta  + \tau_{\theta} - \H \partial_a(a{\cal S}_\alpha)\right].
\end{align}
Here $\delta_1({\bf k})$ is the growing mode initial condition, and $D_{1}(a)$ is the linear growth factor
\begin{align}
D_1(a) = \frac{5}{2}\H_0^2 \Omega_m^0  \frac{\H}{a} \int_{a_{in}}^a \frac{\text{d} a'}{\H^3(a')}\ ,
\end{align} 
which in Einstein-de Sitter reduces to $D_1(a)=a/a_{in}$, where $a$ is the scale factor. This equation can be solved perturbatively in terms of the linear solution, yielding
\begin{align}
\delta({\bf k},a)=\sum_{n=1}^{\infty} \delta_{(n)}^{\text{SPT}}({\bf k},a)+ \delta_{(n)}^{\text{c}}({\bf k},a),
\end{align} 
where $\delta_{(n)}^{\text{SPT}}$ are the standard perturbation theory (SPT) terms (see \cite{Bernardeau:2001qr}), sourced solely by $\cal{S}_{\alpha,\beta}$:
\begin{align}
\delta_{(n)}^{{\text{SPT}}}({\bf k},a) &\,\approx \, D_{1}^{n}(a)\int_{{\bf k}_1} \ldots \int_{{\bf k}_n} (2\pi)^3 \delta_D\big({\bf k} - {\bf k}_{1  \ldots n}\big) \, F_n({\bf k}_1,\ldots, {\bf k}_n) \, \delta_1({\bf k}_1) \ldots \delta_1({\bf k}_n);
\end{align}
and $\delta_{(n)}^{\text{c}}$ is the `counterterm' contribution, i.e. the terms proportional to one of the free parameters in \eqref{stresstensor}, which we write as  
\begin{align}\label{equ:ngcounter}
\delta^c_{(n)}({\bf k},a) &= \int_{{\bf k}_1}\ldots\int_{{\bf k}_n}(2\pi)^3\delta_D\big({\bf k} - {\bf k}_{1  \ldots n}\big)\,F_n^{c}({\bf k}_1,\ldots,{\bf k}_n|a)\,\delta_{(1)}({\bf k}_1,a)\ldots\delta_{(1)}({\bf k}_n,a)
\nonumber\\
&\  + f_{\text{NL}}\int_{{\bf k}_1}\ldots\int_{{\bf k}_n}(2\pi)^3\delta_D\big({\bf k} - {\bf k}_{1  \ldots n}\big)\,H_n^{c}({\bf k}_1,\ldots,{\bf k}_n|a)\,\psi({\bf k}_1)
\ldots\delta_{(1)}({\bf k}_n,a)  \ .
\end{align}
Note that we have not specified the time dependence of the free parameters yet, which is why the counterterm kernels are still time dependent. It turns out that at first order in the perturbations, this is not really an issue, as the time dependence is just given by an integral of the Green's function over some unknown function of time, which yields some other unknown function. However, in order to get the momentum dependence right at second order, we have to make an assumption about the first order terms. A convenient ansatz is 
\begin{align}\label{equ:lotimedep}
d^2(a) &= [\H(a)f(a)]^2[D_1(a)]^{m_d+1}\,\bar d^{\hskip 1.5pt 2}\ , \\[4pt]
g(a) &= [\H(a)f(a)]^2[D_1(a)]^{m_g+1}\,\bar g\ , \label{eq:timeansatz}
\end{align}
which in Einstein-de Sitter reduces to $(d^{2},g) \propto a^{m_d}$. 
Then the expressions for the counterterm kernels at second order are as follows. We split up the kernels in the following way 
\begin{align}
F_2^{c}({\bf k}_1,{\bf k}_2|a) &= F_2^{\tau}({\bf k}_1,{\bf k}_2|a)+ F_{2}^{\alpha\beta}({\bf k}_1,{\bf k}_2|a) + F_2^{\delta}({\bf k}_1,{\bf k}_2|a),
\end{align} 
where the terms coming from the new counterterms at second order are
\begin{align}
F_2^{\tau}({\bf k}_1,{\bf k}_2|a)&= -\sum_{i=1}^3\epsilon_i(a) E_{i}({\bf k}_1,{\bf k}_2) , \text{ with} \\
E_1({\bf k}_1,{\bf k}_2)&\equiv {\bf k}_{12}^2 ,\\
E_2({\bf k}_1,{\bf k}_2)&\equiv {\bf k}_{12}^2\left[\frac{({\bf k}_1\cdot{\bf k}_2)^2}{k_1^2k_2^2}-\frac{1}{3}\right] ,\\
E_3({\bf k}_1,{\bf k}_2)&\equiv \left[-\frac{1}{6}{\bf k}_{12}^2+\frac{1}{2}{\bf k}_1\cdot{\bf k}_2\left[\frac{{\bf k}_{12}\cdot{\bf k}_2}{k_2^2}+\frac{{\bf k}_{12}\cdot{\bf k}_1}{k_1^2}\right]\right] , \text{ and} \\
\epsilon_i(a)&\equiv-\frac{1}{[D_1(a)]^2}\int_{a_{in}}^a\text{d} a'\ G_\delta(a,a')\, [D_1(a')]^2\,e_i(a')\ . 
\end{align}
Furthermore, the terms coming from plugging the lowest order counterterm back into the equations of motion are
	\begin{align}
	F_2^{\alpha\beta}({\bf k}_1,{\bf k}_2|a)&=-\xi(a)E_{\alpha\beta}({\bf k}_1,{\bf k}_2) , \text{ with}\\
E_{\alpha\beta}({\bf k}_1,{\bf k}_2)&\equiv\frac{1}{2m_{d}+9}\bigg[2\beta({\bf k}_1,{\bf k}_2)(k_1^2+k_2^2) \\
&\hskip 25pt+\frac{2m_{d}+7}{2(m_{d}+2)}\Big(\alpha({\bf k}_1,{\bf k}_2)\big(k_2^2+(m_{d}+2)k_1^2\big)
+\{1\leftrightarrow 2\}\Big)\bigg] , \text{ and}\\
\xi(a) &= \frac{2}{(m_{d}+1)(2m_{d}+7)}[D_1(a)]^{m_{d}+1}\,\bar d^{\hskip 1.5pt 2}\ .\label{eq:gamma}
\end{align}
Finally, the term from the second order contribution to the density is
\begin{align}
F_2^{\delta}({\bf k}_1,{\bf k}_2|a)& = -\xi(a)E_{\delta}({\bf k}_1,{\bf k}_2), \text{ with}\\
E_{\delta}({\bf k}_1,{\bf k}_2) &=\frac{(m_{d}+1)(2m_{d}+7)}{(m_{d}+2)(2m_{d}+9)}\,{\bf k}_{12}^2\,F_2({\bf k}_1,{\bf k}_2)\ .
\end{align}
Similarly, for the non-Gaussian kernels we have
\begin{align}
H_2^{c}({\bf k}_1,{\bf k}_2|a)=H_2^{\tau}({\bf k}_1,{\bf k}_2|a)+H_2^{\alpha\beta}({\bf k}_1,{\bf k}_2|a)+H_2^{\Psi}({\bf k}_1,{\bf k}_2|a)\ ,\label{eq:H2c}
\end{align}
where second order counterterm contribution is
\begin{align}
H_2^{\tau}({\bf k}_1,{\bf k}_2|a)&=-\sum_{i=1}^2\gamma_i(a) G_{i}({\bf k}_1,{\bf k}_2) , \text{ with}\\
G_1({\bf k}_1,{\bf k}_2)&={\bf k}_{12}^2 ,\\
G_2({\bf k}_1,{\bf k}_2)&=\frac{({\bf k}_{12}\cdot{\bf k}_{2})^2}{k_2^2}-\frac{1}{3}{\bf k}_{12}^2 , \text{ and}\\
\gamma_i(a) & \equiv-\frac{1}{D_1(a)}\int_{a_{in}}^a\text{d} a'\ G_\delta(a,a')\,D_1(a')g_i(a') .
\end{align}
Subsequently, the convoluted first order counterterms give 
\begin{align}
H_2^{\alpha\beta}({\bf k}_1,{\bf k}_2|a)	&=-\gamma(a) G_{\alpha\beta}({\bf k}_1,{\bf k}_2) ,
\text{ with} \\
G_{\alpha\beta}({\bf k}_1,{\bf k}_2) &\equiv \frac{4}{2m_g+7}\,\beta({\bf k}_1,{\bf k}_2)k_1^2 \\
&\quad +\frac{2m_g+5}{(m_g+1)(2m_g+7)}\big[(m_g+1)\alpha({\bf k}_1,{\bf k}_2)+\alpha({\bf k}_2,{\bf k}_1)\big]k_1^2 , \text{ and}\\
\gamma(a)& = \frac{2}{m_g(2m_g+5)}[D_1(a)]^{m_g+1}\,\bar g\ .\label{eq:beta1}
\end{align}
Lastly, the second order contribution to the lowest order counterterm is given by
\begin{align}
H_2^{\Psi}({\bf k}_1,{\bf k}_2|a)	&=-\gamma(a) G_{\Psi}({\bf k}_1,{\bf k}_2) , \text{ with}\\
G_{\Psi}({\bf k}_1,{\bf k}_2)& = \frac{m_g(2m_g+5)}{(m_g+1)(2m_g+7)}\,\left[{\bf k}_{12}^2\frac{{\bf k}_1\cdot{\bf k}_2}{k_2^2}-{\bf k}_{12}\cdot{\bf k}_1\right]\ .
\end{align}


\subsection{One Loop Bispectrum}

Having obtained perturbative solutions for the evolved density contrast in terms of the initial field, we can compute correlation functions using the statistical properties of the initial distribution. Along the lines of the discussion above, we decompose the bispectrum as
\begin{align}
B_{\text{tot}}=B_{\text{SPT}}^{\text{G}}+B_{\text{EFT}}^{\text{G}}+f_{\text{NL}}\left(B_{\text{SPT}}^{\text{NG}}+B_{\text{EFT}}^{\text{G}}\right).
\end{align}
The expressions for the Gaussian part of the bispectrum at one loop were given in \cite{Baldauf:2014qfa} and \cite{Angulo:2014tfa}, and read
\begin{align}
B^{\rm G_{\phantom 1}}_{\rm SPT}&=B_{112} + \Big[B_{114}+B_{123}^{\rm (I)}+B_{123}^{\rm (II)}+B_{222}\Big] , \label{eq:BGloop}\\
B^{\rm G_{\phantom 1}}_{\rm EFT} &=  \xi B_\xi^{\rm G_{\phantom 1}} + \sum_{i=1}^3 \epsilon_i B_{\epsilon_i} , \label{BG} 
\end{align}
where
\begin{subequations}
	\begin{align}
	B_{112} =&  2F_2(\v{k}_1,\v{k}_2)P_{11}(k_1)P_{11}(k_2) +  \text{2 cycl. perms} \\
	B_{222} =& 8\int_\v{p} F_{2}(-{\bf p},{\bf p}+{\bf k}_1)F_{2}({\bf p}+{\bf k}_1,-{\bf p}+{\bf k}_2)F_{2}({\bf k}_2+-{\bf p},{\bf p})\\
	& \quad P_{11}(p)P_{11}(|{\bf p}+{\bf k}_1|)P_{11}(|{\bf p}-{\bf k}_2|), \nonumber \\
	B^{{(\rm I)}}_{321} =& 6P_{11}(k_3)\int_\v{p} F_{3}(-{\bf p},\v{p}-\v{k}_2,-\v{k}_3)F_2(\v{p},\v{k}_2-\v{p})P_{11}(p)P_{11}(|\v{p}-\v{k}_2|) + \text{5 perms} , \label{equ:B113-v2}\\
		B^{{(\rm II)}}_{321} =& 6F_2(\v{k}_2,\v{k}_3)P_{11}(k_2)P_{11}(k_3)\int_\v{p} F_{3}(-{\bf k}_3,\v{p},-\v{p})P_{11}(p) + \text{5 perms}, \\
	B_{411} =& 12P_{11}(k_2)P_{11}(k_3)\int_\v{p} F_{4}(\v{p},-\v{p},-\v{k}_2,-\v{k}_3)P_{11}(p) + \text{2 cycl. perms}. \label{equ:B122-v2}
	\end{align}
\end{subequations} 
Here $P_{11}(k)$ is the linear power spectrum, whose time dependence is implied. Furthermore
\begin{subequations}
	\begin{align}
	B_\xi^{\rm G_{\phantom 1}}&\equiv-2\big[E_{\alpha\beta}({\bf k}_1,{\bf k}_2)+E_{\delta}({\bf k}_1,{\bf k}_2)\big]P_{11}(k_1)P_{11}(k_2) + \text{2 perms} ,\\[4pt]
	B_{\epsilon_i}&\equiv-2\hskip 1ptE_i({\bf k}_1,{\bf k}_2)\hskip 1ptP_{11}(k_1)P_{11}(k_2) + \text{2 perms}.
	\end{align}
\end{subequations}
The non-Gaussian contribution at one loop is
\begin{align}
B^{\rm NG_{\phantom 1}}_{\rm SPT}&=B_{111} + \Big[B_{113}^{\rm (I)}+B_{113}^{\rm (II)}+B_{122}^{\rm (I)}+B_{122}^{\rm (II)}\Big],\label{eq:BNGloop} \\
B^{\rm NG_{\phantom 1}}_{\rm EFT} &= \xi B_\xi^{\rm NG_{\phantom 1}} + \gamma B_\gamma + \sum_{i=1}^2 \gamma_i  B_{\gamma_i}  , \label{BNG}
\end{align}
where
\begin{subequations}
	\begin{align}
	B^{{(\rm I)}}_{113} &= 3P_{11}(k_2)\int_\v{p} F_{3}(\v{k}_1+\v{p},-\v{p},\v{k}_2)B_{111}(k_1,p,|\v{k}_1+\v{p}|) + \text{5 perms}, \label{equ:B113-v1}\\
	B^{{(\rm II)}}_{113} &= 3B_{111}(k_1,k_2,k_3)\int_\v{p} F_{3}(\v{k}_1,\v{p},-\v{p})P_{11}(p) + \text{2 perms} , \\
	B^{{(\rm I)}}_{122} &= 4\int_\v{p} F_2(\v{k}_3+\v{p},-\v{p})F_{2}(\v{p},\v{k}_2-\v{p})B_{111}(k_1,|\v{k}_3+\v{p}|,|\v{k}_2-\v{p}|)P_{11}(p) + \text{2 perms}, \label{equ:B122-v1}\\
	B^{{(\rm II)}}_{122} &= 2F_{2}(\v{k}_1,\v{k}_2)P_{11}(k_2)\int_\v{p} F_{2}(\v{p},\v{k}_1-\v{p})B_{111}(k_1,p,|\v{k}_1-\v{p}|) + \text{5 perms}, 
	\end{align}
\end{subequations} 
and
\begin{subequations}
\begin{align}
B_\xi^{\rm NG_{\phantom 1}} &\equiv-(k_1^2+k_2^2+k_3^2)B_{111}(k_1,k_2,k_3) \label{equ:BxiNG},\\[4pt]
B_{\gamma}&\equiv -\big[G_{\alpha\beta}({\bf k}_1,{\bf k}_2)+G_{\Psi}({\bf k}_1,{\bf k}_2)\big]P_{11}(k_1)P_{1\psi}(k_2)+\text{5 perms} ,\\[4pt]
B_{\gamma_i}&\equiv -G_i({\bf k}_1,{\bf k}_2)P_{11}(k_1)P_{1\psi}(k_2)+\text{5 perms}.
\end{align}
\end{subequations}
Again, the time dependence of the correlation functions is implied. Finally, we have to specify the type of non-Gaussianity. In this work we consider the primordial bispectra given in \ref{subsec:shapesPNG}. The corresponding correlation between the linear density field and the first order non-Gaussian counterterm $\psi$ (see \eqref{equ:ngcounter}) is given by
\begin{align}
P_{1\psi}(k)\equiv \frac{(k/\mu)^\Delta}{M(k)}P_{11}(k) ,\label{eq:psidelta}
\end{align}
where $\mu$ is some arbitrary momentum scale, introduced for dimensional reasons, which cancels when multiplied with the EFT parameter in the full contribution to the bispectrum. We therefore set it to unity in the numerical evaluation. In our case the $k$-dependence is respectively given by $\Delta=\{0,1,2\}$ for local, quasi-single field and equilateral type non-Gaussianities. The transfer function was defined in \eqref{eq:transferfunction}. For this work we choose the time dependence of the lowest order counterterms \eqref{equ:lotimedep} to match the divergence it is supposed to cancel, which corresponds to the choice $m_{d}=m_{g}=1$. This is was argued for in \cite{Baldauf:2015aha}. Moreover, our main results do not depend much on this assumption.  

\subsection{Ansatz Two Loop Bispectrum}

As an ansatz for the two loop bispectrum we compute the two reducible two loop diagrams, given by \cite{Baldauf:2014qfa}, 
\begin{subequations}
	\begin{align}
	B_{332}^{{\rm I}} =&  2F_2(\v{k}_1,\v{k}_2)\frac{P_{13}(k_1)}{2}\frac{P_{13}(k_2)}{2} +  \text{2 cycl. perms} \\
	B^{{\rm II}}_{332} =& 6\frac{P_{13}(k_3)}{2}\int_\v{p} F_{3}(-{\bf p},\v{p}-\v{k}_2,-\v{k}_3)F_2(\v{p},\v{k}_2-\v{p})P_{11}(p)P_{11}(|\v{p}-\v{k}_2|) + \text{5 perms} , 
	\end{align}
\label{eq:B332}
\end{subequations} 
with
\begin{align}
P_{13}(k) =  6P_{11}(k)\int_\v{p} F_{3}({\bf k},\v{p},-\v{p})P_{11}(p).
\end{align}	
As an estimate for the theoretical error we use
\begin{align}
B_{332}=|B_{332}^{{\rm I}}|+|B_{332}^{{\rm II}}|.
\end{align}


\section{Theoretical noise}
\label{app:theoreticalnoise}
 This appendix contains the details of the implementation of the theoretical error and further investigates some issues related to it. First, we give the intermediate steps to derive \eqref{eq:effectivefishermatrix} and provide an alternative derivation of the effective Fisher matrix in the presence of theoretical error. Then, we study the effect of the correlation length, both by means of a toy model and by running the analysis for several correlations lengths. Finally, we discuss in more detail the effect of the two possible ans\"atze for $\Ber$. 

\subsection{Derivation of \eqref{eq:effectivefishermatrix}}
Let us first show how to go from equations \eqref{eq:fullfishermatrix}, \eqref{eq:fisheralphabeta} and \eqref{eq:fisheralphai} to the effective Fisher matrix given in \eqref{eq:effectivefishermatrix}. We will need to use the Woodbury matrix identity several times, which relates the inverse of sums of matrices to their individual inverses
\begin{equation}
(A+B)^{-1} = A^{-1}-A^{-1}\left(A^{-1}+B^{-1}\right)^ {-1}A^{-1}.
\label{eq:woodburyidentity}
\end{equation}
By using this identity, we can rewrite \eqref{eq:fisheralphabeta} as
\begin{equation}
F_{\alpha\beta}^{-1}  = (N_{\alpha\beta}^{-1}+D_{\alpha\beta})^{-1}= D_{\alpha\beta}^{-1} -D_{\alpha\gamma}^{-1}\left(N+D^{-1}\right)^{-1}_{\gamma\delta}D_{\delta\beta}^{-1}.
\end{equation}
This allows us to compute
\begin{align*}
F_{i\gamma} F^{-1}_{\gamma\delta} F_{\delta j} &\ = \sum_{\bin{k},\bin{p}}B_i(\bin{k}) C^{-1}(\bin{k},\bin{k}_\gamma) \Ber(k_\gamma) \left(D^{-1}-D^{-1}(N+D^{-1})^{-1}D^{-1}\right)_{\gamma\delta} \Ber(k_\delta) C^{-1}(\bin{k}_\delta, \bin{p})B_j(\bin{p})  \\
&\ = \sum_{\bin{k},\bin{p}} B_i(\bin{k})\delta_{\bin{k},\bin{k}_\alpha} \left(C^{-1}(\bin{k}_\alpha,\bin{k}_\beta)  - \frac{1}{ \Ber(k_\alpha)}(N+D^{-1})^{-1}_{\alpha\beta}\frac{1}{ \Ber(k_\beta)}\right)\delta_{\bin{k}_\beta, \bin{p}} B_j(\bin{p})\\
&\ = \sum_{\bin{k}_\alpha,\bin{k}_\beta} \frac{B_i(\bin{k}_\alpha)}{\Ber(k_\alpha)} \left(D - (N+D^{-1})^{-1}\right)_{\alpha\beta} \frac{B_i(\bin{k}_\beta)}{\Ber(k_\beta)} \\
&\ = \sum_{\bin{k}_\alpha,\bin{k}_\beta} \frac{B_i(\bin{k}_\alpha)}{B_{\alpha}}\left(D(N^{-1}+D)^{-1}D\right)_{\alpha\beta}\frac{B_i(\bin{k}_\beta)}{B_{\beta}} .
\end{align*}
Here summation over the Greek indices is understood.
Upon applying the Woodbury identity again, the effective Fisher matrix then becomes
\begin{align}
\begin{split}
F_{ij}^{\text{eff}} &\ =   \sum_{\bin{k}_\alpha,\bin{k}_\beta} \frac{B_i(\bin{k}_\alpha)}{B_{\alpha}} \left(D-D(N^{-1}+D)^{-1}D\right)_{\alpha\beta}  \frac{B_j(\bin{k}_\beta)}{B_{\beta}}+\left(C^{-1}_{\Theta}\right)_{ij}\\
&\ = \sum_{\bin{k}_\alpha,\bin{k}_\beta} \frac{B_i(\bin{k}_\alpha)}{B_{\alpha}} \left(N+D^{-1}\right)^{-1}_{\alpha\beta}  \frac{B_j(\bin{k}_\beta)}{B_{\beta}}+\left(C^{-1}_{\Theta}\right)_{ij}\\
&\ =\sum_{\bin{k}_\alpha,\bin{k}_\beta} B_i(\bin{k}_\alpha) \left( N^{\text{eff}}+C_B \right)^{-1}_{\alpha\beta}  B_j(\bin{k}_\beta)+\left(C^{-1}_{\Theta}\right)_{ij},
\end{split}
\end{align}
with $N_{\alpha\beta}^{\text{eff}} =  \Ber(k_\alpha) N_{\alpha\beta}  \Ber(k_\beta)$.


\subsection{Alternative derivation of the effective Fisher matrix}
Next, we present a slightly different derivation of the effective Fisher matrix, by marginalizing at the level of the likelihood function. 
Let us first expand $\chi^2 = -2\log(\L)$ in the nuisance parameters $\Theta_\alpha$. We would like to expand about some value $\bar{\Theta}_\alpha$ to get
\begin{equation}
 \chi^2(\Theta_i, \Theta_\alpha) = \chi^2(\Theta_i,\bar{\Theta}_\alpha) +  (\Theta_\alpha-\bar{\Theta}_\alpha)\chi^2_\alpha(\Theta_i,\bar{\Theta}_\alpha) +\frac{1}{2} (\Theta_\alpha-\bar{\Theta}_\alpha)(\Theta_\beta-\bar{\Theta}_\beta) \chi^2_{\alpha\beta}(\Theta_i,\bar{\Theta}_\alpha),
\end{equation}
where summation over repeating indices is understood, and the index $\alpha$ on $\chi^2$ denotes a derivative with respect to the corresponding nuisance parameter. It is an equality, since the variables are Gaussian distributed. We can rewrite this expression in more compact notation as
\begin{equation}
\chi^2 = \chi^2 _0 + \delta \Theta^\alpha X_\alpha + \frac{1}{2}\delta \Theta^\alpha \delta \Theta^\beta Y_{\alpha \beta},
\end{equation}
where $\chi^2 _0 $ is the chi-squared we would get if we ignored the presence of the nuisance parameters $\Theta_\alpha$. By completing the square and adding some prior information on the nuisance parameters (i.e. a covariance matrix), we can integrate them out to get an effective chi-squared. In other words, we would like to evaluate the following integral
\begin{equation}
\int d^N \Theta_\alpha \exp \left( -\frac{1}{2}\left(\chi^2 _0 + \delta \Theta^\alpha X_\alpha + \frac{1}{2} \delta \Theta^\alpha \delta \Theta^\beta Y_{\alpha \beta}\right) \right) \exp \left(-\frac{1}{2} \delta \Theta^\alpha (N^{-1})_{\alpha \beta} \delta \Theta^{\beta} \right).
\end{equation}
where $N_{\alpha \beta}$ is the covariance matrix of the theoretical error parameters. The integration results into
\begin{equation}
\sqrt{(2\pi)^N\cdot \det((\tfrac{1}{2}Y+N^{-1})^{-1})}\ \exp\left( -\frac{1}{2}\chi^2 _0 +\frac{1}{4} X_\gamma \left(Y+ 2 N^{-1}\right)^{-1}_{\gamma\delta} X_\delta \right),
\end{equation}
and therefore,
\begin{equation}
\chi^2_{\text{eff}} = \chi^2 _0 - \frac{1}{2} X_\gamma \left(Y+ 2 N^{-1}\right)^{-1}_{\gamma\delta} X_\delta  + \ln\left(\det((\tfrac{1}{2}Y+N^{-1}))\right).
\end{equation}
Please note that all these terms do in general depend on $\Theta_i$. Since the joint probability distribution of all parameters is a multivariate Gaussian, we know that $Y$ is independent of $\Theta_i$ and $X$ only depends linearly on $\Theta_i$. In that case, we get
\begin{equation}
\left(\chi^2_{\text{eff}}\right)_{ij} = \left(\chi^2_0\right)_{ij} - \frac{1}{2} X_{i \gamma} \left(Y+ 2 N^{-1}\right)^{-1}_{\gamma\delta} X_{\delta j}.
\label{eq:chisquaredeffective}
\end{equation}
The full Fisher matrix is given by
\begin{equation}
F_{\mu\nu} 
=
\begin{pmatrix}
\tfrac{1}{2} Y_{\alpha\beta} +  N^{-1}_{\alpha\beta} & \tfrac{1}{2} X_{\alpha j} \\
 \tfrac{1}{2} X_{i \beta} & F_{ij}
 \end{pmatrix}
\end{equation}
where we have to evaluate the matrices at the maximum likelihood value of the parameters. This means that the effective chi-squared is given by
\begin{equation}
\left(\chi^2_{\text{eff}}\right)_{ij}  = \left(\chi^2_0\right)_{ij}  - 2 F_{i \gamma} F^{-1}_{\gamma\delta} F_{\delta j},
\end{equation}
or, in other words, the effective Fisher matrix for the theoretical parameters is given by
\begin{equation}
F^{\text{eff}}_{ij}  = F_{ij}  -  F_{i \gamma} F^{-1}_{\gamma\delta} F_{\delta j}.
\end{equation}
This is what we found before in the main text. \\

Alternatively, starting from \eqref{eq:chisquaredeffective} we can write down immediately the expression for the effective likelihood
\begin{align}
\L^{\text{eff}}&\ = \frac{1}{\sqrt{\det\left(\tfrac{1}{2}Y+N^{-1}\right)}}\exp\left[-\frac{1}{2}\left(\chi^2_0-X_\gamma(\tfrac{1}{2}Y+N^{-1})^{-1}_{\gamma\delta}X_\delta\right)\right]\nonumber\\
&\ = \frac{\exp\left[-\frac{1}{2}\sum_{\bin{k},\bin{p}}\Delta B(\bin{k})\left(C_B^{-1}(\bin{k},\bin{p})-C_B^{-1}(\bin{k},\bin{k}_\gamma) B_{2L}(k_\gamma)(D+N^{-1})_{\gamma\delta}^{-1} \Ber(k_\delta) C_B^{-1}(\bin{k}_\delta,\bin{p}) \right)\Delta B(\bin{p})\right]}{\sqrt{\det\left(D+N^{-1}\right)}}\nonumber\\
&\ = \frac{1}{\sqrt{\det\left(D+N^{-1}\right)}}\exp\left[-\frac{1}{2}\sum_{\bin{k}_\alpha,\bin{k}_\beta}\Delta B(\bin{k}_\alpha)\left(C_B+N^{\text{eff}} \right)^{-1}_{\alpha\beta}\Delta B(\bin{k}_\beta)\right].
\label{eqn:effectivelikelihood}
\end{align}
Here the difference between the data and theory vector $\Delta B$ is evaluated at the fiducial values for the nuisance parameters $\bar{\Theta}_\alpha$ (i.e. at zero in our case). Taking now two derivatives with respect to the remaining theoretical parameters $\Theta_i$, we find the effective Fisher matrix
\begin{equation}
F_{ij}^{\text{eff}}  = \sum_{\bin{k}_\alpha,\bin{k}_\beta} B_i(\bin{k}_\alpha)\left(C_B + N^{\text{eff}}\right)^{-1}_{\alpha\beta} B_j(\bin{k}_\beta)
\end{equation}
with $N_{\alpha\beta}^{\text{eff}} =\Ber(k_\alpha) N_{\alpha\beta} \Ber(k_\beta) $.


\subsection{Theoretical error - a toy model}
\label{appsub:toymodel}

In our approach, the theoretical error on the value of the bispectrum is modeled in the following way. For every bin, we introduce a nuisance parameter that is drawn from a Gaussian distribution with average zero and variance set by the estimated size of the theoretical error for that bin. Importantly, we allow for non-vanishing correlations among these nuisance parameters, i.e. we allow for a non-diagonal covariance matrix for them. The purpose of this appendix is to show that both the limit of zero and maximal correlation among the parameters have a clear interpretation, neither of which resembles the way we think the theoretical error should act. To be more precise, we prove, by means of a simple toy model that still captures the essence of the real analysis, the intuitive statements that:
\begin{itemize}
	\item  for zero correlation length, the theoretical error just acts as shot noise per bin;
	\item  for maximal correlation length, the theoretical error acts as some free coefficient multiplying a fixed shape function, which by definition we think is the wrong function.
\end{itemize}

\subsubsection{Toy model}

We consider measuring some observable $d$ a total of $N$ times and collecting the data $d_{i}$. Our model is $d_{i}=x+e_{i}$, with $x$ a Gaussian random variable with variance $\sigma_{x}^{2}$, whose average, $\bar{x}$, we would like to determine as well as possible. The $e_{i}$ are additional Gaussian variables that represent the systematic error or theoretical uncertainty in every measurement. Their averages and variances are $\bar{e_{i}}$ and $\sigma_{e_{i}}^{2}$, respectively. One can think of this scenario as determining the average weight of a group of people, knowing that their weights are Gaussian distributed with variance $\sigma_{x}^{2}$, where we use a different weighing scale with a systematic error $\bar{e_{i}}$ and some uncertainty in the measurement characterized by $\sigma_{m_{i}}^{2}$ every time we weigh someone. Since the $e_{i}$ are uncorrelated with $x$, this leads to the likelihood
   \begin{align}
   \log L=-\sum_{i=1}^{N}\frac{(d_{i}-\bar{x}-\bar{e_{i}})^{2}}{2\sigma^{2}_{d}},
   \end{align}    
where $\sigma_{d}^{2}=\sigma_{x}^{2}+\sigma_{m}^{2}$, assuming for convenience that $\sigma_{m_{i}}=\sigma_{m}$ (the arguments below do not depend on this assumption). Without any prior information on the systematic errors, they are completely degenerate with $\bar{x}$, so we do not expect to be able to learn anything about $\bar{x}$ in this case. This can be verified using a Fisher analysis. We have
 \begin{align}
 F_{ab}= \begin{pmatrix}
 \frac{N}{\sigma_{d}^{2}} & \frac{1}{\sigma^{2}_{d}} \overrightarrow{1}^{T}  \\
 \frac{1}{\sigma_{d}^{2}} \overrightarrow{1}& \frac{1}{\sigma^{2}_{d}}1_{N\times N}  \\
 \end{pmatrix},
 \end{align}
where $a,b=\bar{x},\bar{e_{i}}$. Since we are ignorant about the systematic errors, we compute the marginalized error on $\bar{x}$,
\begin{align}
\sigma_{\bar{x},marg}^{2}=\left(F^{-1}\right)_{\bar{x}\bar{x}},
\end{align}
which can be computed using the block matrix inversion formula (see also \eqref{eq:fullfishermatrix}): 
\begin{align}\label{eq:blocks}
\text{let} \quad
F= \begin{pmatrix}
A & B^{T}  \\
B & D  \\
\end{pmatrix},
\end{align}
then 
\begin{align}
\sigma_{\bar{x},marg}^{2}=(A-B^{T}D^{-1}B)^{-1}=\left(\frac{N}{\sigma_{d}^{2}}-\frac{1}{\sigma_{d}^{2}}\overrightarrow{1}^{T}1_{N\times N}\overrightarrow{1} \right)^{-1}=\frac{1}{0},
\end{align}
as expected. In a realistic situation we do have some prior information about the systematic errors. Here, and in the paper, we assume they are also Gaussian random variables with some variance $\sigma_{\bar{e_{i}}}^{2}$. Moreover, we allow for nontrivial correlations among the $\bar{e_{i}}$, which for the scales could mean they were produced by the same machine for instance. This means we obtain the updated likelihood 
\begin{align}
\log L=-\sum_{i=1}^{N}\frac{(d_{i}-\bar{x}-\bar{e_{i}})^{2}}{2\sigma^{2}_{d}}-\bar{e_{i}}\left(C^{-1}\right)_{ij}\bar{e_{j}},
\end{align}
where 
\begin{align}
C_{ij}=<\bar{e_{i}}\bar{e_{j}}>.
\end{align}
In the following, we investigate the effect of zero and maximal correlation length on $\sigma_{\bar{x},marg}^{2}$. 
 
\subsubsection{Zero correlation}

Let us first assume zero correlation among the systematic errors, leading to a diagonal covariance matrix, 
\begin{align}
C_{ij}=\sigma_{i}^{2}\delta_{ij}.
\end{align}
In terms of the weighing scales this could mean all scales really come from different companies with uncorrelated systematic errors. We now show that in this case the ignorance about the systematic errors acts as shot noise per bin; it simply updates the variance of the measurements $\sigma_{d}^{2} \to \sigma_{d_{i}}^{2}$. For any covariance matrix, the Fisher matrix is 
 \begin{align}
 F= \begin{pmatrix}
 \frac{N}{\sigma_{d}^{2}} & \frac{1}{\sigma^{2}_{d}} \overrightarrow{1}^{T}  \\
 \frac{1}{\sigma_{d}^{2}} \overrightarrow{1}& \frac{1}{\sigma^{2}_{d}}\delta_{ij}+(C^{-1})_{ij}  \\
 \end{pmatrix}.
 \end{align}
Then, using the block matrix inversion formula, zero correlation leads to an error
\begin{align}
\sigma_{\bar{x},marg}^{2}&=\left(\frac{N}{\sigma_{d}^{2}}-\frac{1}{\sigma_{d}^{4}}\overrightarrow{1}^{T}\frac{1}{\frac{1}{\sigma^{2}_{d}}+\frac{1}{\sigma_{i}^{2}}}\delta_{ij}\overrightarrow{1} \right)^{-1}\\
&=\left[\sum_{i=1}^{N}\left(\frac{1}{\sigma_{d}^{2}}-\frac{1}{\sigma_{d}^{4}}\frac{1}{\frac{1}{\sigma^{2}_{d}}+\frac{1}{\sigma_{i}^{2}}}\right) \right]^{-1}\\
&=\left[\sum_{i=1}^{N}\left(\frac{1}{\sigma_{d}^{2}+\sigma_{i}^{2}}\right) \right]^{-1}\equiv\left[\sum_{i=1}^{N}\frac{1}{\sigma_{d_{i}}^{2}} \right]^{-1},
\end{align}
which is the same error one gets from assuming the likelihood function
 \begin{align}
 \log L=-\sum_{i=1}^{N}\frac{(d_{i}-\bar{x})^{2}}{2\sigma^{2}_{d_{i}}}.
 \end{align} 
This shows that indeed for zero correlation the systematic errors acts as shot noise per bin. In particular, this means the error on $\bar{x}$ can be made arbitrarily small by increasing the number of measurements (if the $\sigma_{\bar{e_{i}}}$ do not grow too fast for additional measurements). The intuitive reason is of course that in this model we expect the systematic errors to average out to zero in the long run. This is clearly not what is expected of the theoretical error in the measurement of the bispectrum.  

\subsubsection{Maximal correlation}
\label{app:correlations}  

Next we assume maximal correlation, which by definition means
\begin{align}
C_{ij}=<\bar{e_{i}}\bar{e_{j}}>=\sqrt{<\bar{e_{i}}^{2}>}\sqrt{<\bar{e_{j}}^{2}>}=\sigma_{i}\sigma_{j}. 
\end{align}
Since this matrix has rank one (all columns are multiples of the same vector), it is not invertible in more than one dimension. One way to deal with this is to introduce a regulator, such as a small matrix $\epsilon \delta_{ij}$, to break the degeneracy. Using our block matrix inversion formula, this is however not necessary. In the notation of (\eqref{eq:blocks}), we wish to compute $D^{-1}$. Let us write $D=S+C^{-1}$, where $S=\frac{1}{\sigma^{2}_{d}}\delta_{ij}$. The Woodbury identity then gives
\begin{align}
(S+C^{-1})^{-1}=S^{-1}-S^{-1}(S^{-1}+C)^{-1}S^{-1}.
\end{align}
Hence, we have to compute the inverse of $S^{-1}+C$, where $S^{-1}=\sigma^{2}_{d}\delta_{ij}$, and $C=\sigma_{i}\sigma_{j}$. Conveniently, since $C$ is of the form $\overrightarrow{\sigma}(\overrightarrow{\sigma})^{T}$, we can use the Sherman-Morrison formula to compute the inverse
\begin{align}
(S^{-1}+C)^{-1}=S-\frac{S(\sigma_{i}\sigma_{j})S}{1+\sigma_{i}S_{ij}\sigma_{j}}.
\end{align}    
 Plugging this into the previous formula, we find
 \begin{align}
 D^{-1}=(S+C^{-1})^{-1}=S^{-1}-S^{-1}+\frac{\sigma_{i}\sigma_{j}}{1+\frac{\left(\sum\sigma_{i}^{2}\right)}{\sigma_{d}^{2}}} = \frac{\sigma_{i}\sigma_{j}}{1+\frac{\left(\sum\sigma_{i}^{2}\right)}{\sigma_{d}^{2}}}.
 \end{align}  
 This finally leads to the error on $\bar{x}$:
 \begin{align}
\sigma_{\bar{x},marg}^{2}=\left[\frac{N}{\sigma_{d}^{2}}-\frac{\frac{\left(\sum\sigma_{i}\right)^{2}}{\sigma_{d}^{2}}}{\sigma_{d}^{2}+\left(\sum\sigma_{i}^{2}\right)}\right]^{-1}.
 \end{align}
 In order to interpret this result, let us rewrite this expression as follows
 \begin{align}
 \sigma_{\bar{x},marg}^{2}=\left[\frac{N}{\sigma_{d}^{2}}-\frac{\left(\sum\frac{\sigma_{i}}{\sigma}\right)^{2}}{\sigma_{d}^{4}}\frac{1}{\frac{1}{\sigma^{2}}+\frac{\sum\left(\frac{\sigma_{i}}{\sigma}\right)^{2}}{\sigma_{d}^{2}}}\right]^{-1},
 \end{align}
 where we have introduced the dimensionful parameter $\sigma$ to keep the dimensions clean. Now observe that we get exactly the same error on $\bar{x}$ from the following likelihood function
 \begin{align}
 \log L=-\sum_{i=1}^{N}\frac{(d_{i}-\bar{x}-\frac{\sigma_{i}}{\sigma}\bar{e})^{2}}{2\sigma^{2}_{d}}-\frac{\bar{e}^{2}}{2\sigma^{2}}.
 \end{align}
 whose Fisher matrix is
 \begin{align}
 F= 
 \begin{pmatrix}
 \frac{N}{\sigma_{d}^{2}} & \frac{1}{\sigma^{2}_{d}} \sum\frac{\sigma_{i}}{\sigma}  \\
\frac{1}{\sigma^{2}_{d}} \sum\frac{\sigma_{i}}{\sigma} & \frac{1}{\sigma^{2}}+\frac{\sum\left(\frac{\sigma_{i}}{\sigma}\right)^{2}}{\sigma_{d}^{2}}  \\
 \end{pmatrix},
 \end{align}
 This means that the maximal correlation case is equivalent to having a single, unknown parameter multiplying a known `shape' function $\sigma_{i}/\sigma$. In terms of the weighing scales this would mean that we know in advance exactly the ratios between the systematic errors of the scales. In terms of the bispectrum this would mean that we claim to know the theoretical error is exactly some number times the two loop estimate we put in, which it is clearly not. Finally note that if we choose all $\sigma_{i}$ to be equal, which for convenience we take to be $\sigma$, we find
  \begin{align}
  \sigma_{\bar{x}}^{2}=\left[\frac{N}{\sigma_{d}^{2}}-\frac{N^{2}}{\sigma_{d}^{4}}\frac{1}{\frac{1}{\sigma^{2}}+\frac{N}{\sigma_{d}^{2}}}\right]^{-1}=\frac{\sigma_{d}^{2}}{N}+\sigma^{2},
  \end{align}
 meaning the error on $\bar{x}$ can never get below the uncertainty in the degenerate parameter $\bar{e}$. In terms of the weighing problem this makes perfect sense, as this case is equivalent to simply using one and the same scale for every measurement. In this case we never expect to beat the unknown systematic error in the scale. In terms of the bispectrum this shows the importance of the relation between the shapes of the non-Gaussian signal and the theoretical error. In fact, in the maximal correlation limit we treat the theoretical error exactly the same as the EFT terms. 
 
 \subsubsection{Conclusions}
 
 From the above example it is clear that in neither limit the implementation of the theoretical error is completely satisfactory. Moreover, if the shapes are not too similar, the estimates from both limits are probably too optimistic. For this reason we recommend a conservative use of the method. In particular, we choose to use the correlation length that gives the weakest constraints on $f_{\text{NL}}$, as we show in the next subsection.  


\subsection{Choice of correlation length}
\begin{SCfigure}
  \centering
    \includegraphics[width=0.7\textwidth]{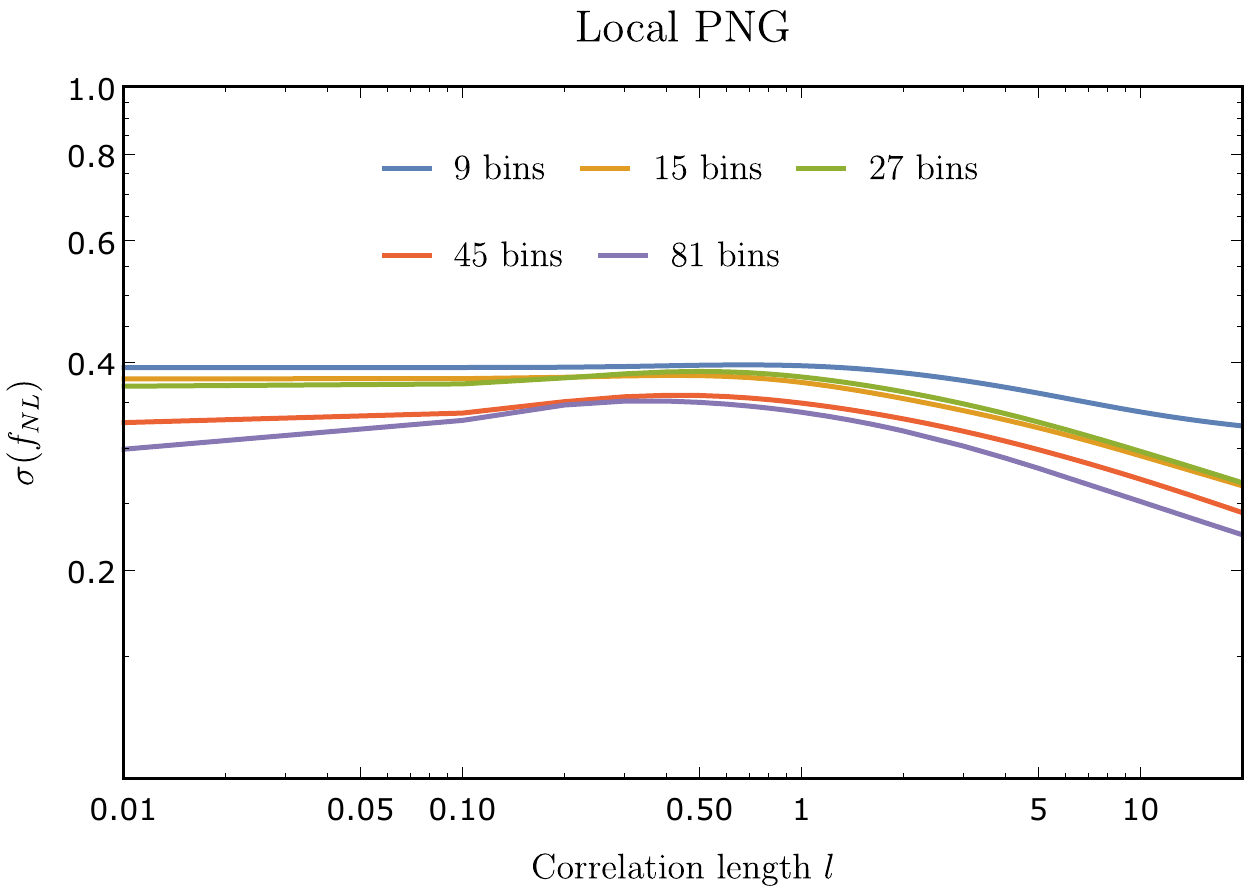}
  \caption{A test computation of $\sigma(f_{NL})$ at redshift $z=0$ including theoretical error as function of correlation length. We choose $k_{\text{min}}=0.001\ h \text{Mpc}^{-1}$ and $k_{\text{max}}=1 \ h \text{Mpc}^{-1}$ where we divide the $k$-range in 9 (blue), 15 (orange), 27 (green), 45 (red) and 81 (purple) bins.}
\label{fig:correlationlength}
\end{SCfigure}

In order to find the most conservative correlation length to work with, we ran a test computation of $\sigma(f_{\text{NL}})$. We did this at redshift zero with $k_{\text{min}}=0.001\ h \text{Mpc}^{-1}$ and $k_{\text{max}}=1 \ h \text{Mpc}^{-1}$, where we divide the $k$-range in 9 (blue), 15 (orange), 27 (green), 45 (red) and 81 (purple) bins. We find that the weakest constraints are obtained for $l\approx 0.5$, see Figure \ref{fig:correlationlength}. This is therefore the value we take for the analysis in the paper. 

Remarkably, the error is actually very insensitive to the correlation length, despite the very different nature of the effect of small and large correlation length. We believe the reason for this to be the fact that our ansatz for the theoretical error is a much steeper function of $k$ than the non-Gaussian signal. The transition from the $k$'s for which the error is negligible to the region where it is completely dominant is therefore very small, and the shape of the error is therefore not very important in this case. 

Another observation is that the error keeps increasing as we increase the correlation length beyond 10 decades, whereas the $k$'s we consider only run over a couple of decades. This makes the nuisance parameters almost maximally correlated for all these large correlation lengths. At the moment, we have no good explanation for the fact that the error seems to keep improving, other than it being a numerical fluke, perhaps related to the inversion of the correlation matrix.   

\subsection{Ans\"atze for higher loop corrections}
\label{appsub:comparisonansatze}
\begin{figure}[h]
\begin{subfigure}{0.42\textwidth}
\includegraphics[width=\linewidth]{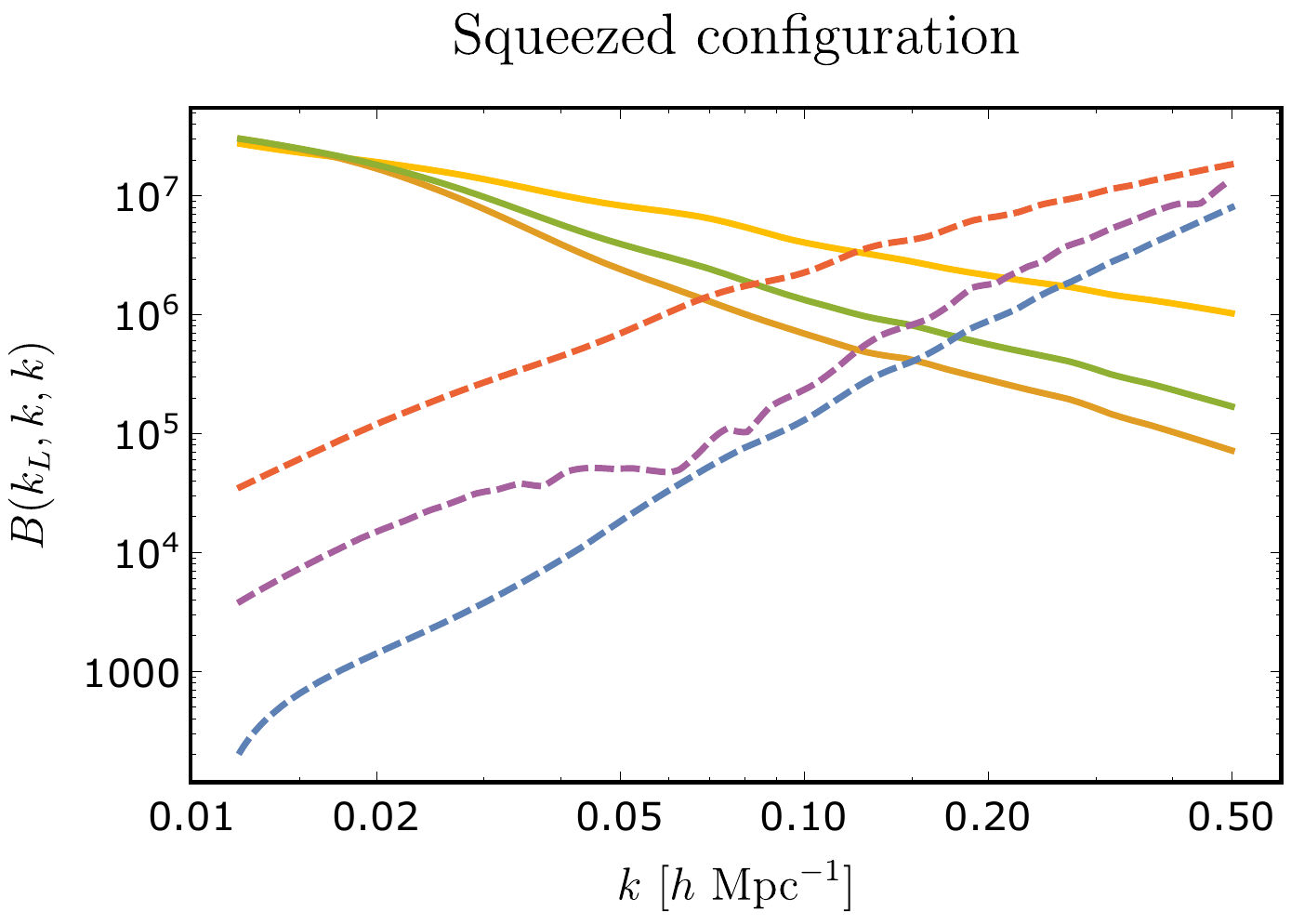} 
\end{subfigure}
\begin{subfigure}{0.42\textwidth}
\includegraphics[width=\linewidth]{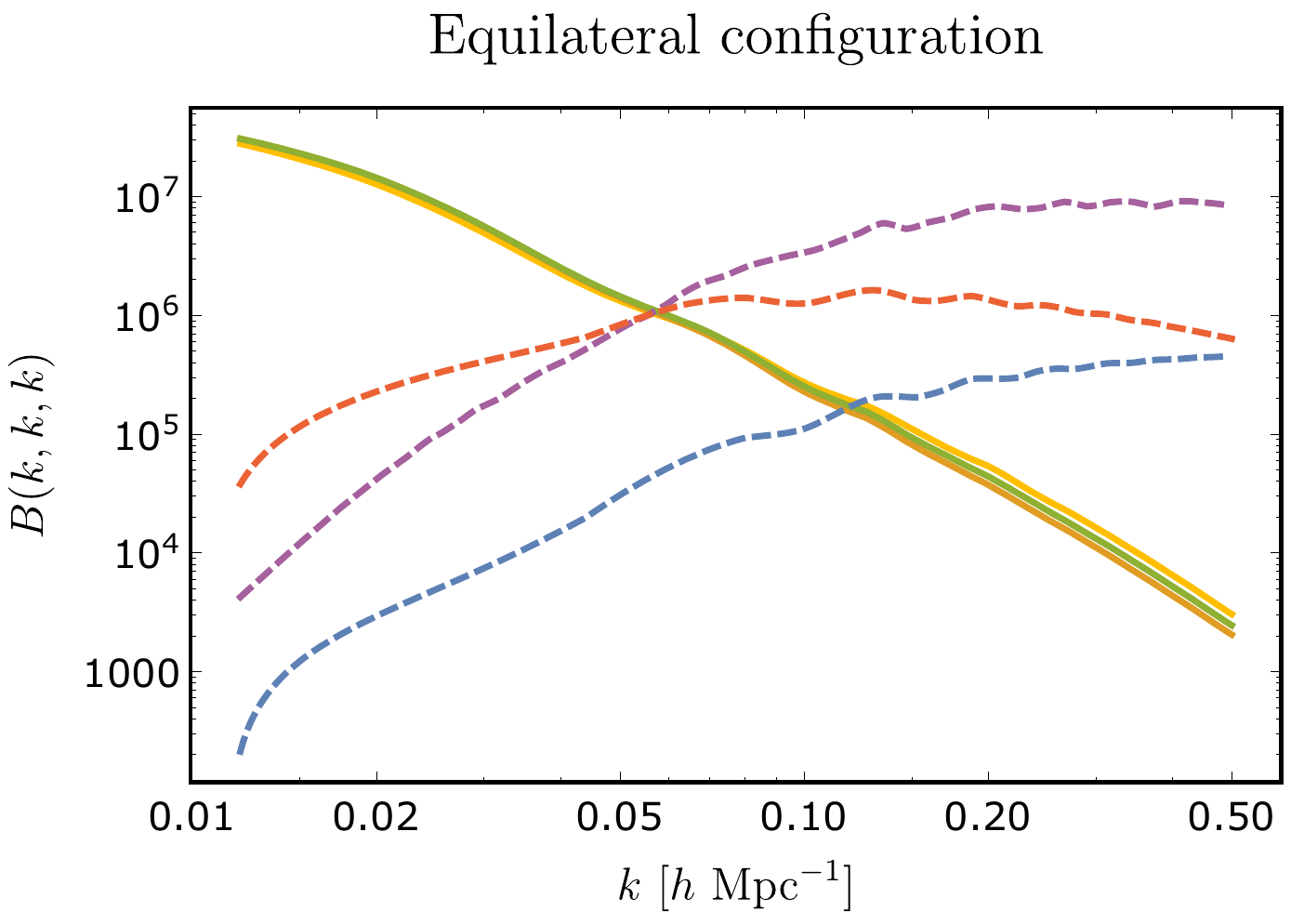} 
\end{subfigure}
\begin{subfigure}{0.14\textwidth}
\includegraphics[width=\linewidth]{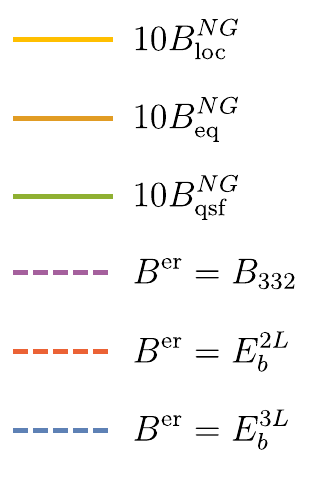}
\end{subfigure}
 \caption{Comparison of the ans\"atze for the higher loop corrections. We plot $E_b$ from equation  \eqref{eqn:theoreticalnoise2loopansatz} for two and three loops (yellow and green dashed lines) versus $B_{332}$ (green dashed line) defined in equation \eqref{eq:B332}. For the two loop ansatz using $E_b$ we take $n=-1.4$, $k_\text{NL} = 0.45 h \text{Mpc}^{-1}$ and $l=2$ and for the three loop ansatz we use $n=-1.5$, 
$k_\text{NL} = 0.50 h \text{Mpc}^{-1}$ and $l=3$. We compare these ans\"atze with the non-Gaussian contribution to the bispectrum up to one loop with $f_{NL} =10$ for local, equilateral and quasi-single-field PNG (red, blue and purple solid lines). In the left panel we compare the different contributions in the configuration $B(k_L, k ,k)$ where we varied $k$ and fixed $k_L = 0.012  h \text{Mpc}^{-1}$. The smaller $k$ the more squeezed the configuration is. In the right panel we show the equilateral configuration $B(k, k ,k)$.}
\label{fig:comparisonansatze}
\end{figure}

We compare the ans\"atze for the higher loop corrections in Figure \ref{fig:comparisonansatze}. It is a zoom-in of Figure \ref{fig:bispectrumplot}, where we now show in addition the ansatz used in \cite{Baldauf:2016sjb}, for the two and three loop contribution to the bispectrum. Please note that in \cite{Assassi:2015jqa}, we only showed one of the two reducible two loop diagrams contributing to $B_{332}$. Therefore, the plots look different now, in particular in the squeezed configuration of the bispectrum. We see that in the squeezed configuration, the ansatz $B_{332}$ is an order of magnitude smaller than $E_b$. This explains why we have to multiply $B_{332}$ by a factor 10 in section \ref{subsec:integratingouttheoreticalerror} to get reliable results. Furthermore, we note that at redshift zero, $B_{332}$ allows one to go to higher $k_\text{max}$ in the squeezed configuration, whereas $E_b$ allows one to go further in the equilateral configuration (for $f_{NL}$ bigger than 10). This explains why using $B_{332}$ as an ansatz gives more optimistic results for local PNG, whereas $E_b$ gives more optimistic results for equilateral PNG (see section \ref{subsec:constraintsasfunctionofzmax}). Keep in mind also that the time dependence of the theoretical error terms is different from the signal, making the signal stronger at higher redshifts.

\begin{figure}[h]
\begin{subfigure}{0.44\textwidth}
\includegraphics[width=\linewidth]{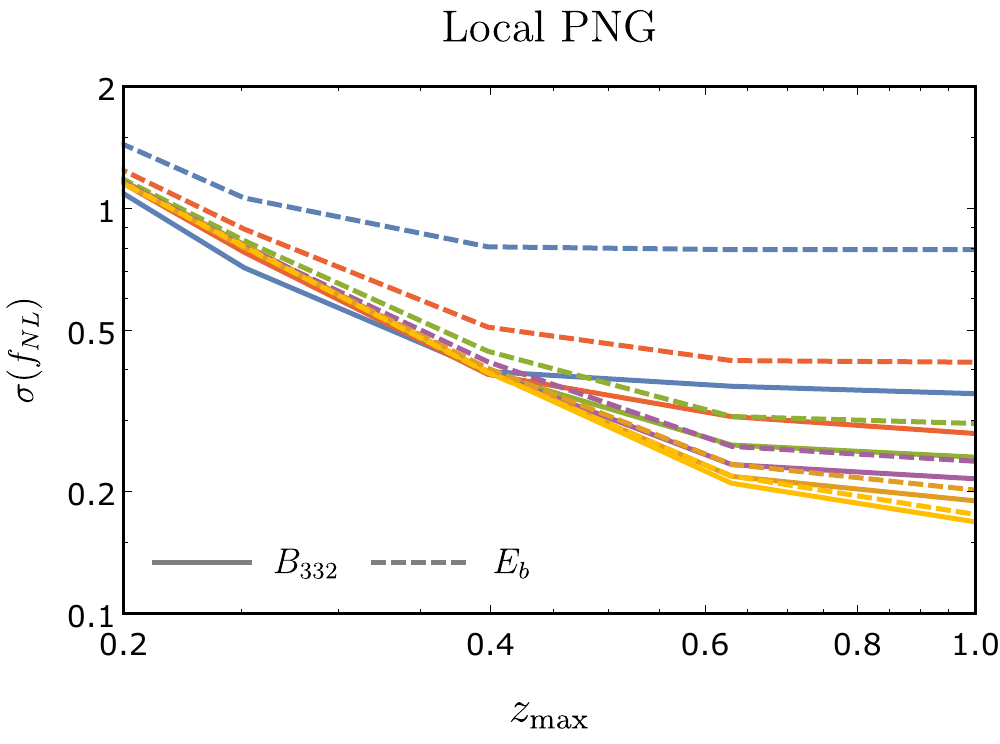} 
\end{subfigure}
\begin{subfigure}{0.44\textwidth}
\includegraphics[width=\linewidth]{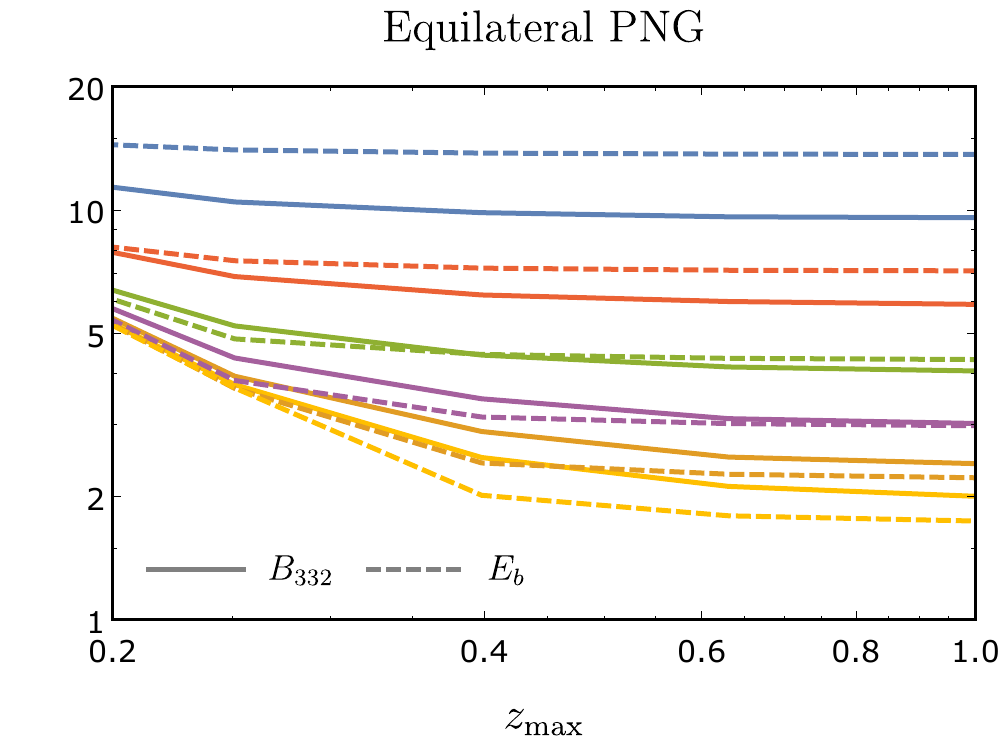} 
\end{subfigure}
\begin{subfigure}{0.10\textwidth}
\includegraphics[width=\linewidth]{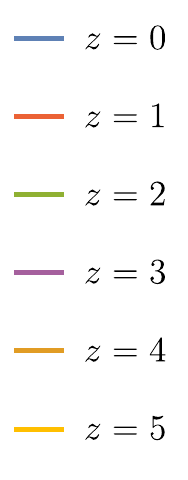}
\end{subfigure}
\caption{We show $\sigma(f_{NL})$ as function of $k_\text{max}$ using $B_{332}$ and $E_b$ as ansatz for the two loop corrections (solid and dashed lines). In the left panel we show the results for local PNG and in the right panel for equilateral PNG. The redshift takes values between $z=0$ and $z=5$. We use $k_\text{min} = 0.001 h \text{Mpc}^{-1}$ and $V=(2\pi/k_\text{min})^3$ at each redshift.}
\label{fig:freezeoutsigma}
\end{figure}

Next, we consider $\sigma(f_{NL})$ as function of $k_\text{max}$ at various redshifts in Figure \ref{fig:freezeoutsigma}. We do not include shotnoise, but we integrate out the theoretical error. The result for local PNG is shown in the left panel. We see that at redshift zero, the signal freezes out at some $k_\text{max} < 1 h \text{Mpc}^{-1}$. Furthermore, in agreement with what we expect from Figure \ref{fig:comparisonansatze}, we see that using $B_{332}$ as ansatz for the two loop corrections gives more optimistic results. More specifically, $\sigma(f_{NL})$ is a factor 5 smaller. At higher redshifts, we find that $\sigma(f_{NL})$ does not freeze when we reach $k_\text{max} = 1 h \text{Mpc}^{-1}$. We think this is due to the fact that we keep gaining information as we go to more and more squeezed configurations. This is also important for equilateral PNG, shown in the right panel, even though, $\sigma(f_{NL})$ does freeze out in this case. Interestingly, compared to scaling estimates for $k_\text{max}$ for equilateral PNG (see for instance \cite{Baldauf:2016sjb}) we find that we can go to much smaller scales than naively thought. The squeezed limit allows us to extract more information, also for equilateral PNG. The fact that $k_\text{max} = 1 h \text{Mpc}^{-1}$ is not large enough to ensure that $\sigma(f_{NL})$ is frozen when we ignore shotnoise explains the results we find in section \ref{subsec:constraintsasfunctionofzmax}. 

\section{Choice of binning and volume of the bins}
\label{app:choicebinning}
 In this appendix, we motivate the decision of section \ref{subsec:covariancematrix} to use logarithmic binning and exactly computed values of $V_{123}$.


\subsection{Exact computation of $V_{123}$}
We will now explain how to compute $V_{123}$ exactly and systematically by dividing all the bins in `interior' and `edge' bins. Moreover, the selection of bins is now determined by whether it contains at least some valid triangles instead of the usual selection rule that the central point should be a triangle. \\
Recall that $V_{123}$ is defined as 
\begin{equation}
V_{123}= \int_{\v{q}_1}\int_{\v{q}_2}\int_{\v{q}_3}\ \delta_D(\v{k}_1+\v{k}_2+\v{k}_3).
\end{equation}
We choose logarithmic binning, i.e. we have
\begin{equation}
q_i\equiv |\v{q}|_i \in \left[k_i e^{-\tfrac{1}{2}\Delta \ln k},k_i e^{\tfrac{1}{2}\Delta \ln k}\right].
\end{equation}
The integrand above only depends on the relative orientations of the vectors and their lengths. Fixing $\v{q}_1$ along the $\hat{z}$-direction and $\v{q}_2$ to be in the $(x,z)$-plane, their relative orientation is given by $\theta_{12}=\theta_2$. Now the lengths of these vectors, together with $c_{12}$, the cosine of $\theta_{12}$, completely determine $\v{q}_3$. The length of $\v{q}_3$ is then restricted to be in $\left[k_3 e^{-\tfrac{1}{2}\Delta \ln k},k_3 e^{\tfrac{1}{2}\Delta \ln k}\right]$, which means
\begin{equation}
c_{12} \in \left[-1, 1\right] \cap \left[\frac{\left(k_3 e^{-\tfrac{1}{2}\Delta \ln k}\right)^2-q_1^2-q_2^2}{2q_1q_2}, \frac{\left(k_3 e^{\tfrac{1}{2}\Delta \ln k}\right)^2-q_1^2-q_2^2}{2q_1q_2}\right],
\end{equation}
where $q_1$ and $q_2$ also take values within their bin. Then, if $\left[-1, 1\right]$ contains the range on the right for all values of $q_1$ and $q_2$, we are dealing with an `interior bin', and we get
\begin{equation}
\int dc_{12} dq_1 dq_2\ q_1^2q_2^2 = k_1^2 k_2^2 k_3^2 \sinh^3(\Delta \ln k).
\end{equation}
Finally, accounting for the fact that we fixed $\theta_1$, $\phi_{1,2}$ and the factors of $(2\pi)^3$ we find
\begin{equation}
V_{123} = \frac{1}{(2\pi)^9} 8 \pi^2 k_1^2 k_2^2 k_3^2 \sinh^3(\Delta \ln k).
\end{equation}
This approximation breaks down when the two ranges are partly overlapping, in which case we have an `edge bin'. \\
let us evaluate $V_{123}$ more precisely. We have seen that the integral simplifies to
\begin{equation}
V_{123}= \frac{ 8 \pi^2}{(2\pi)^9} \int_{q_1} \int_{q_2} \int_{c_{12}} \ q_1^2q_2^2
\end{equation}
where the $c_{12}$ is restricted to be in the range given above.
 This integral can therefore be rewritten as
\begin{align*}
V_{123} = \frac{ 8 \pi^2}{(2\pi)^9} \int_{q_1} \int_{q_2} \ \tfrac{1}{2} q_1q_2 \max\left[0, \left(\min\left[(q_1+q_2)^2,k_3^2 e^{\Delta \ln k}\right]-\max\left[(q_1-q_2)^2, k_3^2 e^{-\Delta \ln k}\right]\right)\right].
\end{align*}
In other words we integrate over the overlap 
\begin{equation}
\left[(q_1-q_2)^2,(q_1+q_2)^2\right]\cap\left[k_3^2 e^{-\Delta \ln k},k_3^2 e^{\Delta \ln k}\right].
\end{equation}
There are multiple possibilities:
\begin{itemize}
 \item The overlap is always zero.\\
 This happens whenever $|q_1 + q_2|_{\text{max}} \leq k_3 e^{-\tfrac{1}{2}\Delta \ln k}$ or $|q_1 - q_2|_{\text{min}} \geq k_3 e^{\tfrac{1}{2}\Delta \ln k}$. This means we should exclude the cases $k_3 \geq (k_1 + k_2) e^{\Delta \ln k}$ and $k_3 < k_2$. The latter is already excluded since we have $k_1 \leq k_2 \leq k_3$. The first leads to a constraint to select the bins, namely
\begin{equation}
k_3 < (k_1 + k_2) e^{\Delta \ln k}.
\end{equation}
\item The first range always contains the second range.\\
This happens when $|q_1 - q_2|_{\text{max}} \leq k_3 e^{-\tfrac{1}{2}\Delta \ln k}$ {\'a}nd $|q_1 + q_2|_{\text{min}} \geq k_3 e^{\tfrac{1}{2}\Delta \ln k}$. So we need both
\begin{equation}
\left\{ \begin{matrix} k_3 \geq k_2 e^{\Delta \ln k}-k_1 \\  k_3 \leq (k_1+k_2) e^{-\Delta \ln k} \end{matrix}\right. .
\end{equation}
In this case the volume takes the simple form 
\begin{equation}
V_{123} =  \frac{ 8 \pi^2}{(2\pi)^9} k_1^2 k_2^2 k_3^2 \sinh^3(\Delta \ln k).
\end{equation}
\item Any other type of overlap.\\
For the other cases we have to compute the actual volume of the bin. We will numerically perform the integral given above. This is when one of the two options below is satisfied
\begin{equation}
\left\{ \begin{matrix} k_3 < k_2 e^{\Delta \ln k}-k_1 \\  k_3 > (k_1+k_2) e^{-\Delta \ln k} \end{matrix}\right.
\end{equation}
Not for all these edge bins the central point has to be a triangle, since there are some cases for which $k_3 > k_1 + k_2$, considering the second inequality. Thus, we can either decide to define another point in the bin to represent the central triangle or we can merge these bins with one of their neighbors. In the first case a valid central triangle in the bin $(k_1, k_2, k_3)$ is given by
\begin{equation}
\left(k_1 e^{ \frac{k_3}{2(k_2+k_1)}}, k_2 e^{ \frac{k_3}{2(k_2+k_1)}}, k_3e^{-\frac{k_3}{2(k_2+k_1)}} \right)
\end{equation}
The other option is to merge (i.e. we add the volumes) the bin with one of its neighbors, which has the advantage that we never have to change the representing triangle of a given bin. For practical reasons, we choose this option. We implement this by merging each bin $\v{k}$ for which $k_3 > k_1 +k_2$ with the bin $\v{p}$ which has $p_1 = k_1$, $p_2 = k_2$ and $p_3$ the biggest value below or equal to $k_1 + k_2$.   
\end{itemize}

\subsection{Logarithmic versus linear binning}
\begin{figure}[!ht]
\centering
 \begin{subfigure}[b]{0.45\textwidth}
    \includegraphics[width=\textwidth]{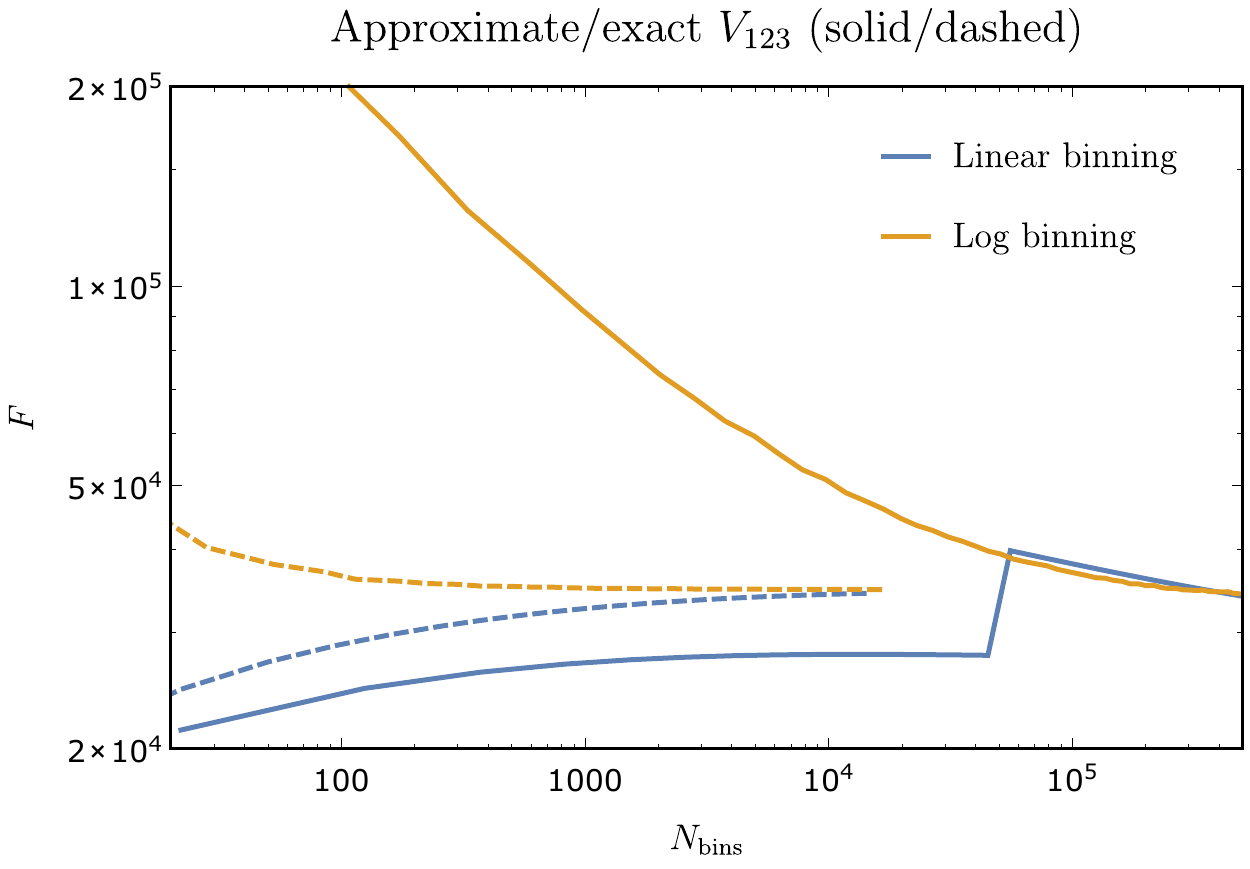}
\caption{Local function $f^{\text{loc}}$}
\label{fig:V123loc}
    \end{subfigure}
 \begin{subfigure}[b]{0.45\textwidth}
    \includegraphics[width=\textwidth]{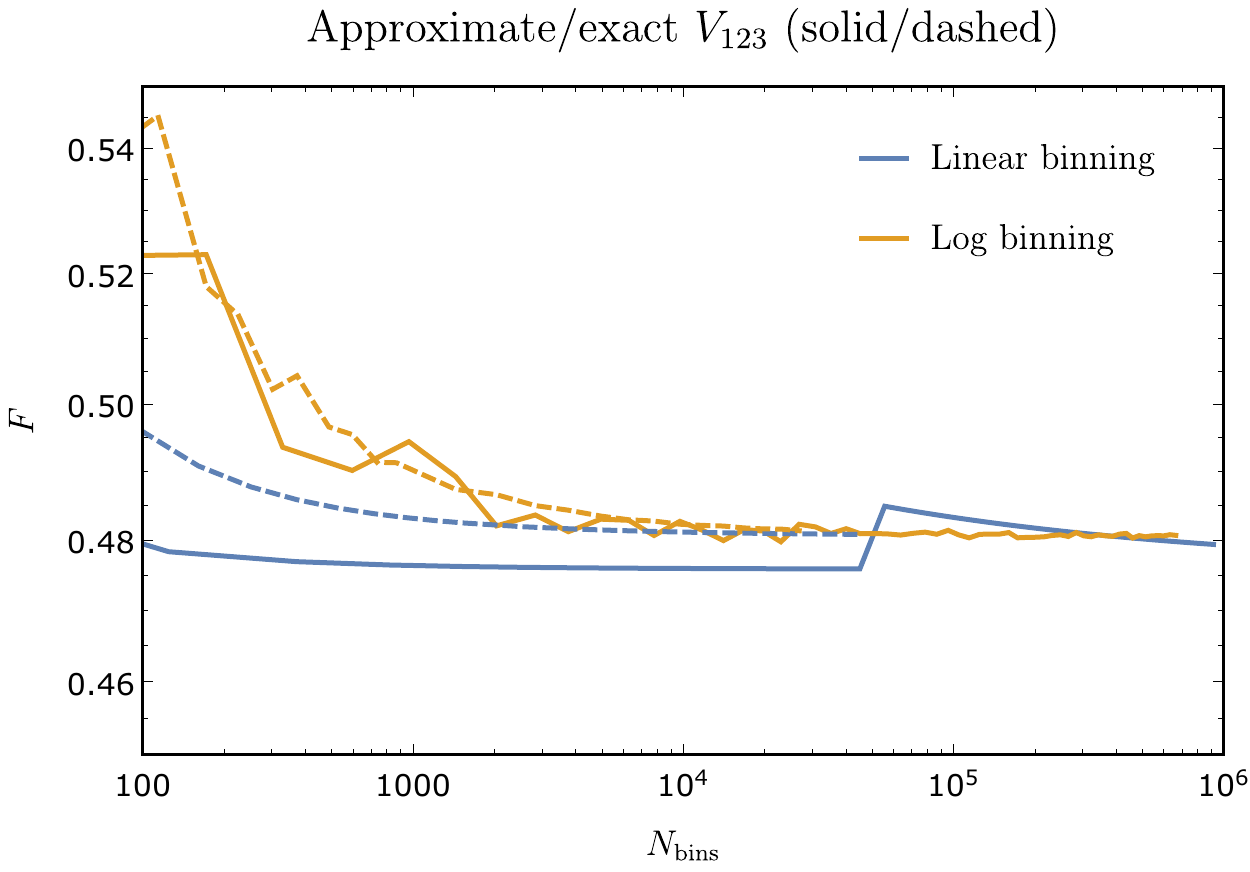}
\caption{Equilateral function $f^{\text{eq}}$}
\label{fig:V123eq}
    \end{subfigure}
  \caption{Computation of $F$ as function of the number of triangles $N_\text{bins}$ for (a) $f^{\text{loc}}$ and (b) $f^{\text{eq}}$ as given in equations \eqref{eqn:floc} and \eqref{eqn:feq}. We show the results for both the approximate (solid) and exact (dashed) expression for $V_{123}$. Moreover we denote the results from linear binning with a blue line and logarithmic binning with a orange line.}

\label{fig:V123}
\end{figure}

Let us now compare logarithmic binning with linear binning. We will show two examples of a computation of a Fisher matrix and show that the linear binning might cause problems. \\
We assume the following form for the Fisher matrix 
\begin{equation}
F = \sum_{k_1, k_2, k_3} f(k_1, k_2, k_3) \frac{V_{123}}{s_{123}},
\end{equation}
for some function $f(k_1, k_2, k_3)$. We will consider a `local'-type function $f^{\text{loc}}$ and an `equilateral'-type function $f^{\text{eq}}$. The local function corresponds to assuming the late time power spectrum scales as $P(k) \sim k^{-3}$, where $F$ represents the $(f_{NL}, f_{NL})$-component of the Fisher matrix for local PNG. Forgetting about the right normalization, this gives 
\begin{equation}
f^{\text{loc}}(k_1, k_2, k_3)  = \left(\frac{k_1^{3/2}}{k_2^{3/2}k_3^{3/2}}+\frac{k_2^{3/2}}{k_1^{3/2}k_3^{3/2}}+\frac{k_3^{3/2}}{k_2^{3/2}k_1^{3/2}}\right)^2.
\label{eqn:floc}
\end{equation}
Similarly, we can define a function that corresponds to equilateral PNG
\begin{equation}
f^{\text{eq}}(k_1, k_2, k_3)  = \frac{k_1 k_2 k_3}{(k_1 +k_2 +k_3)^6}.
\label{eqn:feq}
\end{equation}
We compute $F$ over a range $k \in [0.003, 0.5]\  h \text{Mpc}^{-1}$ for both logarithmic and linear bins and for both the approximate and exact computation of $V_{123}$. In Figure \ref{fig:V123} we plot $F$ as function of number of bins (number of triangles) considered. \\
First of all, in Figure \ref{fig:V123loc}, restricting ourselves to about $10.000$ bins, we see that if we use the approximate value for $V_{123}$ (the usual assumption), the linear bins seem to converge quickly to the asymptotic value. The logarithmic bins seem to converge much slower. However, as we keep increasing the number of triangles, suddenly the graph of the linear bins jumps to the graph of the logarithmic bins. This shows that if we would have trusted the linear binning for a smaller number of bins we would have gotten the wrong result. This is quite unexpected and alarming as it seems we cannot always trust linear binning! If we now change to the exact $V_{123}$ we see that both linear and logarithmic binning converge much faster and both to the same value. In fact, it turns out that they reach one percent agreement for about $15.000$ triangles. Let us try to understand why this happens. As we are summing over a function which peaks in the squeezed limit we do in fact get most signal from $k$-triplets which satisfy $k_3 \sim k_2  \gg k_1$. In particular the edge bins will contribute an important part to the final result. We know that  precisely for these bins the approximate value for $V_{123}$ does not work, which is probably why the results improve dramatically when using the exact $V_{123}$. Now one can still wonder why the linear binning performs so badly in this case. A reason might be that we are sampling the values for $k_1$ much better in case of logarithmic binning. However one could argue exactly the opposite, namely that linear binning samples the values of $k_2$ and $k_3$ much better. We have not found a convincing argument why linear binning fails, this remains an open question. As we are also studying the Fisher matrix for local PNG in our paper, we decided to stick to logarithmic binning. Even with the exact value of  $V_{123}$ the result converges quite slowly for the local function. In order to be within a couple of percent of the actual outcome of the Fisher analysis we need quite some triangles. For the analysis therefore we divide the $k$-axis over three logarithmic decades in 45 bins. By this we mean that if for instance each $k_i$ from the triplet can take values in the range $[0.001, 1]\  h \text{Mpc}^{-1}$, it can take one of the 45 logarithmically separated values.\\
We did the same analysis for the equilateral function. In Figure \ref{fig:V123eq} we see again a jump of the graph corresponding to linear binning. This time we do not expect to gain most signal from the edge bins. However, when we use the exact value for $V_{123}$ everything seems to be fine again. The jump takes place at a comparable value of $N_{\text{bins}}$. The graph of the logarithmic binning remains a bit wiggly, but we find one percent agreement between linear and logarithmic binning already for $1000$ triangles. 
For equilateral PNG we therefore divide the $k$-axis over three logarithmic decades in either 27 or 45 bins.

\newpage

\section{Table of parameters}

\label{app:tableparameters}

\begin{table}[h!]
\begin{tabular}{|>{\centering\arraybackslash}p{\dimexpr 0.09\textwidth}  | p{\dimexpr 0.26\textwidth-4\tabcolsep} | p{\dimexpr 0.535\textwidth-4\tabcolsep} |}
\hline
Symbol    & Relation &	Meaning 	\\ \hline\hline
$a$	  & 	& scale factor 			\\ \hline
$\tau$ & $a \, \d\tau=\d t$ & conformal time  \\ \hline
$\H$ &$\equiv d \ln (a)/ d\tau$ & conformal Hubble parameter  \\ \hline
$\H_0$ && present value of $\H$  \\ \hline
$\x$ &  & comoving coordinate  \\ \hline
$\k$ &  & momentum  \\ \hline
$\Omega_m$ &  & matter density in units of the critical density \\ \hline
$\Omega_\Lambda$ &  & dark energy density \\ \hline
$h$ &  & dimensionless Hubble constant  \\ \hline
$ \rho $ & & dark matter density  \\ \hline
$ \delta $ & $ \equiv \delta \rho/\rho$ & dark matter density contrast   \\ \hline
$ \theta $ & $ \equiv \partial_i v^i$ & velocity divergence   \\ \hline
$ \delta_{(n)}  $ & & density contrast in SPT at order $ n $ \\ \hline
$F_{n}$ & & kernel function in $\delta_{(n)}$ \\ \hline
$ P_{mn}$ &$\equiv \vev{\delta_{(m)} \delta_{(n)}}'$ & power spectrum in SPT  \\ \hline
$ B_{lmn}$ &$\equiv \vev{\delta_{(l)} \delta_{(m)} \delta_{(n)}}'$ & bispectrum in SPT  \\ \hline
$ \phi $ & & Newtonian potential \\ \hline
$ \Phi $ & $\bigtriangleup \Phi=\delta  $  & rescaled Newtonian potential   \\ \hline
$ \varphi $ & $\phi = T(k)\varphi$  & primordial potential  \\ \hline
$ \varphi_g $ &  & Gaussian primordial potential  \\ \hline
$ T(k)$ &   & transfer function \\ \hline
$ M(k)$ &   & transfer function in the Poisson equation  \\ \hline
$D_1$ &  & linear growth factor \\ \hline
$f$ &  $\equiv d \ln D_1/\ln a$ & growth rate  \\ \hline
$ P_{\varphi}$ & & primordial power spectrum \\ \hline
$ B_{\varphi}$ & & primordial bispectrum \\ \hline 
$n_s$ & & scalar spectral index  \\ \hline
$ \psi $ & & correlation in the initial conditions  \\ \hline
$ \Psi $ & $ \Psi(\x)\equiv \psi(\q(\x)) $ & Eulerian definition of $\psi$  \\ \hline
$\Delta$ & & scaling dimension in $ K_{\rm NL} $   \\ \hline
$ f_{\rm NL} $ & & amplitude of the primordial bispectrum  \\ \hline
 $d^2$ & $\equiv c_s^2 + f (c_{vis}^2+\hat c_{vis}^2)$  & parameter in $\tau_v$ \\ \hline
 $e_i$, $g$, $g_i$ &  & parameters in $\tau_v$ \\ \hline
 $\xi$ &  & parameter in $\delta^c_{(1)}$  \\ \hline
 $\gamma$ &  & parameter in $\delta^c_{(1)}$ \\ \hline
 $\epsilon_i$ &  &  parameter in $\delta^c_{(2)}$ \\ \hline
 $\gamma_i$ &  & parameter in $\delta^c_{(2)}$ \\ \hline
\end{tabular}
\end{table}


\newpage
\begin{table}[h!]
\begin{tabular}{|>{\centering\arraybackslash}p{\dimexpr 0.09\textwidth}  | p{\dimexpr 0.26\textwidth-4\tabcolsep} | p{\dimexpr 0.535\textwidth-4\tabcolsep} |}
\hline
Symbol    & Relation &	Meaning 	\\ \hline\hline
$ \mathcal{S}_{\alpha,\beta} $ & & SPT quadratic source terms \\ \hline
$ \tau_{\theta} $ & & EFT source in Euler equation \\ \hline
$\delta^c_{(n)}$ & & viscosity counterterm at order $n$ \\ \hline
$F^{c}_{n}$ & & kernel function in $\delta^c_{(n)}$ \\ \hline
$H^{c}_{n}$ & & kernel function in $\delta^c_{(n)}$ \\ \hline
$G_\delta$ &  & Green's function for $\delta$ \\ \hline
${\cal D}_\delta$ &  & evolution operator in the fluid equation  \\ \hline
$P_{1\psi}$ & $\equiv \vev{\delta_{(1)} \psi}'$  & correlation of $\delta_{(1)}$ and $\psi$ \\ \hline
$B_{\rm SPT}^{\rm G}$ & & Gaussian SPT contributions to $B_\delta$\\ \hline
$B_{\rm SPT}^{\rm NG}$ & & non-Gaussian SPT contributions to $B_\delta$   \\ \hline
$B_{\rm EFT}^{\rm G}$ & & sum of Gaussian EFT counterterms  \\ \hline
$B_{\rm EFT}^{\rm NG}$ & & sum of non-Gaussian EFT counterterms     \\ \hline
\end{tabular}
\end{table}

\newpage


\bibliographystyle{hieeetr}
\bibliography{bibfile}{}

\begin{thebibliography}{10}

\bibitem{Dalal:2007cu}
N.~Dalal, O.~Dore, D.~Huterer, and A.~Shirokov, ``{The imprints of primordial
  non-gaussianities on large-scale structure: scale dependent bias and
  abundance of virialized objects},'' {\em Phys. Rev.}, vol.~D77, p.~123514,
  2008, 0710.4560.

\bibitem{Dore:2014cca}
O.~Doré {\em et~al.}, ``{Cosmology with the SPHEREX All-Sky Spectral
  Survey},'' 2014, 1412.4872.

\bibitem{Tellarini:2016sgp}
M.~Tellarini, A.~J. Ross, G.~Tasinato, and D.~Wands, ``{Galaxy bispectrum,
  primordial non-Gaussianity and redshift space distortions},'' 2016,
  1603.06814.

\bibitem{Baumann:2010tm}
D.~Baumann, A.~Nicolis, L.~Senatore, and M.~Zaldarriaga, ``{Cosmological
  Non-Linearities as an Effective Fluid},'' {\em JCAP}, vol.~1207, p.~051,
  2012, 1004.2488.

\bibitem{Bernardeau:2001qr}
F.~Bernardeau, S.~Colombi, E.~Gaztanaga, and R.~Scoccimarro, ``{Large scale
  structure of the universe and cosmological perturbation theory},'' {\em
  Phys.Rept.}, vol.~367, pp.~1--248, 2002, astro-ph/0112551.

\bibitem{Scoccimarro:2003wn}
R.~Scoccimarro, E.~Sefusatti, and M.~Zaldarriaga, ``{Probing primordial
  non-Gaussianity with large - scale structure},'' {\em Phys. Rev.}, vol.~D69,
  p.~103513, 2004, astro-ph/0312286.

\bibitem{Sefusatti:2007ih}
E.~Sefusatti and E.~Komatsu, ``{The bispectrum of galaxies from high-redshift
  galaxy surveys: Primordial non-Gaussianity and non-linear galaxy bias},''
  {\em Phys. Rev.}, vol.~D76, p.~083004, 2007, 0705.0343.

\bibitem{Baldauf:2010vn}
T.~Baldauf, U.~Seljak, and L.~Senatore, ``{Primordial non-Gaussianity in the
  Bispectrum of the Halo Density Field},'' {\em JCAP}, vol.~1104, p.~006, 2011,
  1011.1513.

\bibitem{Jeong:2009vd}
D.~Jeong and E.~Komatsu, ``{Primordial non-Gaussianity, scale-dependent bias,
  and the bispectrum of galaxies},'' {\em Astrophys. J.}, vol.~703,
  pp.~1230--1248, 2009, 0904.0497.

\bibitem{Tasinato:2013vna}
G.~Tasinato, M.~Tellarini, A.~J. Ross, and D.~Wands, ``{Primordial
  non-Gaussianity in the bispectra of large-scale structure},'' {\em JCAP},
  vol.~1403, p.~032, 2014, 1310.7482.

\bibitem{Roth:2012yy}
N.~Roth and C.~Porciani, ``{Can we really measure $f_{NL}$ from the galaxy
  power spectrum?},'' {\em Mon. Not. Roy. Astron. Soc.}, vol.~425,
  pp.~L81--L85, 2012, 1205.3165.

\bibitem{Baldauf:2014qfa}
T.~Baldauf, L.~Mercolli, M.~Mirbabayi, and E.~Pajer, ``{The Bispectrum in the
  Effective Field Theory of Large Scale Structure},'' {\em JCAP}, vol.~1505,
  no.~05, p.~007, 2015, 1406.4135.

\bibitem{Angulo:2014tfa}
R.~E. Angulo, S.~Foreman, M.~Schmittfull, and L.~Senatore, ``{The One-Loop
  Matter Bispectrum in the Effective Field Theory of Large Scale Structures},''
  {\em JCAP}, vol.~1510, no.~10, p.~039, 2015, 1406.4143.

\bibitem{Assassi:2015jqa}
V.~Assassi, D.~Baumann, E.~Pajer, Y.~Welling, and D.~van~der Woude,
  ``{Effective Theory of Large-Scale Structure with Primordial
  Non-Gaussianity},'' 2015, 1505.06668.

\bibitem{Baldauf:2016sjb}
T.~Baldauf, M.~Mirbabayi, M.~Simonović, and M.~Zaldarriaga, ``{LSS constraints
  with controlled theoretical uncertainties},'' 2016, 1602.00674.

\bibitem{Ade:2015ava}
P.~A.~R. Ade {\em et~al.}, ``{Planck 2015 results. XVII. Constraints on
  primordial non-Gaussianity},'' 2015, 1502.01592.

\bibitem{Baumann:2014cja}
D.~Baumann, D.~Green, and R.~A. Porto, ``{B-modes and the Nature of
  Inflation},'' {\em JCAP}, vol.~1501, no.~01, p.~016, 2015, 1407.2621.

\bibitem{Alvarez:2014vva}
M.~Alvarez {\em et~al.}, ``{Testing Inflation with Large Scale Structure:
  Connecting Hopes with Reality},'' 2014, 1412.4671.

\bibitem{Lewis:2002ah}
A.~Lewis and S.~Bridle, ``{Cosmological parameters from CMB and other data: a
  Monte- Carlo approach},'' {\em Phys. Rev.}, vol.~D66, p.~103511, 2002,
  astro-ph/0205436.

\bibitem{Carrasco:2012cv}
J.~J.~M. Carrasco, M.~P. Hertzberg, and L.~Senatore, ``{The Effective Field
  Theory of Cosmological Large Scale Structures},'' {\em JHEP}, vol.~09,
  p.~082, 2012, 1206.2926.

\bibitem{Mercolli:2013bsa}
L.~Mercolli and E.~Pajer, ``{On the velocity in the Effective Field Theory of
  Large Scale Structures},'' {\em JCAP}, vol.~1403, p.~006, 2014, 1307.3220.

\bibitem{Bartolo:2001cw}
N.~Bartolo, S.~Matarrese, and A.~Riotto, ``{Nongaussianity from inflation},''
  {\em Phys. Rev.}, vol.~D65, p.~103505, 2002, hep-ph/0112261.

\bibitem{Creminelli:2003iq}
P.~Creminelli, ``{On non-Gaussianities in single-field inflation},'' {\em
  JCAP}, vol.~0310, p.~003, 2003, astro-ph/0306122.

\bibitem{Chen:2009zp}
X.~Chen and Y.~Wang, ``{Quasi-Single Field Inflation and Non-Gaussianities},''
  {\em JCAP}, vol.~1004, p.~027, 2010, 0911.3380.

\bibitem{Pajer:2013jj}
E.~Pajer and M.~Zaldarriaga, ``{On the Renormalization of the Effective Field
  Theory of Large Scale Structures},'' {\em JCAP}, vol.~1308, p.~037, 2013,
  1301.7182.

\bibitem{Sefusatti:2006pa}
E.~Sefusatti, M.~Crocce, S.~Pueblas, and R.~Scoccimarro, ``{Cosmology and the
  Bispectrum},'' {\em Phys. Rev.}, vol.~D74, p.~023522, 2006, astro-ph/0604505.

\bibitem{Baldauf:2015aha}
T.~Baldauf, L.~Mercolli, and M.~Zaldarriaga, ``{Effective field theory of large
  scale structure at two loops: The apparent scale dependence of the speed of
  sound},'' {\em Phys. Rev.}, vol.~D92, no.~12, p.~123007, 2015, 1507.02256.

\bibitem{Foreman:2015lca}
S.~Foreman, H.~Perrier, and L.~Senatore, ``{Precision Comparison of the Power
  Spectrum in the EFTofLSS with Simulations},'' 2015, 1507.05326.

\bibitem{Carrasco:2013mua}
J.~J.~M. Carrasco, S.~Foreman, D.~Green, and L.~Senatore, ``{The Effective
  Field Theory of Large Scale Structures at Two Loops},'' {\em JCAP},
  vol.~1407, p.~057, 2014, 1310.0464.

\bibitem{Heavens:2009nx}
A.~Heavens, ``{Statistical techniques in cosmology},'' 2009, 0906.0664.

\bibitem{Pajer:tobepublished}
E.~Pajer and Y.~Welling, ``{Time dependence in correlation functions},''

\bibitem{Scoccimarro:1997st}
R.~Scoccimarro, S.~Colombi, J.~N. Fry, J.~A. Frieman, E.~Hivon, and A.~Melott,
  ``{Nonlinear evolution of the bispectrum of cosmological perturbations},''
  {\em Astrophys. J.}, vol.~496, p.~586, 1998, astro-ph/9704075.

\bibitem{randomfield_notes}
C.~E. Powell, ``{Generating Realisations of Stationary Gaussian Random Fields
  by Circulant Embedding}.''
  \url{https://www.nag.co.uk/doc/techrep/pdf/tr1_14.pdf}.

\bibitem{Laureijs:2011gra}
R.~Laureijs {\em et~al.}, ``{Euclid Definition Study Report},'' 2011,
  1110.3193.

\bibitem{Dawson:2012va}
K.~S. Dawson {\em et~al.}, ``{The Baryon Oscillation Spectroscopic Survey of
  SDSS-III},'' {\em Astron. J.}, vol.~145, p.~10, 2013, 1208.0022.

\bibitem{Dawson:2015wdb}
K.~S. Dawson {\em et~al.}, ``{The SDSS-IV extended Baryon Oscillation
  Spectroscopic Survey: Overview and Early Data},'' {\em Astron. J.}, vol.~151,
  p.~44, 2016, 1508.04473.

\bibitem{Levi:2013gra}
M.~Levi {\em et~al.}, ``{The DESI Experiment, a whitepaper for Snowmass
  2013},'' 2013, 1308.0847.

\bibitem{Garny:2015oya}
M.~Garny, T.~Konstandin, R.~A. Porto, and L.~Sagunski, ``{On the Soft Limit of
  the Large Scale Structure Power Spectrum: UV Dependence},'' {\em JCAP},
  vol.~1511, no.~11, p.~032, 2015, 1508.06306.

\bibitem{Sefusatti:2004xz}
E.~Sefusatti and R.~Scoccimarro, ``{Galaxy bias and halo-occupation numbers
  from large-scale clustering},'' {\em Phys. Rev.}, vol.~D71, p.~063001, 2005,
  astro-ph/0412626.

\bibitem{Seljak:2008xr}
U.~Seljak, ``{Extracting primordial non-gaussianity without cosmic variance},''
  {\em Phys. Rev. Lett.}, vol.~102, p.~021302, 2009, 0807.1770.

\bibitem{Verde:2001pf}
L.~Verde and A.~F. Heavens, ``{On the trispectrum as a Gaussian test for
  cosmology},'' {\em Astrophys. J.}, vol.~553, p.~14, 2001, astro-ph/0101143.

\bibitem{Bertolini:2016bmt}
D.~Bertolini, K.~Schutz, M.~P. Solon, and K.~M. Zurek, ``{The Trispectrum in
  the Effective Field Theory of Large Scale Structure},'' 2016, 1604.01770.

\bibitem{Cooray:2006km}
A.~Cooray, ``{21-cm Background Anisotropies Can Discern Primordial
  Non-Gaussianity},'' {\em Phys. Rev. Lett.}, vol.~97, p.~261301, 2006,
  astro-ph/0610257.

\bibitem{Heitmann:2008eq}
K.~Heitmann, M.~White, C.~Wagner, S.~Habib, and D.~Higdon, ``{The Coyote
  Universe I: Precision Determination of the Nonlinear Matter Power
  Spectrum},'' {\em Astrophys. J.}, vol.~715, pp.~104--121, 2010, 0812.1052.

\bibitem{Schneider:2015yka}
A.~Schneider, R.~Teyssier, D.~Potter, J.~Stadel, J.~Onions, D.~S. Reed, R.~E.
  Smith, V.~Springel, F.~R. Pearce, and R.~Scoccimarro, ``{Matter power
  spectrum and the challenge of percent accuracy},'' {\em JCAP}, vol.~1604,
  no.~04, p.~047, 2016, 1503.05920.

\end{thebibliography}

\end{document}